\documentclass[fleqn,usenatbib]{mnras}

\usepackage{newtxtext,newtxmath}
\usepackage{color,ulem}

\usepackage[T1]{fontenc}
\usepackage{ae,aecompl}

\usepackage{graphicx}	
\usepackage{amsmath}	
\usepackage{amssymb}	

\title[The VISCACHA survey - I. Overview and First Results]{The VISCACHA survey - I. Overview and First Results\thanks{Based on observations obtained at the Southern Astrophysical Research (SOAR) telescope (projects SO2015A-013, SO2015B-008, SO2016B-015, SO2016B-018, SO2017B-014), which is a joint project of the Minist\'erio da Ci\^encia, Tecnologia, e Inova\c c\~ao (MCTI) da Rep\'ublica Federativa do Brasil, the U.S. National Optical Astronomy Observatory (NOAO), the University of North Carolina at Chapel Hill (UNC), and Michigan State University (MSU).} }

\author[F. Maia et al.]{Francisco F.S. Maia$^{1}$$^,$\thanks{E-mail: ffsmaia@usp.br}, 
Bruno Dias$^{2,3}$,
Jo\~ao F.C. Santos Jr.$^{4}$,
Leandro de O. Kerber$^{1,5}$,
\newauthor
Eduardo Bica$^{6}$,
Andr\'es E. Piatti$^{7,8}$,
Beatriz Barbuy$^{1}$,
Bruno Quint$^{9,10}$,
Luciano Fraga$^{11}$,
\newauthor
David Sanmartim$^{10}$,
Mateus S. Angelo$^{12}$,
Jose A. Hernandez-Jimenez$^{1,3}$,
\newauthor
Orlando J. Katime Santrich$^{5}$,
Raphael A. P. Oliveira$^{1}$,
Angeles P\'erez-Villegas$^{1}$
\newauthor
Stefano O. Souza$^{1}$,
Rodrigo G. Vieira$^{1}$,
Pieter Westera$^{13}$
\\
$^{1}$Universidade de S\~ao Paulo, IAG, Rua do Mat\~ao 1226, 05508-090, Brazil \\
$^{2}$European Southern Observatory, Alonso de C\'ordova 3107, Vitacura 19001, Chile \\
$^{3}$Departamento de Ciencias F\'isicas, Universidad Andres Bello, Fernandez Concha 700, Las Condes, Santiago, Chile \\
$^{4}$Universidade Federal de Minas Gerais, ICEx, Av. Ant\^onio Carlos 6627, 31270-901, Brazil \\
$^{5}$Universidade Estadual de Santa Cruz, Depto. de Ci\^encias Exatas e Tecnol\'ogicas, Rodovia Jorge Amado km 16, 45662-900, Brazil \\
$^{6}$Universidade Federal do Rio Grande do Sul, Instituto de F\'isica, Av. Bento Gon\c calves 9500, 91501-970 ,Brazil \\ 
$^{7}$Consejo Nacional de Investigaciones  Cient\'{\i}ficas y T\'ecnicas, Av. Rivadavia 1917, 
C1033AAJ, Buenos Aires, Argentina\\
$^{8}$Observatorio Astron\'omico de C\'ordoba, Laprida 854, 5000, C\'ordoba, Argentina\\
$^{9}$SOAR Telescope, c/o AURA - Casilla 603, La Serena, Chile\\
$^{10}$Gemini Observatory, c/o AURA - Casilla 603, La Serena, Chile\\
$^{11}$Laborat\'orio Nacional de Astrof\'isica, Rua Estados Unidos 154, 37504-364, Brazil\\
$^{12}$Centro Federal de Educa\c c\~ao Tecnol\'ogica de Minas Gerais, Av. Monsenhor Luiz de Gonzaga, 103, 37250-000, Brazil \\
$^{13}$Universidade Federal do ABC, Centro de Ci\^encias Naturais e Humanas, Avenida dos Estados, 5001, 09210-580, Brazil}

\date{Accepted XXX. Received YYY; in original form ZZZ}

\pubyear{2018}

\begin{document}
\label{firstpage}
\pagerange{\pageref{firstpage}--\pageref{lastpage}}
\maketitle

\begin{abstract}
The VISCACHA (VIsible Soar photometry of star Clusters in tApii and Coxi HuguA) Survey is an ongoing project based on deep photometric observations of Magellanic Cloud star clusters, collected using the SOuthern Astrophysical Research (SOAR) telescope together with the SOAR Adaptive Module Imager. Since 2015 more than 200 hours of telescope time were used to observe about 130 stellar clusters, most of them with low mass (M < 10$^4$ M$\odot$) and/or located in the outermost regions of the Large Magellanic Cloud and the Small Magellanic Cloud. With this high quality data set, we homogeneously determine physical properties from statistical analysis of colour-magnitude diagrams, radial density profiles, luminosity functions and mass functions. Ages, metallicities, reddening, distances, present-day masses, mass function slopes and structural parameters for these clusters are derived and used as a proxy to investigate the interplay between the environment in the Magellanic Clouds and the evolution of such systems. 
In this first paper we present the VISCACHA Survey and its initial results, concerning the SMC clusters 
AM3, K37, HW20 and NGC796 and the LMC ones KMHK228, OHSC3, SL576, SL61 and SL897, chosen to compose a representative subset of our cluster sample. The project's long term goals and legacy to the community are also addressed.

\end{abstract}




\section{Introduction}

The gravitational disturbances resulting from interactions between the Large Magellanic Cloud (LMC) 
and the Small Magellanic Cloud (SMC) and between these galaxies and the Milky Way (MW) are probably imprinted 
on their star formation histories, as strong tidal effects are known to trigger star 
formation across dwarf galaxies \citep{Kennicutt:1996}. Gas dynamics simulations of galaxy collision and 
merging have shown that the properties of tidally induced features such as the Magellanic Stream and 
Bridge can be used to gather information about the collision processes and to infer the history of 
the colliding galaxies \citep{Olson:1990}. When applied to model the Magellanic System, 
present-day simulations have been able to reproduce several of the observed features of the interacting 
galaxies such as shape, mass and the induced star formation rates. However, it is still not clear 
whether the Magellanic Clouds are on their first passage, or if they have been orbiting the MW 
for a longer time \citep[e.g.][]{Mastropietro:2005, besla+07, Diaz:2012, kallivayalil+13}.

\cite{putman+98} confirmed the existence of the Leading Arm, which is the counterpart of 
the trailing Magellanic Stream. The existence of both gas structures most likely has a tidal 
origin. Because of that, it is also expected that the Magellanic Stream, the Leading Arm and 
the Magellanic Bridge should have a stellar counterpart of the tidal effects within the 
Magellanic System \citep[e.g.][]{Diaz:2012}. Besides, the close encounters among SMC, LMC 
and the MW should trigger star formation at specific epochs \citep{Harris:09}, presumably 
imprinted in the age and metallicity distribution of field and cluster stars. 

In the context of interacting galaxies, it is well known that the tidal forces have a direct 
impact over the dynamical evolution and dissolution of stellar clusters and that the intensity 
of these effects typically scale with galactocentric distances \citep{Bastian:2008}. The outcome 
of these gravitational stresses imprinted on the stellar content of these systems can be diagnosed 
by means of the clusters structural parameters \citep{Werchan:2011,Miholics:2014} and mass 
distribution \citep{Glatt:2011}. In a similar fashion, the effects of the galactic 
gravitational interactions in the Magellanic System should also be seen in the structural, 
kinematical and spatial properties of their stellar clusters, particularly on those on the 
peripheries of the LMC and SMC. Whether or not they are affected by significant disruption during 
their lifetime is an open question and subject of current debate 
\citep{Casetti-Dinescu:2014}. Comparing these properties at different locations across the 
Magellanic Clouds is the key to unveiling the role of tidal forces over the cluster's evolution 
and to map crucial LMC and SMC properties at projected distances usually not covered
by previous surveys. Given the complexity of the cluster dynamics in the outer LMC
and SMC, additional kinematic information might be required (e.g. radial velocities) to 
constraint their orbits and address the issue of possible cluster migration, both in a galactic
context and between the Clouds, as such behaviour has already been seen in their stellar content \citep{Olsen:11}. 

Fortunately, most of the star clusters fundamental parameters such as age, metallicity, distance, 
reddening and structural parameters can be inferred from photometry using well established 
methodologies such as simple stellar population models, N-body simulations, stellar evolution models
and colour-magnitude diagrams (CMDs). These parameters, in turn, can be used to probe the 3D
structure of the Magellanic Clouds and Bridge, to sample local stellar populations and also to map
their chemical gradients and evolutionary history. When combined with proper motions from
\citet{Gaia:2018} and with radial velocities and metallicities from a spectroscopic follow-up they
can provide a wealth of additional information such as the radial metallicity gradients, still 
under discussion for these galaxies, the internal dynamical status and evolutionary timescales of
the clusters and their 3D motions and orbits, which constrain the mass of the LMC and SMC.

Some efforts have been made to collect heterogeneous data from the literature and study the topics 
above (e.g. \citealp{Pietrzynski+00}, \citealp{rafelski+05}, \citealp{glatt+10}, \citealp{piatti+11}, 
\citealp{palma+16}, \citealp{perren+17}, \citealp{parisi+09,parisi+14,parisi+15}, 
\citealp{Dias:2014,Dias:2016}, \citealp{nayak+16}, \citealp{pieres+16} etc). However, the dispersion 
in the parameters due to different data qualities, analysis techniques and photometric bands used 
do not put 
hard constraints on the history of the SMC and LMC star cluster populations. This is usually one of 
the most compelling arguments to carry out a survey in the Magellanic Clouds. 


After \citet{putman+98}, the investigation of some of these subjects has greatly benefited from several
photometric surveys, some dedicated exclusively to the Magellanic Clouds.
We describe the main surveys covering the Magellanic Clouds in Table \ref{tab:surveys}. It can be 
seen that they complement each other in terms of sky coverage, filters, photometric depth, and 
spatial resolution. All of them give preference to large sky coverage over photometric depth at 
the expense of good photometry of low-mass stars in star clusters. The Hubble Space Telescope 
(HST) is suitable to explore this niche, but only for a few selected massive clusters given the 
time limitations implied in observing hundreds of low-mass ones.

Our VISCACHA (VIsible Soar photometry of star Clusters in tApii and Coxi HuguA\footnote{LMC
and SMC names in the Tupi-Guarani language}) survey exploits the unique niche of deep photometry 
of star clusters and a good spatial resolution throughout the LMC, SMC, and Magellanic Bridge. 
In order to observe a large sample, including the numerous low-mass clusters we 
need large access to a suitable ground-based facility. 
These conditions are met at the 4.1-m
Southern Astrophysical Research (SOAR) telescope combined with the
SOAR Telescope Adaptive Module (SAM) using ground-layer adaptive optics (GLAO).
The VISCACHA team can 
access a large fraction of nights at SOAR (Brazil: 31\%, Chile: 10\%) to cover hundreds of star clusters 
in the Magellanic System during a relatively short period, with improved photometric depth and spatial 
resolution. This combination allows us to generate precise CMDs especially for the oldest, compact 
clusters immersed in dense fields, which is not possible with large surveys. A more detailed description
of the survey is given in Section \ref{sec:viscacha}.

\begin{table*}
\centering
\scriptsize
\caption{Summary of photometric surveys covering the Magellanic System. Future surveys (LSST, Euclid), the ones with marginal cover of the Magellanic System or that are photometrically shallow are not listed (DSS, 2MASS, Pan-STARRS, MagLiteS, ATLAS, MAGIC). Spectroscopic surveys are not listed either (APOGEE-2, Local Volume Mapper, {\it Gaia}, 4MOST).}
\label{tab:surveys}
\setlength{\tabcolsep}{5pt}
\begin{tabular}{p{1.7cm} p{1.3cm} p{2.5cm} p{1.1cm} p{0.9cm} p{1cm} p{0.8cm} p{2cm} p{2.5cm} p{0.8cm}}
\noalign{\smallskip}
\hline
\noalign{\smallskip}
survey & period & telescope/ & typical & filters & mag.lim. & scale & total sky & main goals & main\\
 (PI)  & (observ.) & instrument & seeing &       &          & ($\arcsec$/px) & coverage &  & refs.\\
\noalign{\smallskip}
\hline \hline
\noalign{\smallskip}
MCPS (Zaritsky) & 1996-1999 (+2001) & 1m Swope @ LCO, Great circle camera (drift-scan) & 1.2-1.8$\arcsec$ & UBVI & $V<$21$^a$ & 0.7 & 64$\deg^2$ (LMC) 18$\deg^2$ (SMC) & field SFH SMC/LMC, cluster census, reddening map & 1, 2, 3, 4\\
\noalign{\smallskip}
\hline
\noalign{\smallskip}
VMC (Cioni) & 2009-2018 & 4m VISTA @ ESO, VIRCAM (1$^{\circ}$x1$^{\circ}$) & 0.8-1.2$\arcsec$ & YJK$_{\rm s}$ & $J<$21.9$^a$ & 0.34 & 116$\deg^2$ (LMC) 45$\deg^2$ (SMC) 20$\deg^2$ (Bridge) 3$\deg^2$ (Stream) & spatially-resolved SFH, 3D structure, stellar variability & 5, 6, 7, 8, 9\\
\noalign{\smallskip}
\hline
\noalign{\smallskip}
OGLE-IV (Udalski) & 2010-2014 & 1.3m Warsaw @ LCO ($\diameter\sim$ 1.5$^{\circ}$) & 1.0-2.0$\arcsec$ & (B)VI & $I<$21.7 ($I<$20.5$^b$) & 0.26 & 670$\deg^2$ (SMC, LMC, Bridge) & Stellar variability & 10, 11, 12, 13 \\
\noalign{\smallskip}
\hline
\noalign{\smallskip}
STEP (Ripepi) & 2011+ & 2.6m VST @ ESO OmegaCAM (1$\deg^2$) & 1.0-1.5$\arcsec$ & griH$\alpha$ & $g<$23.5$^a$ & 0.21 & 74$\deg^2$ (SMC main body) 30$\deg^2$ (Bridge) 2$\deg^2$ (Stream) & visible complement of VMC, SFH of SMC down to oldest populations & 14 \\
\noalign{\smallskip}
\hline
\noalign{\smallskip}
SMASH (Nidever) & 2013-2016 & 4m Blanco @ CTIO DECam, NOAO (3$\deg^2$) & 1.0-1.2$\arcsec$ & ugriz & $g<$22.5$^a$ & 0.27 & 480$\deg^2$ (Leading arm, SMC, LMC cores) & stellar counterpart of Leading Arm, spatially resolved SFH LMC/SMC & 15, 16, 17, 18 \\
\noalign{\smallskip}
\hline
\noalign{\smallskip}
DES (Frieman$^c$) & 2013-2018 & 4m Blanco @ CTIO DECam, NOAO (3$\deg^2$)& 0.8-1.2$\arcsec$  & grizY & $g<$23.7$^b$ & 0.27 & 5000$\deg^2$ (Stream plus large area unrelated to SMC/LMC) & Magellanic Stream, tidal dwarf galaxies & 19, 20, 21\\
\noalign{\smallskip}
\hline
\noalign{\smallskip}
{\it Gaia} (Prusti$^d$) & 2013-2019 & 1.49m$\times$0.54m ($\times 2$) {\it Gaia} @ ESA (space) & $>0.1\arcsec^{ e}$ & G (blue, red photometer) & $G<$20.7$^f$ & 0.06 $\times$ 0.18 & all sky & proper motion of brightest stars, stellar variability, SFH  & 22, 23, 24, 25 \\
\noalign{\smallskip}
\hline
\noalign{\smallskip}
Skymapper (Da Costa) & 2014-2020 & 1.35m SSO @ ANU (2.4$\times$2.3$\deg^2$) & 1.2-1.8$\arcsec$ & uvgriz & $g<$18$^f$ (g$<22^g$) & $\sim$0.5 & all Southern sky & outskirts of LMC/SMC, origin of Stream at the Bridge & 26 \\
\noalign{\smallskip}
\hline
\noalign{\bigskip}
{\bf VISCACHA} (Dias) & 2015+ & 4.1m SOAR @ Cerro Pachon / SAMI with GLAO ($3\arcmin\times3\arcmin$) & 0.8-1.0$\arcsec$ {\bf (AO$\sim$0.5$\arcsec$)} & (B)VI & $V<24^a$ & {\bf 0.09} (binned) & only star clusters & star clusters of all ages, LMC, SMC, bridge, tidal effects on clusters, precise CMDs & 27, 28, 29, 30 \\ 
\noalign{\bigskip}
\hline
\noalign{\smallskip}
\end{tabular}
\begin{flushleft}
Based on the presentation by M.R. Cioni at ESO2020 workshop in 2015, updated with more surveys and details: \url{https://www.eso.org/sci/meetings/2015/eso-2020/program.html}
(a) Completeness at 50\% using artificial star tests in the crowded regions.
(b) Completeness at 95-100\%.
(c) Director.
(d) Project scientist.
(e) {\it Gaia} is able to separate two point sources that are $>0.1\arcsec$ apart, but this is only a reference, it cannot be directly compared with ground-based telescope FWHM or resolving power. Another parameter is that {\it Gaia} can resolve stars up to a density of 0.25 star/$\arcsec^2$.
(f) Hard limit, large uncertainty, low completeness.
(g) DR1 only contains shallow survey. The full survey is expected to reach 4 mag deeper.
(1) \cite{zaritsky+96};  
(2) \cite{Zaritsky:1997}; 
(3) \cite{Zaritsky:2002}; 
(4) \cite{Zaritsky:2004}; 
(5) \cite{Cioni:2011}; 
(6) \cite{piatti+15vmc}; 
(7) \cite{subramanian+17}; 
(8) \cite{niederhofer+18}; 
(9) \cite{rubele+18}; 
(10) \cite{udalski+15}; 
(11) \cite{skowron+14}; 
(12) \cite{Jacyszyn-Dobrzeniecka+16}; 
(13) \cite{Sitek+17}; 
(14) \cite{ripepi+14}
(15) \cite{Nidever+2017} 
(16) \cite{nidever+18} 
(17) \cite{choi+18a} 
(18) \cite{choi+18b} 
(19) \cite{abbott+18} 
(20) \cite{pieres+16} 
(21) \cite{Pieres:2017} 
(22) \cite{prusti+16} 
(23) \cite{brown+16} 
(24) \cite{vandermarel+16} 
(25) \cite{helmi+18} 
(26) \cite{wolf+18} 
(27) \cite{Dias:2014}
(28) \cite{Dias:2016}
(29) \cite{Maia:2014}
(30) \cite{Bica:2015}.
\end{flushleft}
\end{table*}

Among the topics that the VISCACHA data shall allow to address and play an important role, we list: 
(i) position dependence structural parameters of clusters, 
(ii) age-metallicity relations of star clusters and radial gradients, 
(iii) 3D structure of the Magellanic System in contrast with results from variable stars, 
(iv) star cluster formation history, 
(v) dissolution of star clusters,
(vi) initial mass function for high- and low-mass clusters, 
(vii) extended main-sequence turnoffs in intermediate-age clusters,
(viii) combination with kinematical information to calculate orbits, 
among others.

This paper is organized as follows. In Section \ref{sec:viscacha} we present an overview of 
the VISCACHA survey. In Sections \ref{sec:obs} and \ref{sec:data} we describe the observations 
and data reduction. The analysis we will perform on the whole data set is presented in Section 
\ref{sec:analysis}, and the first results are shown in Section \ref{sec:results}.
Conclusions and perspectives are summarised in Section \ref{sec:conclusions}.

\section{The VISCACHA survey}
\label{sec:viscacha}

Photometric studies of Magellanic Clouds clusters are usually limited to those with the main sequence 
turn-off above the detection limits \citep{Chiosi:2006}, which is directly related to the depth 
of the observations. Furthermore, crowding can also hamper the studies of many compact clusters and those immersed in rich backgrounds such as the LMC bar. This limits the sample 
to massive, young to intermediate-age clusters, while leaving the much more numerous low mass ones
largely unexplored.

The VISCACHA survey\footnote{\url{ http://www.astro.iag.usp.br/~viscacha/}} is performing a comprehensive study of the outer regions of the 
Magellanic Clouds by collecting deep, high quality images of its stellar clusters using the 
4.1~m SOAR telescope and its SAM Imager (SAMI). 

When compared with other surveys on the Magellanic Clouds, the VISCACHA survey is reaching 
$>$2mag deeper than previous studies (largely based on the 2MASS, MCPS or the VMC surveys),
attaining S/N $\approx$ 10 at V $\approx$ 24, which is slightly better than those achieved by 
SMASH ($z \sim 23.5$, $g \sim 22.5$). Furthermore, while SMASH aims to search and identify low surface brightness 
stellar populations across the Magellanic Clouds, the VISCACHA survey will provide local high 
quality data of specific 
targets enabling the most complete characterization of their populations. Due to the employment of 
the adaptive optics system, the spatial resolution achieved by VISCACHA (FWHM $\approx$ 0.5\arcsec, $V$ band) 
is higher than that of any other survey on the Magellanic Clouds, enabling the deblending of the 
stellar sources 
down to very crowded scenarios. Even though HST photometry \citep[e.g.][]{Glatt:2008} is still 
deeper than ground based photometry, the spatial coverage of the VISCACHA survey greatly surpasses 
those with appropriate field of view and resolution, allowing for a larger cluster sample and 
a more complete understanding of these galaxy properties. 

On a short term, the VISCACHA survey will deliver a high quality, homogeneous database of star 
clusters in the  Magellanic Clouds, providing reliable physical parameters such as 
core and tidal radii, ellipticities,
distances, ages, metallicities, mass distributions as derived from standard data reduction and 
analysis processes. The effects of the local tidal field over their evolution will be quantified 
through the analysis of their structural parameters, dynamical times, and positions within the Galactic 
system. Comparison of these results with models \citep[e.g.][]{Marel:2009, 
Baumgardt:2013} will provide important constraints to understand the evolution of the Magellanic 
Clouds. 

Once a significant sample has been collected, a study of the star formation history 
and chemical enrichment of the star clusters located at the periphery of these galaxies will be 
carried out to probe the local galactic properties.
Based on this dataset, several aspects concerning the evolution of these galaxies will 
be revisited, such as spatial dependence of age-metallicity relationship \citep{Dobbie:2014}, the 
``V"-shaped metallicity and age gradients found in the SMC \citep{Dias:2014, Dias:2016, parisi+09,
parisi+15}, the 3D cluster distribution, the inclination of the LMC disc, among others.

Finally, our catalogues will be matched against others (e.g. MCPS, VMC, OGLE) 
comprising a more complete panchromatic data set that will serve as reference for future studies 
of star clusters in the Magellanic Clouds. Even though this is not a public survey, it has
a legacy value, therefore we intend to eventually compile an easily accessible on-line 
database, including photometric tables, parameter catalogues, and reduced images.

\section{Observations}
\label{sec:obs}

Historically, the VISCACHA team originated from the merging of two Brazilian teams, one of 
them observing star clusters in the periphery of the LMC looking for structural parameters, 
and the other one observing clusters in the periphery of the SMC looking for age-metallicity 
relation and radial gradients. Both teams started observing with the SOAR optical imager (SOI) 
since its commissioning in 2006, and joined forces to found the VISCACHA collaboration observing 
with the recently commissioned SAMI in 2015. We broadened the science case and the collaboration 
team, having members based in Brazil, Chile, Argentina, and Colombia so far.

\begin{figure*}
\includegraphics[width=\linewidth]{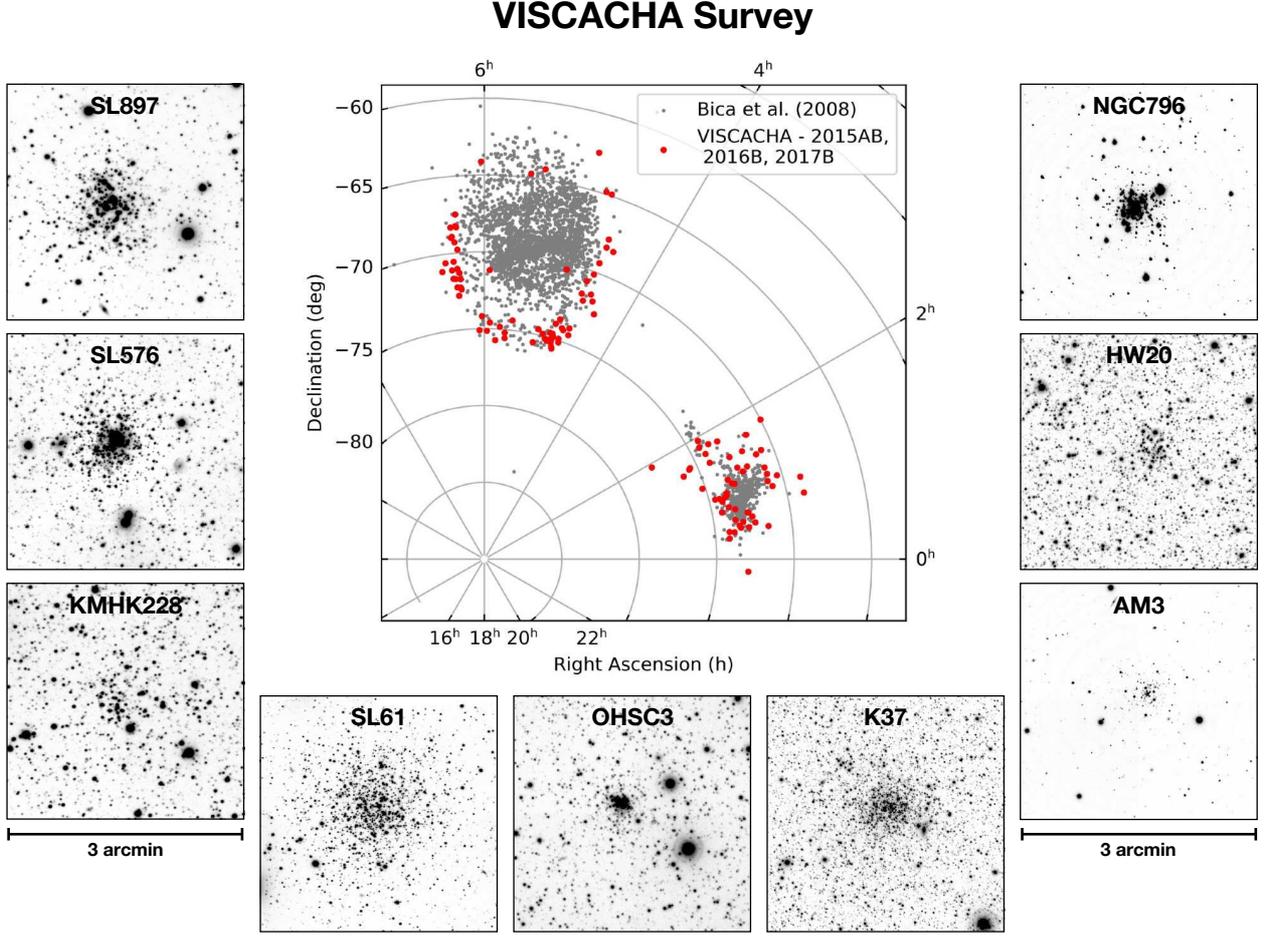}
\caption{Central panel: present VISCACHA sample, including $\sim$130 clusters 
observed through 2015-2017 (red circles). Small black dots correspond to the catalogued objects in the Magellanic System by \citet{Bica:2008}. Surrounding panels: $V$ image of selected targets, representing the variety of cluster types in the survey.}
\label{fig:visclu}
\end{figure*}

Considering the observing runs 2015A, 2015B, 2016B and 2017B we have observed about 130 clusters. 
In order to demonstrate the methods concerning CMDs and cluster structure we use in the present 
study a subsample of 4 SMC and 5 LMC clusters illustrating different concentration, total brightness
and physical parameters. Their $V$ images are shown in Fig.~\ref{fig:visclu} and their observation
log in Tab.~\ref{tab:log}. A list containing the full sample of all observed clusters up to 
the 2017B run is given in the appendix (Table~\ref{tab:clulist}).

\begin{table*}
\caption{Log of observations only for the clusters analysed in this paper.}
\begin{tabular}{lcccccccccc} \hline
Name & RA & Dec & date & filter & exptime & airmass & seeing & IQ & $\tau_0$ & AO? \\ 
 & [h:m:s] & [$^\circ$:$\arcmin$:$\arcsec$] & [DD.MM.YYYY] &  & [sec] &  & [arcsec] & [arcsec] & [ms] & \\ \hline
\hline
\multicolumn{11}{c}{SMC}\\
\hline
AM3     & 23:48:59 & -72:56:43 & 04.11.2016 & V, I & $6\times200$, $6\times300$ & 1.38 & 1.2, 1.1 & 0.5, 0.4 & 7.2, 5.7 & ON \\
HW20    & 00:44:47 & -74:21:46 & 27.09.2016 & V, I & $6\times200$, $6\times300$ & 1.40 & 1.2, 0.9 & 0.6, 0.5 & 4.8, 6.8 & ON \\
K37     & 00:57:47 & -74:19:36 & 04.11.2016 & V, I & $4\times200$, $4\times300$ & 1.44 & 0.8, 0.8 & 0.5, 0.4 & 7.0, 7.2 & ON \\
NGC796  & 01:56:44 & -74:13:10 & 04.11.2016 & V, I & $3\times100$, $4\times100$ & 1.78 & 1.0, 0.9 & 0.6, 0.5 & 5.4, 6.3 & ON \\
\hline
\multicolumn{11}{c}{LMC}\\
\hline
KMHK228 & 04:53:03 & -74:00:14 & 11.01.2016 & V, I & $3\times375$, $3\times560$ & 1.42 & 1.1, 1.0 & 1.1, 1.0 & 3.9, 3.1 & ON \\
OHSC3   & 04:56:36 & -75:14:29 & 02.12.2016 & V, I & $3\times375$, $3\times560$ & 1.45 & 1.0, 1.0 & 1.0, 1.0 & 2.0, 2.0 & OFF\\
SL576   & 05:33:13 & -74:22:08 & 29.11.2016 & V, I & $3\times375$, $3\times560$ & 1.48 & 1.3, 1.0 & 1.2, 1.0 & 4.3, 3.4 & ON \\
SL61    & 04:50:45 & -75:31:59 & 09.01.2016 & V, I & $3\times375$, $3\times560$ & 1.64 & 0.9, 0.8 & 0.7, 0.6 & 7.5, 6.9 & ON \\
SL897   & 06:33:01 & -71:07:40 & 23.02.2015 & V, I & $3\times375$, $3\times560$ & 1.34 & 1.5, 1.4 & 1.1, 0.9 & 3.5, 4.3 & ON \\
\hline 
\end{tabular} 
\label{tab:log}
\end{table*}

\subsection{Strategy}

The overall primary goal of VISCACHA is to further investigate clusters in the outer LMC ring, and to explore the SMC halo and Magellanic Bridge clusters. A panorama of these external LMC and SMC structures and the already collected VISCACHA targets are given in Fig.~\ref{fig:visclu}. In the first outer LMC cluster catalogue \citep{Lynga:1963}, the outer LMC ring could be inferred. It appears to be a consequence of a nearly head-on collision with the SMC, similarly to the Cartwheel scenario \citep{Bica:1998}. This interaction is also responsible for the inflated SMC halo (Fig.~\ref{fig:visclu}). In \citet{Bica:2008} these structures can be clearly seen. In that study they found 3740 star clusters in the Magellanic System.
However, this number does not account for other cluster types such as embedded clusters, small associations \citep{Hodge:1986}, and other types of objects.

The north-east outer LMC cluster distribution has also been recently discussed by \citet{Pieres:2017}.
The outer ring is located from 5 kpc to 7 kpc from the dynamical LMC centre, but well inside 
its tidal radius \citep[$\gtrsim$ 16 kpc -][]{Marel:2014}. 
Since there is a tendency for older clusters to be located in the LMC outer disk regions \citep{Santos:2006}, 
these objects are ideal candidates to be remnants from the LMC formation epoch. In particular, 
such clusters may belong to a sample without a counterpart in our Galaxy due to
the different tidal field strengths, persisting as bound structures for longer times than in the 
Milky Way. 

In the SMC, the galaxy main body can be represented by an inner ellipsoidal region, while its 
outer part can be sectorised as proposed by \cite{Dias:2014,Dias:2016}:  
(i) a wing/bridge, extending eastward towards the Magellanic 
Bridge connecting the LMC and SMC; (ii) a counter-bridge in the northern region, which 
could represent the tidal counterpart of the Magellanic Bridge; (iii) a west halo on the opposite 
side of the bridge. These groups had also been predicted in the stellar distribution
of \citet{Besla:2011} and \citet{Diaz:2012} models and most likely have a tidal origin tied to the dynamical history of the Magellanic Clouds. The wing/bridge clusters 
present distinct age and metallicity gradients 
\citep{parisi+15,Dias:2016} which could be explained by tidal stripping of clusters beyond 4.5 
deg, radial migration, or merging of galaxies. 
The age and metallicity gradients in the west halo were used to propose that these clusters are moving away from the main body \citep{Dias:2016}, as confirmed later by proper motion determinations from VMC survey \citep{niederhofer+18}, HST and {\it Gaia} measurements \citep{zivick+18}. These radial 
trends are crucial to charaterise the SMC tidal structures and to define a more complete picture 
of its history. 


Photometric images with $BVI$ filters were obtained for approximately 130 clusters\footnote{
Eventually, the data acquired between 2006-2013 with the previous generation imager (SOI)
will also be integrated in our database.} in the LMC, SMC and Bridge so far, during the semesters 
of 2015A, 2015B, 2016B and 2017B. Their distribution in the Magellanic System is shown in Fig.~\ref{fig:visclu}.

\subsection{Instrumentation: SAMI data}

Observation of our targets include short exposures to avoid saturation of the brightest stars
($V \sim 16$) and deep exposures 
to sample $V \sim 24$ stars with 
S/N~$\sim$~10. Photometric calibration of individual nights have been done by observing both 
\citet{Stetson:2000} (for extinction evaluation) and MCPS fields (for colour calibration) over
the $B$, $V$ and $I$ filters. 

SAM is a GLAO module using a Rayleigh 
laser guide star at $\sim$7 km from the telescope. SAM was employed with its internal CCD 
detector, SAMI (4K$\times$ 4K CCD), set to a gain of 2.1\, e$^-$/ADU and a readout noise 
of 4.7\,e$^-$ and binned to 2$\times$2 factor, resulting in a plate scale of 
0.091\,arcsec/pixel with the detector covering a field-of-view of 3.1$\times$3.1 arcmin$^2$ on 
the sky. Peak performance of the system produce FWHM $\sim$0.4 arcsec in the $I$ band and 
$\sim$0.5 arcsec in the $V$ band, which still allows for adequate sampling of the point 
spread function (PSF), reaching a minimum size of $\sim$4.4 pixels (FWHM) in those occasions. 

SAM operates at a maximum rate of 440Hz which means it can only correct the effects of ground-layer atmospheric turbulence if the coherence time is $\tau_0 > 2.3$ms. The closer the $\tau_0$ is to this limit the worse is the AO correction. In fact, Table \ref{tab:log} shows that although all clusters were observed under similar seeing and airmass, the delivered image quality (IQ) varied from target to target. The variation is explained by the free-atmosphere seeing variations (above 0.5km) that are not corrected by GLAO. The SMC clusters were observed under better conditions of the free-atmosphere and as a consequence have deeper photometry reaching the goals of the ideal performance for the VISCACHA data.

For the last observation period (2017B), we only took short exposures in the B filter since 
SAM has optimal performance in $V$ and $I$ bands, which decreases towards blue wavelengths. 
This strategy allowed us to increase our number of targets observed with AO, improving the 
efficiency of the survey. It is worth noticing that even for observations with relatively 
high airmass ($X\sim 1.3-1.7$) the instrument performed well, improving the image quality, 
whenever the atmospheric seeing was around 1 arcsec.

\section{Data Reduction}
\label{sec:data}

\subsection{Processing}

The data were processed in a standard way with IRAF, using automated scripts designed to work on 
SAM images. Pre-reduction included bias subtraction and division by skyflats using the 
{\sc ccdred} package and cosmic rays removal with the {\sc crutil} package. Correction of the 
camera known optical distortion was also done, as it is large enough ($\sim$10\%) to shift stellar 
positions by more than 1 arcsec in some image areas. Subsequent astrometric calibration was 
performed with the {\sc imcoords} package, using astrometric references from 2MASS, GSC-2.3 
and MCPS catalogues, and ensuring a typical accuracy better than $\sim0.1$ arcsec for all our 
images. See \citet{Fraga:2013} for further details in the processing and astrometric calibration 
procedures.

The final processing step was to register the repeated long exposures in each filter to a 
common WCS frame and to stack them into a deeper mosaic using the IRAF {\sc immatch} package. 
To preserve image quality of our mosaics the co-added images were weighted according to their 
individual seeing ($\propto FWHM^{-2}$). This, allied with the good quality of our astrometric 
solutions, resulted in very little degradation of the stellar PSF ($<$ 10\%) in the resulting 
mosaics.

\subsection{Photometry}
\label{sec:psf}

\begin{figure}
\includegraphics[width=\linewidth]{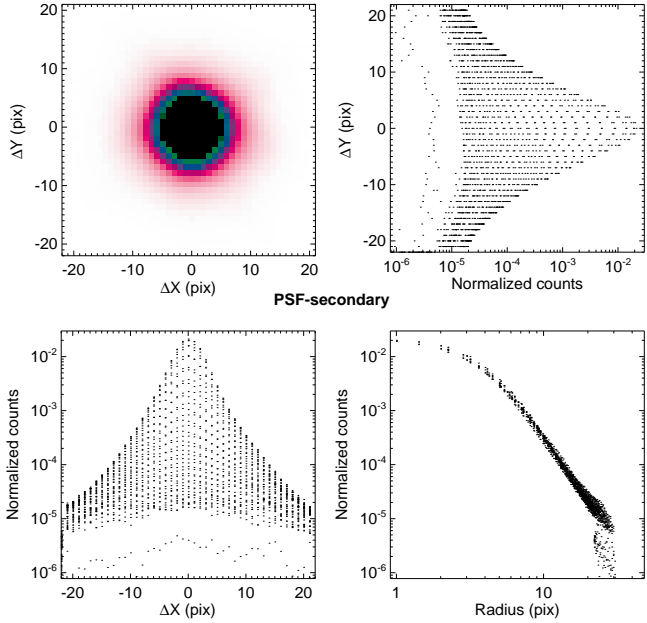}
\caption{Empirical PSF of Kron 37 in the I-band as shown by its image (top-left), 
marginal profile along the X-axis (bottom-left), Y-axis (top-right) and as a function of radius 
(bottom-right). The FWHM of this PSF is about 5 pixels (0.49 arcsec).} 
\label{fig:psf}
\end{figure}

Stellar photometry was done using a modified version of the Starfinder code \citep{Diolaiti:2000}, 
which performs isoplanatic high resolution analysis of crowded fields by extracting an empirical 
PSF from the image and cross-correlating it with every point source detected above a defined threshold. 
The modifications were aimed mainly at automatising the code, minimising the user intervention. 
Modelling of each image PSF was carried out by using 20 to 50 bright, unsaturated stars presenting
no bright neighbour closer than 6 FWHM. This initial PSF was used to model and remove faint 
neighbours around the initially selected stars, which were then reprocessed to generate a 
definitive PSF. Fig.~\ref{fig:psf} shows the resulting PSF of the deep $I$ mosaic of the cluster 
Kron 37 after the subtraction of secondary sources around the model stars. Even though the FWHM 
is only about 5 pixels, the PSF profile is clearly defined up to a distance of 30 pixels ($\sim$6 FWHM), well into the sky region.

\begin{figure}
\includegraphics[width=\linewidth]{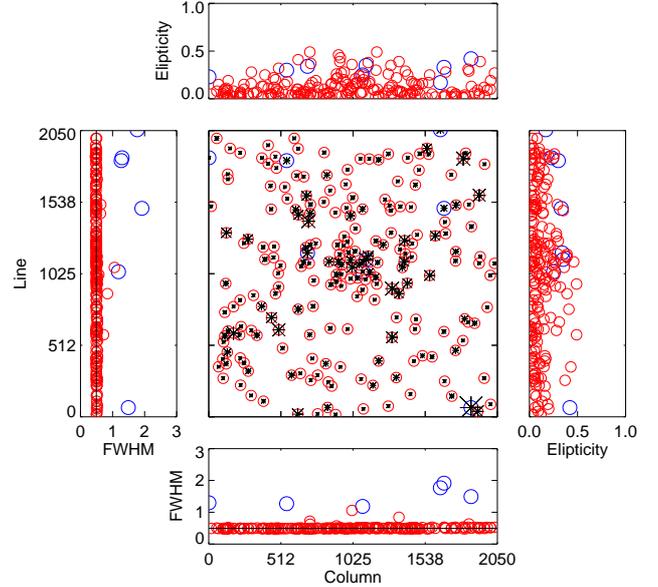}
\caption{PSF quality assessment of Kron 37 I-band image. The stars are represented by asterisks 
with sizes proportional to their brightness on the central sky chart. Marginal distributions of 
the stellar FWHM (left and bottom panels) and ellipticity (right and top panels) are also shown. 
Stars presenting FWHM above the median value are represented by the bigger blue circles; all the 
other ones are marked with smaller red circles.} 
\label{fig:psfqual}
\end{figure}

Quality assessment of the PSF throughout the image was performed with the IRAF {\sc psfmeasure} 
task to derive the empirical FWHM and ellipticity of several bright stars over the image. 
Fig.~\ref{fig:psfqual} shows that the PSF shape parameters (e.g. FWHM, ellipticity), and consequently
the AO performance, are very stable 
through the image, indicating that higher order terms (e.g. quadratically varying PSF) are not
necessary to properly describe the stellar brightness profile on SAM images. 

\subsection{Performance: SAMI vs SOI} 

The members of the VISCACHA project have been acquiring SOAR data for a long time. Before the 
commissioning of the SAM imager, we have extensively used the previous generation imager SOI, 
establishing a considerable expertise with the instrument. The migration to the new imager
after 2013, was an obvious choice given its performance increase over the older instrument. 

\begin{figure}
\centering
\includegraphics[width=0.47\linewidth]{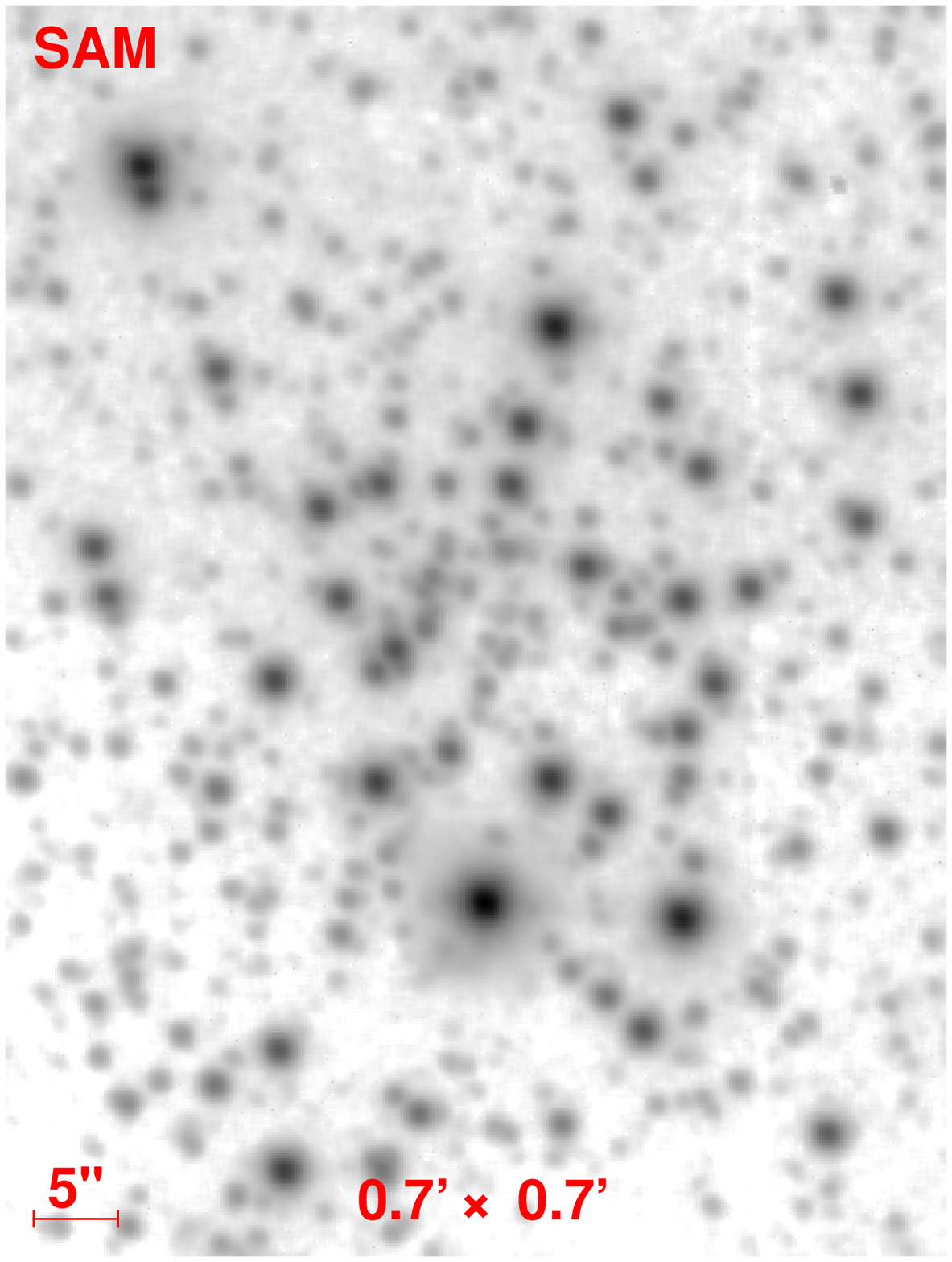}\hspace{0.2cm}
\includegraphics[width=0.47\linewidth]{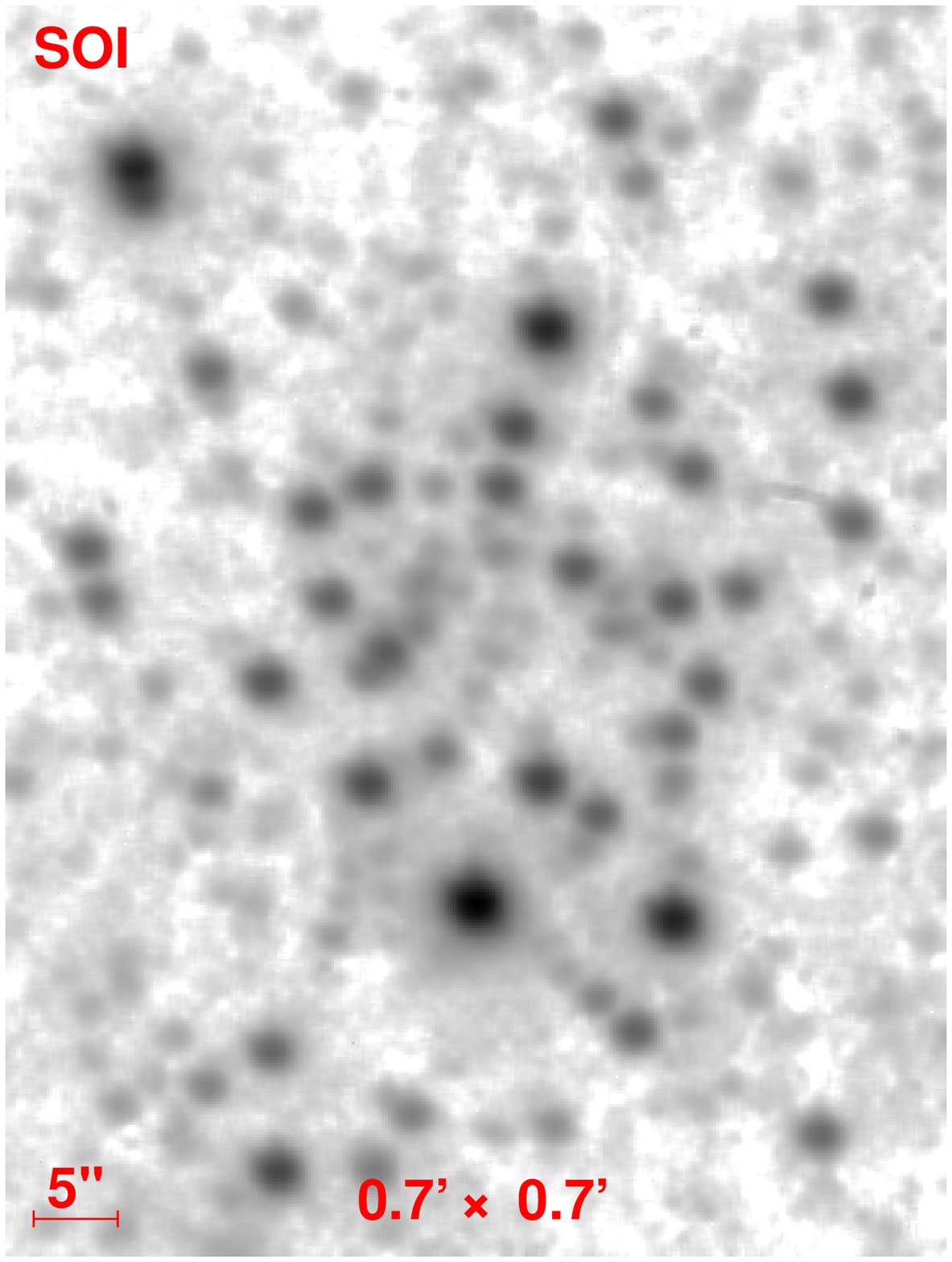}
\caption{$I$ filter images of the centre of HW20 taken with SAM in closed loop (left panel) 
and with SOI (right panel) under comparable conditions. The stellar FWHM in the images are 
0.44\arcsec and 1.19\arcsec respectively.} 
\label{fig:soisam}
\end{figure}

Therefore, we compare the performance of a typical optical imager without AO, 
such as SOI, with SAMI as we observed the cluster HW20 in the night 27/09/2016 
with both instruments. Exposure times were (6$\times$200)\,s in the $V$ filter and 
(6$\times$300)\,s in $I$ filter. Although the Differential Image Motion Monitor (DIMM) 
reported a $0.85\arcsec$ seeing for the observations, the SOI $I$ image attained 
a stellar FWHM of $1.19\arcsec$ and the SAM image reached FWHM of $0.44\arcsec$ on closed 
loop. Fig.~\ref{fig:soisam} compares a section of the SAM and SOI $I$ images 
around the centre of HW20 and shows how the decrease of the seeing by the AO system 
reduces the crowding and effectively improves the depth of the image.

In addition, SOI presents relatively intense fringing in the $I$ filter, requiring correction for 
precision photometry. Since fringe correction requires at least a dozen dithered exposures of non-crowded 
fields, we have used a fringe pattern image we derived from 2012B data to correct the fringes in HW20. 
On the other hand, SAMI $I$ images show negligible to null fringing. 



\begin{figure}
\centering
\includegraphics[width=0.9\linewidth]{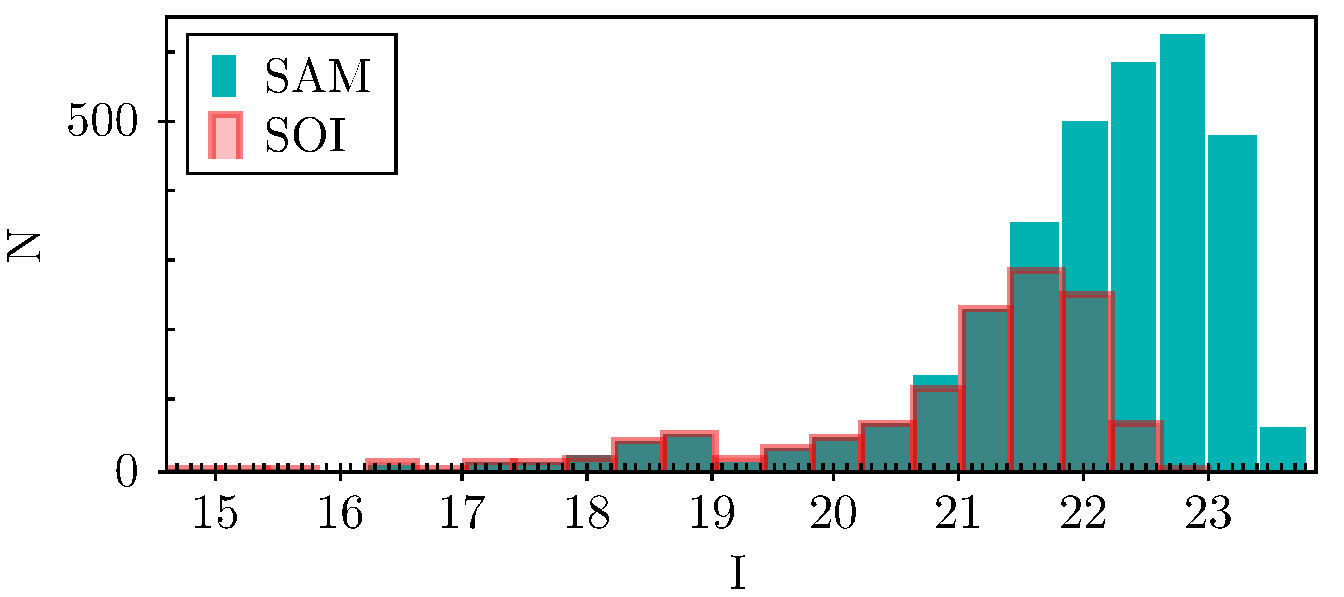}
\includegraphics[width=0.9\linewidth]{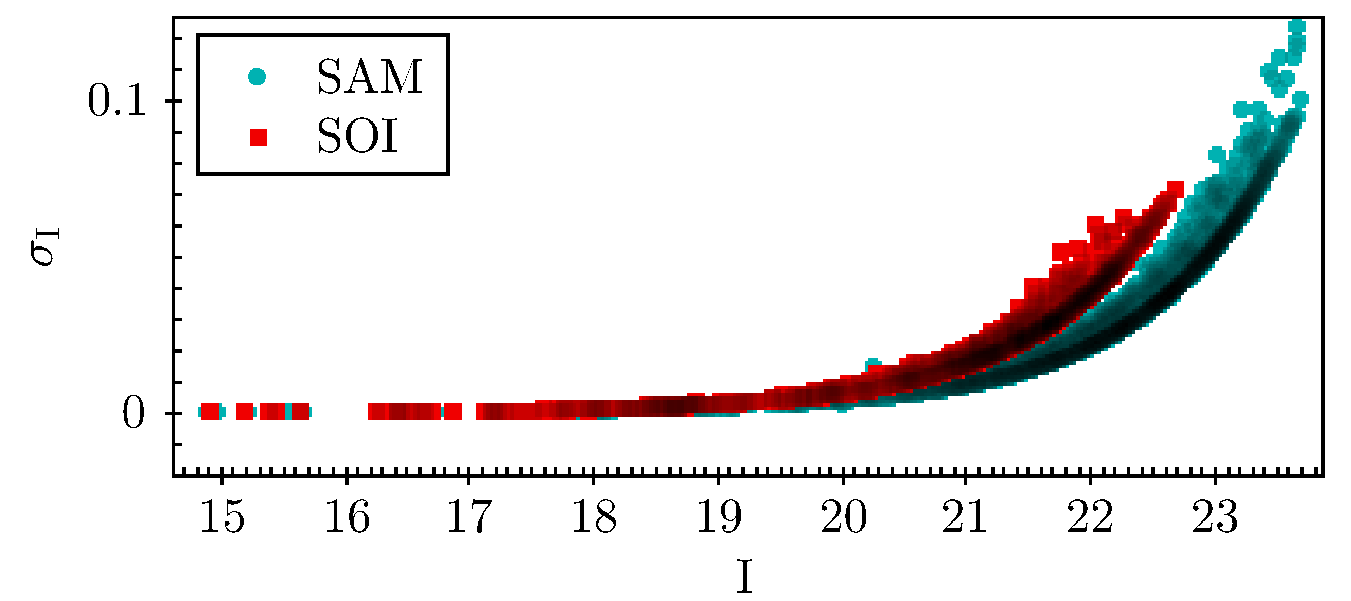}
\caption{Comparison between SOI and SAM photometry of the HW20 cluster in the $I$ filter, showing the
detected objects (top panel) and photometric errors (bottom panel) as function of magnitude. Under
the same conditions SAM exposure reaches about 1.2 mag deeper on a photometric night.}
\label{fig_soiphot}
\end{figure}

Finally, to empirically compare the instruments, we have performed PSF photometry (see Sect.~\ref{sec:psf})
in the fringe-corrected SOI images and SAM images of HW20, subject to the same constraints and 
relative detection thresholds. Given the different fields of view of these instruments, we have restricted 
the analysis to an area of $3\arcmin \times 3\arcmin$ near the cluster centre, equally sampled by both
instruments. Fig.~\ref{fig_soiphot} compares the photometric errors and depth reached 
by each instrument. It can be seen that with the AO system working at its best, SAM images reach more 
than one magnitude deeper than SOI under the same sky conditions. Furthermore, the improved 
resolution also helped detect and deblend more than twice the number of sources found by SOI, 
particularly in the fainter regime ($I \geq 22.0$). 

\subsection{Calibration}

Transformation of the instrumental magnitudes to the standard system was done using at least two 
populous photometric standard fields from \citet{Stetson:2000} (e.g. SN1987A, NGC1904, NGC2298, NGC2818),
observed at 2 to 4 different airmass through each night. Following the suggestions given in \citet{Landolt:2007}, 
the calibration coefficients derived from these fields were calculated in a two-step process: 

\begin{enumerate}
\item[i)] airmass ($X_j$), instrumental ($m_j$) and catalogue ($M_j$) magnitudes in each band ($j$) 
were employed in a linear fit given by Eq.~\ref{eq:calext} to evaluate the extinction coefficients 
($e_j$);
\begin{equation}
m_j - M_j = cte + e_{j} X_j \quad ;
\label{eq:calext}
\end{equation}

\item[ii)] the extra-atmospheric magnitudes ($m_j^\prime = m_j - e_j X_j$) were then used to 
derive colour transformation coefficients ($c_j$) and zero-point coefficients ($z_j$) according to
Eq.~\ref{eq:calcol}:
\begin{eqnarray}
\label{eq:calcol}
 v^\prime - V &=& z_v + c_v(V - I)\nonumber \\ 
(b^\prime -v^\prime) - (B-V) &=& z_{bv} + c_{bv}(B-V)  \\
(v^\prime -i^\prime) - (V-I) &=& z_{vi} + c_{vi}(V-I) \nonumber
\end{eqnarray}
\end{enumerate}

\begin{figure}
\centering
\includegraphics[width=0.85\linewidth]{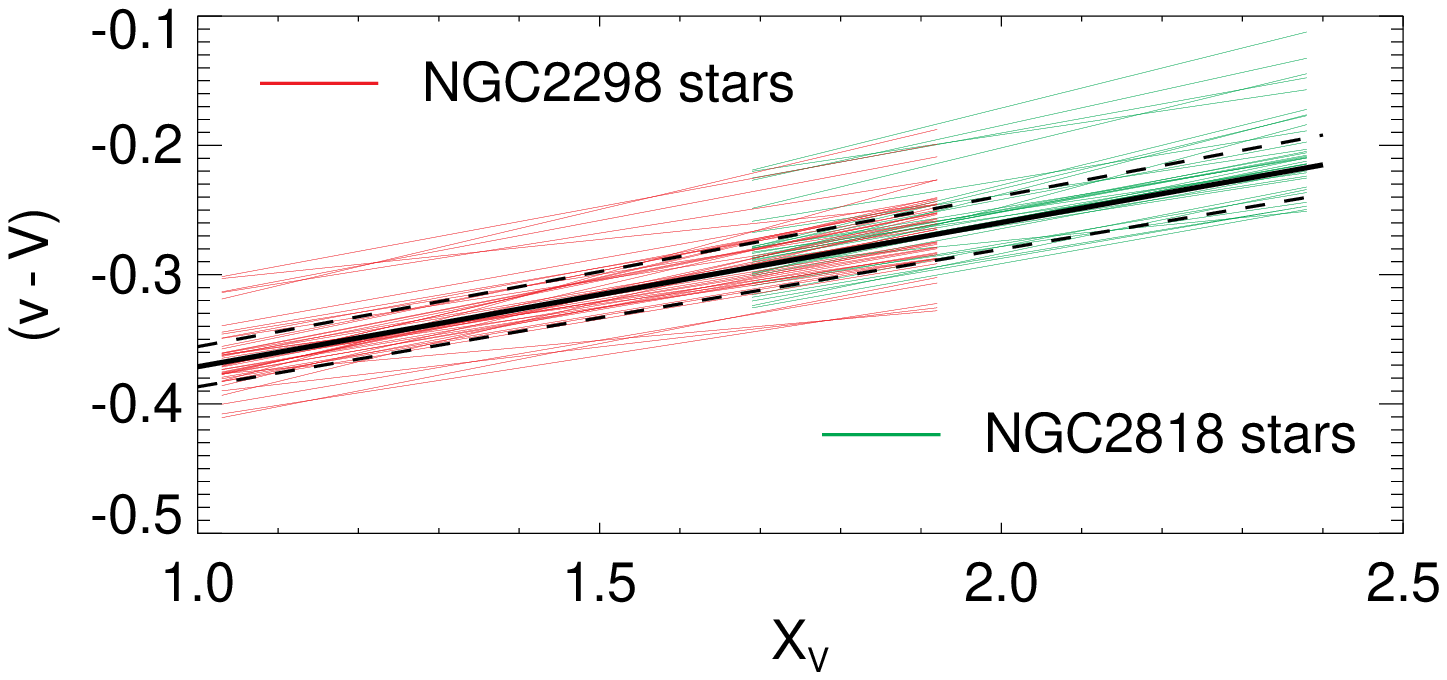} 
\includegraphics[width=0.85\linewidth]{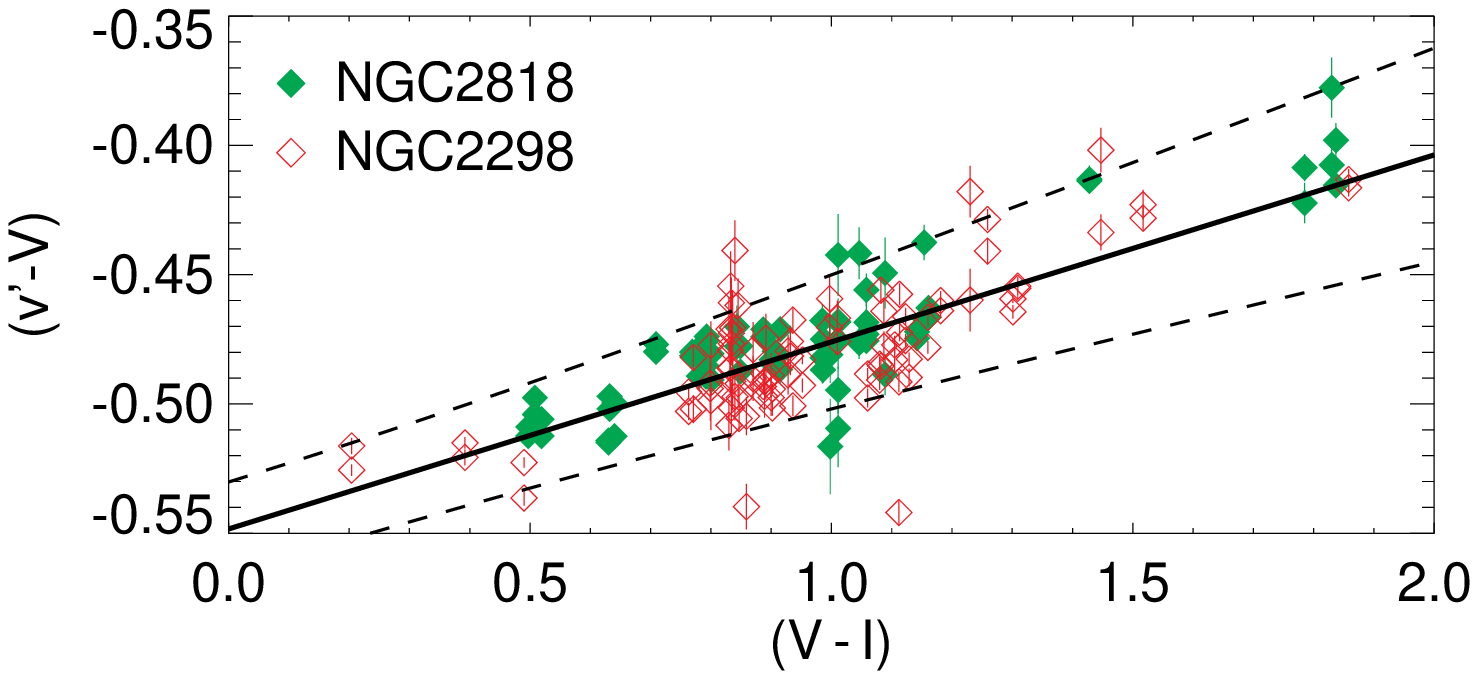}
\caption{Fits in the $V$ filter to determine the extinction coefficients (top) and the colour and 
zero-point coefficients (bottom) for the night of 22-02-2015, using the NGC2818 and NGC2298 standard 
fields. About 70 stars in both fields were used in the determination of the mean extinction coefficient 
and twice that number in the global fit to determine the colour coefficient. The resulting coefficients
and their 6-$\sigma$ uncertainty level are represented by the solid and dashed lines respectively.}
\label{fig:calib}
\end{figure}

Figure~\ref{fig:calib} shows the fit of Eqs. \ref{eq:calext} and \ref{eq:calcol} to determine the 
$V$ filter extinction, zero-point and colour coefficients for stars in the NGC2818 and NGC2298 standard 
fields in the night of February 22, 2015. Since the stars in each standard field were observed more than once 
(typically at 3 different airmass), the fit of Eq.~\ref{eq:calext} was made in a star-by-star basis and 
the final extinction coefficient and its uncertainty determined from the average and deviation of 
the slopes found. This approach offers a better precision than a single global fitting (i.e. carried 
out over all stars simultaneously) such as done by {\sc iraf}, because the intrinsic brightness 
difference between the standard stars (i.e. the spread in the $y-$axis on the upper panel) is factored 
out. On the other hand, the colour and zero-point coefficients were found from a global solution using 
the extra-atmospheric magnitudes for all stars in the two standard fields by means of a robust linear 
fitting method. At this point, the combination of several standard fields in a single fit is advantageous 
as it provides a larger sample and wider colour range to help constrain the fit. 
These fitting procedures were applied to the data calibration from 18 nights observed through semesters 
2015A$-$2016B, resulting in the mean coefficient values and deviations shown in 
Table~\ref{tab:coef}. These values are in excellent agreement with those reported by \citet{Fraga:2013}.

\begin{table}
\caption{Mean calibration coefficients through 2015A$-$2016B}
\begin{tabular}{c r@{$\,\pm\,$}l r@{$\,\pm\,$}l r@{$\,\pm\,$}l r@{$\,\pm\,$}l}
Coef. &\multicolumn{2}{c}{$B$} &\multicolumn{2}{c}{$V$} &\multicolumn{2}{c}{$I$} \\ \hline
$e$ & $0.177$&$0.011$  & $0.106$&$0.006$   & $0.022$&$0.006$ \\
$c$ & $-0.193$&$0.008$ & $0.064$&$0.005$   & $-0.063$&$0.005$ \\
$z^*$ & $-0.138$&$0.006$ & $-0.549$&$0.005$ & $-0.582$&$0.005$ \\ \hline
\end{tabular}

$^*$ relative to the adopted zero point magnitude of 25.
\label{tab:coef}
\end{table}

In order to calculate the photometric errors,
we first write the colour calibration 
equations given by 
Eq.~\ref{eq:calcol} as the following system:

\begin{equation}
\begin{pmatrix}
v - e_v X_v - z_v \\
b-v - e_b X_b + e_v X_v - z_{bv} \\ 
v-i - e_v X_v + e_i X_i - z_{vi}
\end{pmatrix}
= 
\begin{pmatrix}
1 & 0 & c_v \\
0 & 1+c_{bv} & 0 \\
0 & 0 & 1+c_{vi} 
\end{pmatrix}
\cdot
\begin{pmatrix}
V \\ B-V \\ V-I
\end{pmatrix}
\label{eq:syscol}
\end{equation}

\noindent
which can be more easily expressed in matrix notation by:
\begin{equation}
\boldsymbol{m}-\boldsymbol{e}-\boldsymbol{z} = \boldsymbol{C} \cdot \boldsymbol{M} \quad ,
\label{eq:matcol}
\end{equation}

\noindent 
where the instrumental quantities ($v$, $b-v$, $v-i$) and the corrections due to the zero point 
($z$) and extinction ($e$) are now represented by vectors. The calibrated quantities vector ($V$, $B-V$,
$V-I$) can be found by inverting this linear system, which requires only calculating the inverse of the 
colour coefficients matrix ($C$):
\begin{equation}
\boldsymbol{M} = \boldsymbol{C}^{-1} \cdot (\boldsymbol{m}-\boldsymbol{e}-\boldsymbol{z}) \quad .
\label{eq:solcol}
\end{equation}

However, propagating the errors through this solution is more subtle, given that the matrix inversion 
is a nonlinear operation and that the resulting cofactors are often correlated with each other. 
Following the formalism in \citet{Lefebvre:00}, the total uncertainties on the calibrated quantities
($\boldsymbol{\sigma_M}$) can be derived analytically from the uncertainties of the instrumental 
quantities ($\boldsymbol{\sigma_m}$), zero point ($\boldsymbol{\sigma_z}$), extinction 
($\boldsymbol{\sigma_e}$) and colour coefficients ($\boldsymbol{\sigma_C}$) as:
\begin{multline}
\boldsymbol{\sigma_M}^2 = (\boldsymbol{C}^{-1})^2 \cdot [\boldsymbol{\sigma_m}^2 + 
\boldsymbol{\sigma_e}^2 + \boldsymbol{\sigma_z}^2]\ + \\
+ \boldsymbol{\sigma_{C^{-1}}}^2 \cdot (\boldsymbol{m} - \boldsymbol{e} - \boldsymbol{z})^2
\label{eq:calerr}
\end{multline}

\noindent
where the uncertainties in the inverted colour coefficients matrix ($\boldsymbol{\sigma_{C^{-1}}}$) are 
calculated directly from the individual colour coefficients uncertainties as: 
\begin{multline}
\boldsymbol{\sigma_{C^{-1}}}^2 = 
(\boldsymbol{C}^{-1})^2 \cdot \boldsymbol{\sigma_{C}}^2 \cdot (\boldsymbol{C}^{-1})^2 \\
= (\boldsymbol{C}^{-1})^2 \cdot
\begin{pmatrix}
0 & 0 & \sigma_{c_v}^2 \\
0 & \sigma_{c_{bv}}^2 & 0 \\
0 & 0 & \sigma_{c_{vi}}^2
\end{pmatrix}
\cdot (\boldsymbol{C}^{-1})^2
\label{eq:materr}
\end{multline}

\begin{figure}
\centering
\includegraphics[width=0.75\linewidth]{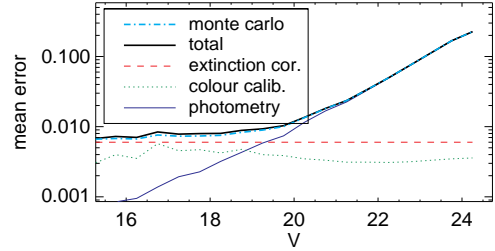}
\caption{Photometric uncertainties as function of $V$
for Kron 37. Contributions from the photometry (thin line), the extinction correction (dashed line) 
and the colour calibration (dotted line) compose the total photometric uncertainty (solid line), which was
also derived using a Monte-Carlo simulation (dot-dashed line).}
\label{fig:calerr}
\end{figure}

According to this prescription the total photometric uncertainty of a source, defined by 
Eq.~\ref{eq:calerr}, can be understood as being composed of three components arising from: 
(i) the PSF photometry (first right hand term), (ii) the extinction correction (second right hand term) 
and (iii) the colour transformation to the standard system (remaining right hand terms), as shown in 
Fig.~\ref{fig:calerr}. In our data these uncertainties are typically dominated by the extinction 
correction and colour calibration contributions for stars brighter than $V\sim$19.5, which is about 
the red clump level of the SMC and LMC clusters, and by the photometric errors 
for stars fainter than that. Typically, we reached a final error of $\sim$0.1 mag for $V$=24 mag, which is 
more accurate than those obtained by surveys without the AO system (e.g SMASH, MCPS). 

A Monte-Carlo simulation was also employed to propagate the uncertainties through the calibration process.
In each step, each coefficient (i.e. zero-point, extinction and colour ones) and instrumental magnitude 
were individually deviated from its assumed value using a random normal distribution of the respective 
uncertainty and the calibrated magnitudes calculated through Eq.~\ref{eq:solcol}. At the end of $10^6$ 
steps, the standard deviation of each calibrated magnitude was computed and assigned as its total 
photometric uncertainty. It can be seen in Fig.~\ref{fig:calerr} that the two solutions for propagating 
the uncertainties are equivalent, with only minor deviations. However, while the Monte-Carlo solution can
be computing intensive, the analytical solution presented in Eq.~\ref{eq:calerr} requires negligible 
computational time.

\subsection{Completeness}
\label{sec:compl}

Artificial star tests were performed in each image of the present sample in order to derive completeness 
levels as function of magnitude and position. The empirical PSF model was used to artificially add 
stars with a fixed magnitude to the image in a homogeneous grid, with a fixed spacing of 6 FWHM 
to prevent overlapping of the artificial star wings and overcrowding the field. Several grids with 
slightly different positioning and with stellar magnitudes ranging from 16 to 25 were simulated, 
generating more than 100 artificial images for each original one. 

\begin{figure}
\centering
\includegraphics[width=0.9\linewidth]{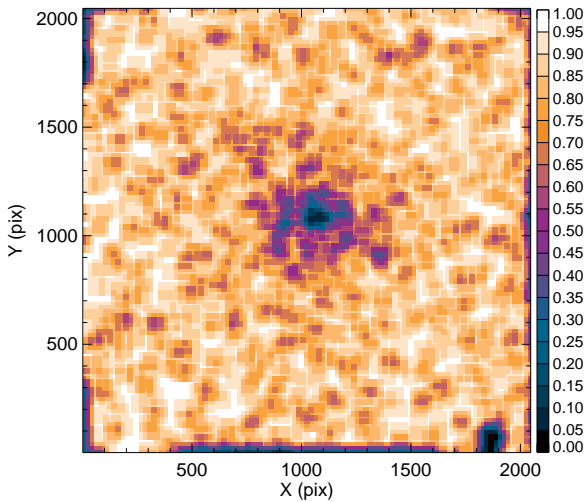}
\caption{Completeness map for Kron\,37, constructed by artificially adding $V = 23$ stars over the 
original image in uniform grids with 6 FWHM spacing, covering the entire image. Even though the average 
completeness over the image is $\sim$85\%, near the centre of the cluster it drops nearly to zero 
($<$ 15\%).}
\label{fig:compl}
\end{figure}

\begin{figure}
\centering
\includegraphics[width=0.9\linewidth]{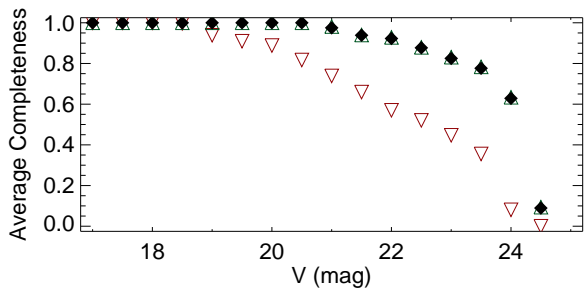}
\caption{Completeness as a function of $V$ magnitude for Kron\,37 over three different regions: the 
region inside the core radius (down facing red triangles), the one outside it (up facing green triangles) and the whole image (filled black diamonds).}
\label{fig:complmag}
\end{figure}

Photometry was carried out over the artificial images using the same PSF and detection thresholds as in
the original one, and the local recovery fraction of the artificially added stars used to construct 
spatially resolved completeness maps, as shown in Fig.~\ref{fig:compl} for Kron\,37 at $V=23$mag.
It can be seen that incompleteness can severely hamper the analysis of the low mass content 
of the cluster, as the local completeness value near the centre ($\lesssim 15\%$) falls much 
more rapidly than the overall field value ($\sim 85\%$). The same trend is clear in
Fig.~\ref{fig:complmag} where average completeness curves are shown for three regions: the whole image,
the cluster core region and the region outside it.
It can be seen that completeness assessments based on an average of the whole image are too
optimistic by a factor of 20-50\% towards the inner regions of the cluster for stars fainter than the 
main sequence turnoff level. Usually, the RGB stars have 100\% completeness and it starts to decrease 
from the turnoff towards fainter stars. Because of that we consider the dependence on the magnitude and on the position when applying photometric completeness corrections, before RDP and CMD fitting. 

\section{Analysis and Methodology}
\label{sec:analysis}

\subsection{Radial profile fitting}
\label{sec:analysis_rdp}


Given the nature of stellar clusters, it is expected that photometry incompleteness will be higher
toward their central regions (see Sect.~\ref{sec:compl}). Therefore, if stellar counts are employed 
to build radial profiles, reliable structural parameters can only be derived after a spatially resolved 
completeness correction is carried out \citep[e.g. as in][]{Maia:2016,Dias:2016}. Alternatively, 
brightness profiles measured directly over the clusters' images can also be used 
\citep{Piatti:2018b}.

Once a reliable radial profile is built, cluster parameters are usually inferred by fitting an 
analytic model which describes its stellar distribution. Although the \citet{King:1962} model has 
long been used in describing Galactic clusters, the EFF model \citep{Elson+87} arguably provides 
better results for young clusters in the LMC, presenting very large halos. In addition, it has the
advantage of also encompassing the \citet{Plummer:1911} profile, largely used in simulations. 

Nevertheless, we preferred the \citet{King:1962} model as it provides a truncation radius to the
cluster, effectively defining its size, whereas the EFF model cluster has no such
parameter. Also, it generally yields best fits than the EFF
model for intermediate-age and old clusters in the Clouds \citep{Werchan:2011, hill+06}.
We note that dynamical models such as the \citet{King:1966} and \citet{Wilson:75} have also 
been successfully used to describe finite Magellanic Clouds clusters with extended 
halos \citep{McLaughlin:05}, being excellent alternatives.

Following this reasoning we have adopted two methods to infer the structural parameters of the 
present sample. First, surface brightness profiles (SBPs) were derived directly from the 
calibrated $V$ and $I$ images. Stellar positions and fluxes were extracted from the reduced 
frames using {\sc DAOPHOT} \citep{Stetson:1987}, considering only sources brighter than 3 $\sigma$
above the sky level. The centre was then determined iteratively by the stars' coordinates
centroid within a visual radius\footnote{A circular region defined by visual inspection 
that encompasses a relevant portion of the cluster.}, starting with an initial guess and adjusted 
for the new centre at each step. Thereafter, the flux median and dispersion were calculated from the 
total flux measured in eight sectors per annular bin around this centre. The sky level, obtained 
from the whole image, was subtracted before the fitting procedure. Although the $I$ band provides 
the best image quality compared with the $V$ band, its enhanced background makes the resulting 
profiles noisier. Since smaller uncertainties were achieved for the $V$ band, it was the one used 
in the present analysis.

\begin{figure}
\centering
\includegraphics[width=\linewidth]{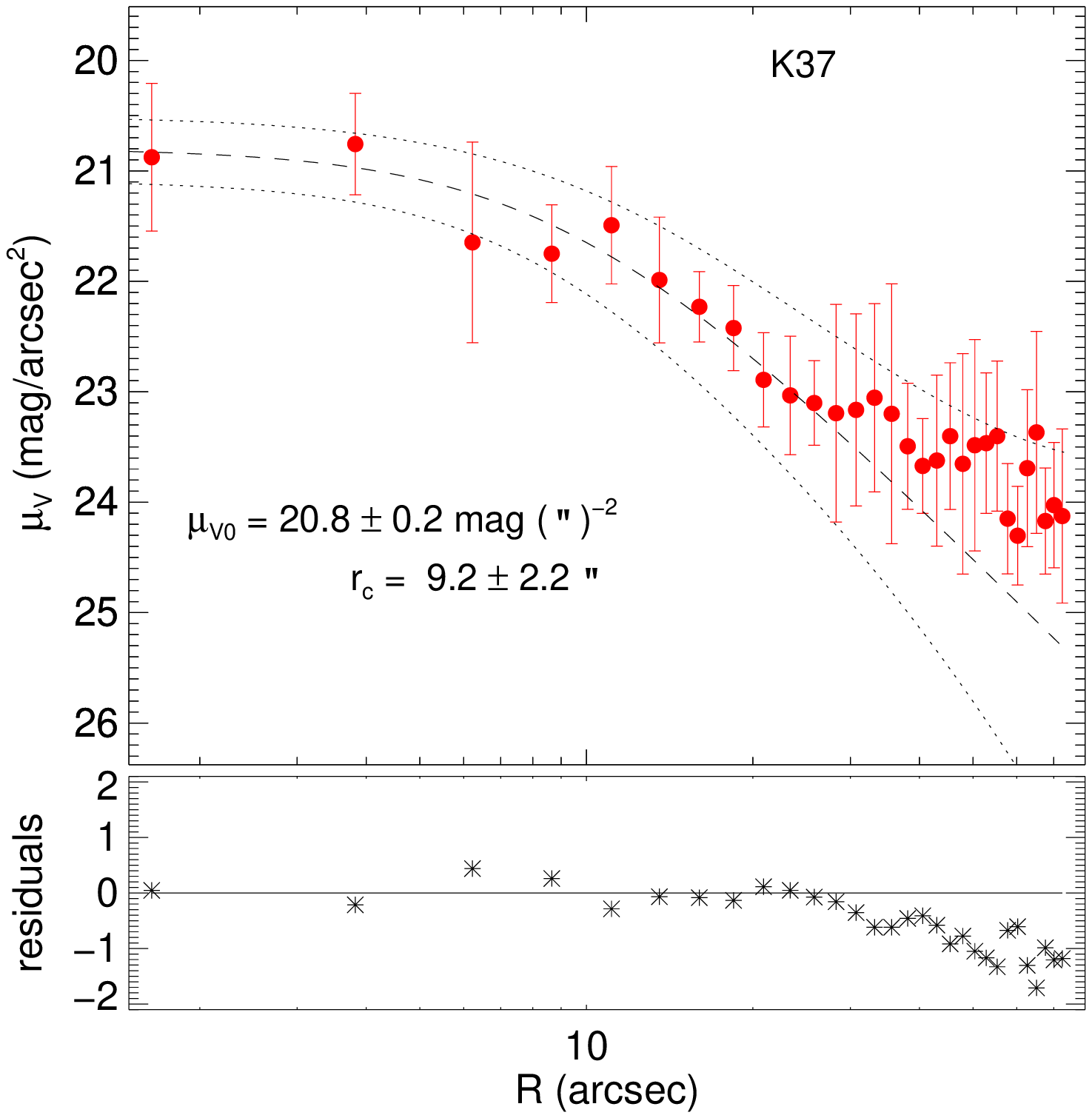}
\includegraphics[width=\linewidth]{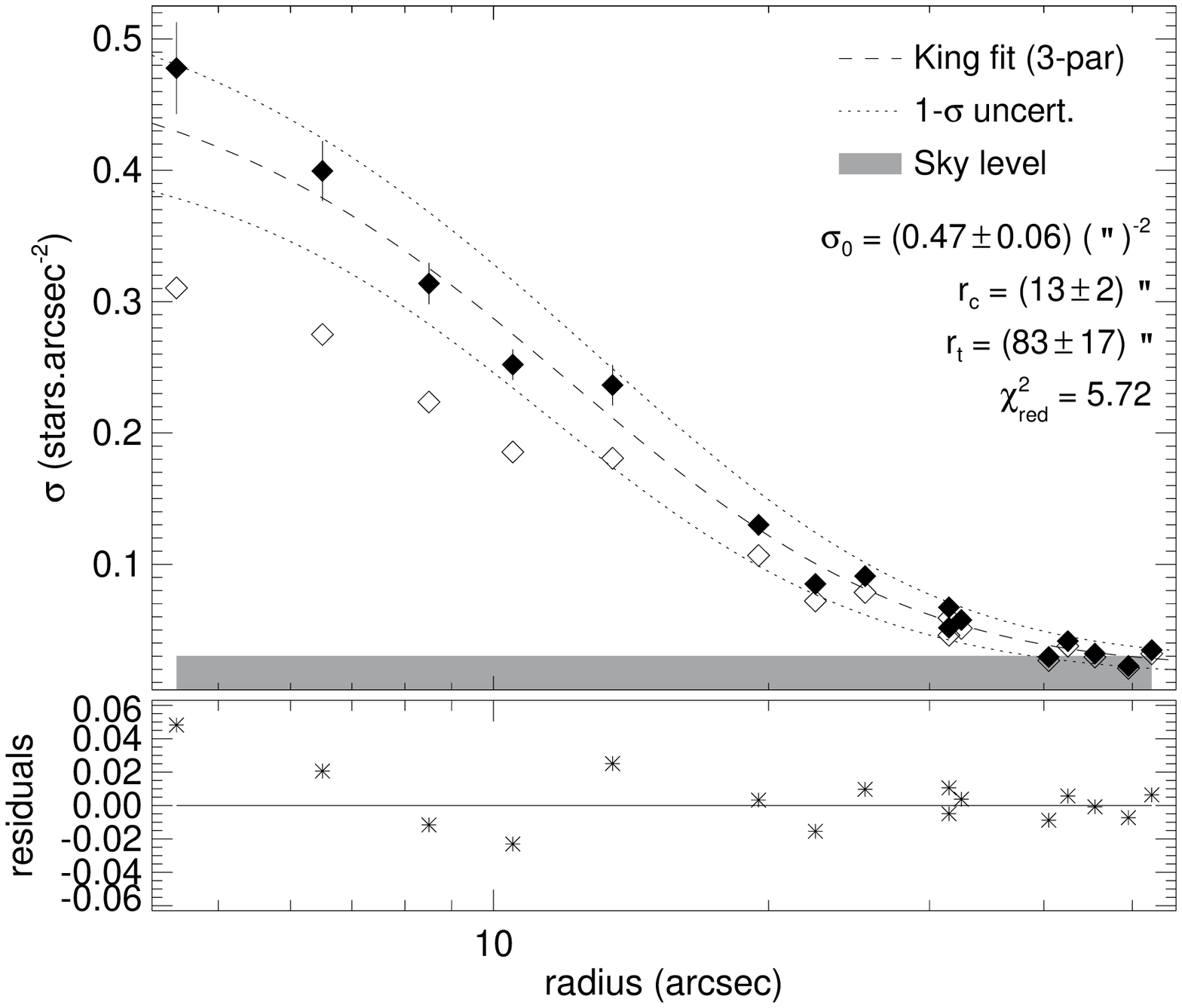}
\caption{Top panel: fit to the SBP of Kron 37 in the $V$ band 
along with the residuals of the fit (bottom sub-panel) and the derived parameters: $\mu_{0}$, $r_c$ and $r_t$. Bottom panel: fit to the completeness corrected stellar density profile of Kron 37 (filled diamonds), for the determination of the parameters $\rho _{0}$, $r_c$ and $r_t$. Residuals of the fit (bottom sub-panel) and the RDP prior to the completeness correction (open diamonds) are also shown. 
}
\label{fig:king}
\end{figure}

The King model \citep{King:1962} parameters --- central surface brightness ($\mu_0$), core radius 
($r_c$) and tidal radius ($r_t$) --- were estimated by fitting the following function to the SBPs:
\begin{equation}
\mu(r)=\mu'_0+5\log\left[\frac{1}{\sqrt{1+(r/r_c)^2}}-\frac{1}{\sqrt{1+(r_t/r_c)^2}} \right]  
\label{eq:king_sbp}
\end{equation}
\noindent
where
\begin{equation}
\mu'_0=\mu_0+5\log\left[1-\frac{1}{\sqrt{1+(r_t/r_c)^2}} \right]  .
\end{equation}
\noindent
The fitting range  was restricted to the cluster limiting radius, defined as the point where the
flux profile reaches an approximately constant level. From the limiting radius outward, the flux 
measurements were used to compute the stellar background/foreground, which was subtracted 
from the profile before fitting. There were cases for which it was not possible to obtain $r_t$ because 
background fluctuations dominate the outer profile. Fig.~\ref{fig:king} (top panel) shows the fit 
of Eq.~\ref{eq:king_sbp} to the SBP of Kron 37. The results for the other clusters in our sample 
can be found in the appendix (Figs.~\ref{fig:SBPs_LMC_king} and \ref{fig:SBPs_SMC_king}).

As a second approach, we have derived the clusters structural parameters from classical radial 
density profiles (RDPs) built from completeness corrected stellar counts \citep[e.g][]
{Maia:2016}, using the King analytical profile:
\begin{equation}
\rho (r) = \rho_0 \left[ \frac{1}{\sqrt{1+(r/r_c)^2}} - \frac{1}{\sqrt{1+(r_t/r_c)^2}} \right]^2 + \rho_\mathrm{bg}.
\label{eq:king_rdp}
\end{equation}
\noindent
Four different bin sizes were used to build the density profile, keeping the smallest bin size at 
about the cluster core radius. The fit for Kron 37 is shown in Fig.~\ref{fig:king} (bottom panel). 
It should be noted that a radial profile without any completeness correction 
(open diamonds) also fits a King profile perfectly well. Although the fit converges, the results 
obtained are not astrophysically meaningful; the tidal radii can be recovered because incompleteness
is not severe there, but the core radii are always in error, usually overestimated by a factor of 2 
or higher. 
The fits for the remaining clusters are shown in Figs.~\ref{fig:RDPs_LMC} and \ref{fig:RDPs_SMC}.

The SBP and the RDP are complementary measurements of cluster structure.
While SBPs are less sensitive to incompleteness than RDPs, a critical issue 
towards the clusters' centre, stochasticity and heterogeneity of field stars 
towards the outer cluster regions make the fluctuations on the SBP background 
much higher than those of the RDP background. Even if this can hinder or even 
make impossible the determination of the tidal radius in SBPs, the problem is 
mitigated in the RDPs, allowing reliable determination of this parameter even 
without completeness correction.

While the SBP uncertainties grow from the cluster centre to its periphery due to progressive 
flux depletion, the RDP uncertainties decrease in this sense as a consequence of the steadily 
rise of the number of stars. By combining the structural parameters obtained from King (1962) 
model fitting to the SBP and to the RDP of the clusters, we expect to minimize such uncertainties
across the entire profile. The parameters' weighted average and uncertainty were calculated as: 
$$  \bar{x}=\dfrac{\sum(x_i/\sigma_i^2)}{\sum(1/\sigma_i^2)} ,$$
\noindent
$$ \sigma_{\bar{x}}=\sqrt{\dfrac{1}{\sum(1/\sigma_i^2)}} .$$
The tidal radii of the clusters K\,37, HW\,20 and KMHK\,228 come only from the RDP because their
fits did not converge for the SBP. Based on the resulting $r_c$ and $r_t$ values, the clusters
concentration parameter $c \equiv \log{(r_t/r_c)}$ \citep{King:1962} was also derived.
Table \ref{tab:structpar} compiles the resulting structural parameters for the present clusters.


\begin{table*}
\caption{Structural parameters of target clusters}
\begin{tabular}{lccc r@{$\pm$}l r@{$\pm$}l r@{$\pm$}l r@{$\pm$}l} \hline
Name & RA & Dec & $\mu_0$ & \multicolumn{2}{c}{$r_c$} & \multicolumn{2}{c}{$r_t$} & \multicolumn{2}{c}{$c$} & \multicolumn{2}{c}{$\sigma_\mathrm{bg}$} \\ 
 & [h:m:s] & [\,$^\circ$\,:\,$\arcmin$\,:\,$\arcsec$\,] & [mag$\cdot$arcsec$^{-2}$] & \multicolumn{2}{c}{[arcsec]} & \multicolumn{2}{c}{[arcsec]} & \multicolumn{2}{c}{ } & \multicolumn{2}{c}{[$10^{-3}\cdot$arcsec$^{-2}$]} \\ \hline
AM3 & 23:48:59 & -72:56:43 & 22.7$\pm$0.3 & 5.6&0.8 & 54&8 & 0.9&0.1 & 1.0&0.1 \\  
HW20 & 00:44:47 & -74:21:46 & 22.6$\pm$0.3 & 10.8&2.0 & 37&11 & 0.5&0.2 & 30.7&9.5 \\ 
K37 & 00:57:47 & -74:19:36 & 20.8$\pm$0.2 & 11.3&1.5 & 83&17 & 0.8&0.1 & 23.6&6.7 \\
NGC796 & 01:56:44 & -74:13:10 & 18.4$\pm$0.3 & 3.2&0.5 & 97&9 & 1.2&0.1 & 1.5&0.5 \\  
KMHK228 & 04:53:03 & -74:00:14 & 23.8$\pm$0.4 & 19.8&5.9 & 68&16 & 0.6&0.2 & 25.6&2.9 \\ 
OHSC3 & 04:56:36 & -75:14:29 & 19.4$\pm$0.7 & 4.3&0.7 & 42&6 & 0.9&0.1 & 12.9&3.7 \\ 
SL576 & 05:33:13 & -74:22:08 & 20.0$\pm$0.2 & 10.6&1.3 & 43&5 & 0.6&0.1 & 30&14 \\ 
SL61 & 04:50:45 & -75:31:59 & 22.1$\pm$0.2 & 26.5&2.6 & 162&44 & 0.8&0.2 & 0.1&6.2 \\ 
SL897 & 06:33:01 & -71:07:40 & 21.2$\pm$0.2 & 12.0&1.7 & 87&9 & 0.9&0.1 & 2.8&0.9 \\ \hline 
\end{tabular} 
\label{tab:structpar}
\end{table*}

\subsection{Isochrone fitting}
\label{sec:analysis_iso}

For the analysis of the photometric data, we initially used the structural parameters to define the cluster and field samples within each observed field. Usually all stars inside the cluster 
tidal radius were assigned to the cluster sample and the ones outside it to the field sample. 
For a few clusters presenting $r_t$ close to or larger than the image boundaries (i.e. leaving no 
field sample), half the tidal radius was employed as a cluster limit instead. 
Integration of the King profiles have shown that depending on the concentration parameter, 75\%
($c\sim$ 0.5) to 99\% ($c \gtrsim$ 1.0) of the cluster population lies within that radius, 
ensuring sufficient source counts in both cluster and field samples. 
The implications of this choice are discussed and accounted for in Sect~\ref{sec:analysis_mf}. 
Then, a decontamination procedure \citep{Maia:2010} 
was applied to statistically probe and remove the most probable field contaminants from the 
cluster region, based on both the positional and the photometric characteristics of the stars, comparing the cluster and field regions defined above.

\begin{figure}
\centering
\includegraphics[width=0.9\linewidth]{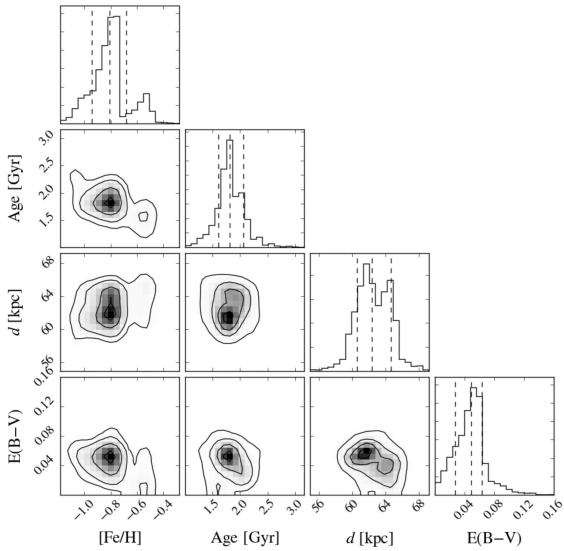}
\caption{Posterior distribution of parameters derived for Kron 37 using a MCMC 
bayesian framework. The derived parameters and their uncertainties are also 
shown.}
\label{fig:bayes}
\end{figure}

\begin{figure}
\centering 
\includegraphics[width=0.9\linewidth]{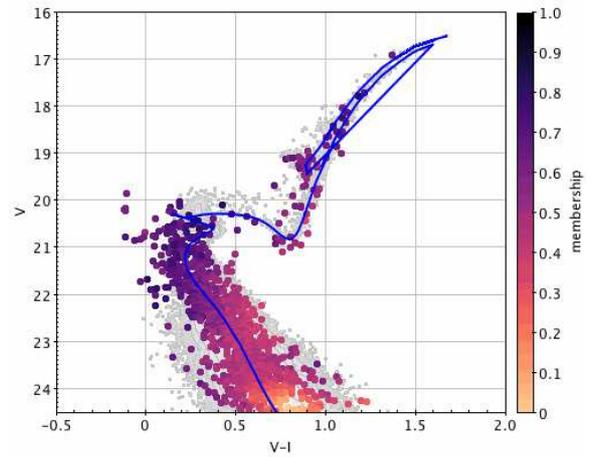}
\caption{Best model isochrone (solid line) and synthetic population (gray dots) 
corresponding to the Kron 37 parameters, superimposed over its field decontaminated CMD. 
}
\label{fig:syntcmd}
\end{figure}

The field decontaminated CMD of the clusters were then used to derive their astrophysical parameters via the Markov-Chain Monte-Carlo technique in a Bayesian framework. The likelihood function was derived using PARSEC isochrones \citep{Bressan:2012} to build synthetic CMDs of simple stellar populations, spanning a wide range of parameters \citep[e.g.][]{Dias:2014}. 
Figure~\ref{fig:bayes} shows the posterior distribution of the determined parameters for Kron 37. Typical uncertainties of the method are about $0.15$ dex in metallicity, $10$-$20\%$ in age, $\sim$2 kpc in distance and $\sim$0.02 mag in colour excess. Figure~\ref{fig:syntcmd} shows the best model isochrone and the synthetic population superimposed over the Kron 37 decontaminated CMD. Respective figures for all other SMC and LMC clusters can be found in Appendix \ref{sec:app2}.

The distance estimates were used to convert the core and tidal radii previously derived in 
Sect.~\ref{sec:analysis_rdp} to physical sizes, thus allowing a more meaningful comparison of 
their values. Most of our targets present core sizes of 2-3 pc, with the exceptions of  
NGC796 and OHSC3 which showed more compact cores and SL61 presenting a very inflated one. Tidal sizes 
were mainly found in the range of 10-20 pc, except for K37, NGC796 and SL61, presenting larger 
tidal domains. Table \ref{tab:results} compiles the resulting astrophysical parameters.

\begin{table*}
\centering
\caption{Astrophysical parameters from isochrone and mass function fits.}
\begin{tabular}{lcccccccccc} \hline
Name & galaxy & $r_c$ & $r_t$ & Age & [Fe/H] & E(B-V) & dist. & M$_\mathrm{obs}$ & M$_\mathrm{int}$ & $\alpha$ \\ 
  &  & [pc] & [pc] & [Gyr] & & & [kpc] & [10$^3$ M$_\odot$] & [10$^3$ M$_\odot$] & \\ \hline
AM3 & SMC & 1.76$\pm$0.26 & 17.0$\pm$2.6 & $5.48^{+0.46}_{-0.74}$ & $-1.36^{+0.31}_{-0.25}$ & $0.06^{+0.01}_{-0.02}$ & $64.8^{+2.1}_{-2.0}$ & 0.23$\pm$0.05 & $--$ & $-$0.27$\pm$0.98 \\  
HW20 & SMC & 3.26$\pm$0.61 & 11.2$\pm$3.3 & $1.10^{+0.08}_{-0.14}$ & $-0.55^{+0.13}_{-0.10}$ & $0.07^{+0.02}_{-0.01}$ & $62.2^{+2.5}_{-1.2}$ & 0.56$\pm$0.10 & 2.06$\pm$0.43 & $-$2.51$\pm$0.61 \\
K37 & SMC & 3.42$\pm$0.47 & 25.1$\pm$5.2 & $1.81^{+0.24}_{-0.21}$ & $-0.81^{+0.13}_{-0.14}$ & $0.05^{+0.01}_{-0.02}$ & $62.4^{+2.3}_{-1.8}$ & 2.58$\pm$0.19 & 9.20$\pm$2.03 & $-$1.97$\pm$0.22 \\
NGC796 & Bridge & 0.94$\pm$0.15 & 28.4$\pm$2.9 & $0.04^{+0.01}_{-0.02}$ & $-0.31^{+0.09}_{-0.12}$ & $0.02^{+0.01}_{-0.01}$ & $60.3^{+2.7}_{-2.4}$ & 1.12$\pm$0.22 & 3.60$\pm$0.70 & $-$2.31$\pm$0.17 \\  
KMHK228 & LMC & 5.8$\pm$1.7 & 19.8$\pm$4.7 & $0.88^{+0.33}_{-0.16}$ & $-0.20^{+0.06}_{-0.06}$ & $0.05^{+0.03}_{-0.01}$ & $60.0^{+1.9}_{-2.4}$ & 0.23$\pm$0.05 & 1.35$\pm$0.30 & $-$2.48$\pm$0.52 \\ 
OHSC3 & LMC & 1.01$\pm$0.17 & 9.8$\pm$1.5 & $1.79^{+0.22}_{-0.20}$ & $-0.70^{+0.13}_{-0.24}$ & $0.07^{+0.02}_{-0.02}$ & $48.3^{+2.0}_{-1.8}$ & 0.44$\pm$0.10 & $--$ & $-$1.18$\pm$0.45 \\
SL576 & LMC & 2.64$\pm$0.34 & 10.7$\pm$1.3 & $0.97^{+0.10}_{-0.11}$ & $-0.39^{+0.08}_{-0.12}$ & $0.02^{+0.03}_{-0.01}$ & $51.3^{+1.9}_{-2.4}$ & 1.81$\pm$0.22 & 5.83$\pm$1.09 & $-$2.14$\pm$0.39 \\ 
SL61 & LMC & 6.55$\pm$0.68 & 40$\pm$11 & $2.08^{+0.27}_{-0.21}$ & $-0.44^{+0.14}_{-0.19}$ & $0.10^{+0.02}_{-0.02}$ & $51.0^{+1.5}_{-1.7}$ & 3.02$\pm$0.25 & 7.00$\pm$1.19 & $-$1.72$\pm$0.30 \\ 
SL897 & LMC & 2.65$\pm$0.39 & 19.2$\pm$2.2 & $1.19^{+0.14}_{-0.12}$ & $-0.32^{+0.11}_{-0.14}$ & $0.09^{+0.02}_{-0.02}$ & $45.6^{+2.4}_{-1.6}$ & 1.17$\pm$0.14 & 5.11$\pm$1.07 & $-$2.49$\pm$0.36 \\ \hline 
\end{tabular} 
\label{tab:results}
\end{table*}

\subsection{Stellar mass function fitting}
\label{sec:analysis_mf}

The distribution of mass in a stellar cluster can yield important information on its evolutionary 
state and on the external environment. As none of the studied objects
show any sign of their pre-natal dust or gas given their ages, their stellar components are the only source of their 
gravitational potential \citep[e.g.][]{lada+03}. Thus, the number of member stars and their concentration will determine, 
in addition to the galaxy potential, for how long clusters survive.

To derive the stellar mass distribution of the target clusters, a completeness corrected $M_I$ 
luminosity function (LF) was first built by applying the distance modulus and extinction corrections 
to the stars' magnitudes. Afterwards, the LF was converted to a mass function (MF) employing the 
mass-$M_I$ relation from the clusters' best-fitted model isochrone, using the procedure described 
in \citet{Maia:2014}. The observed cluster mass ($M_\mathrm{obs}$) is then obtained by adding up the 
contributions of individual bins across the MF. 




%

The MF slope was determined by fitting a power law over the cluster mass 
distribution. Following the commonly used notation, our power law can be written as: 
\begin{equation}
\xi(m) = \frac{dN}{dm} = Am^{\alpha},
\label{eq:mf}
\end{equation}
\noindent
where $\alpha$ is the MF slope and $A$ is a normalisation constant. To avoid discontinuities and
multiple values in the $M_I$-mass relationships, the MF slope fitting procedure was restricted to
main sequence stars, thus excluding giants beyond the turn-off. The masses and the stellar MF slopes
obtained for all clusters are shown in Table~\ref{tab:results}. 

\begin{figure}
\centering 
\includegraphics[width=0.9\linewidth]{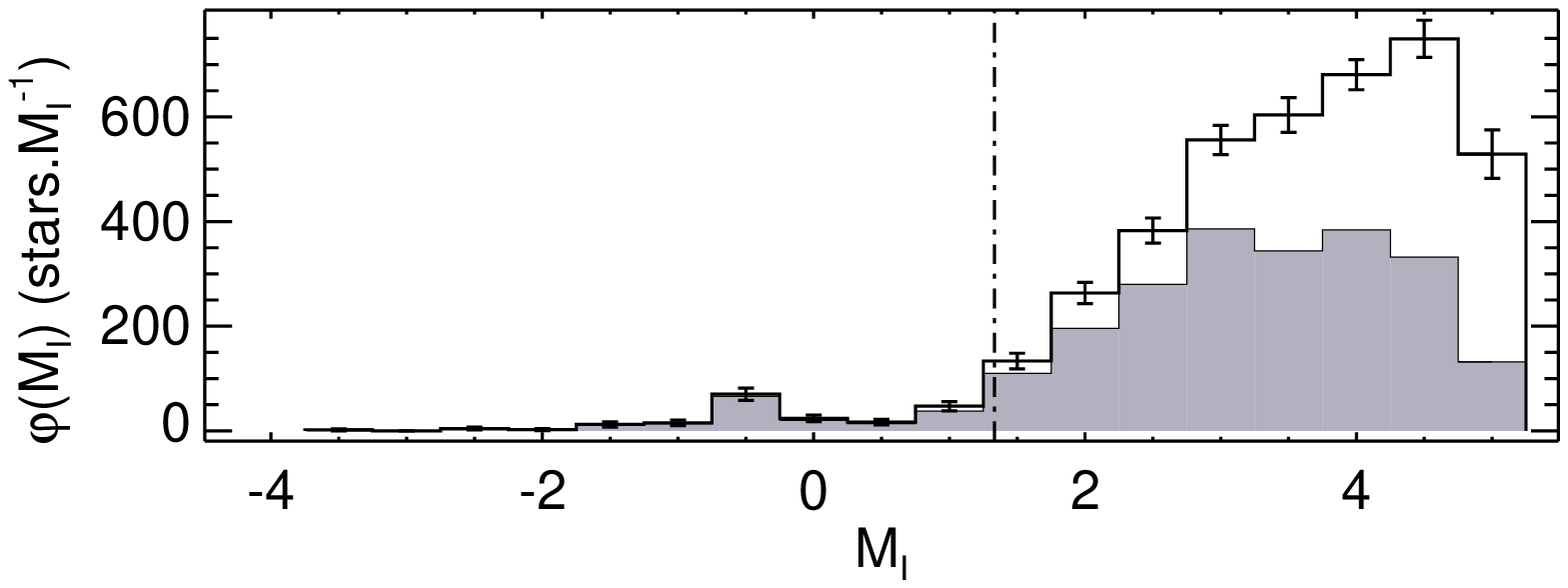}
\includegraphics[width=0.9\linewidth]{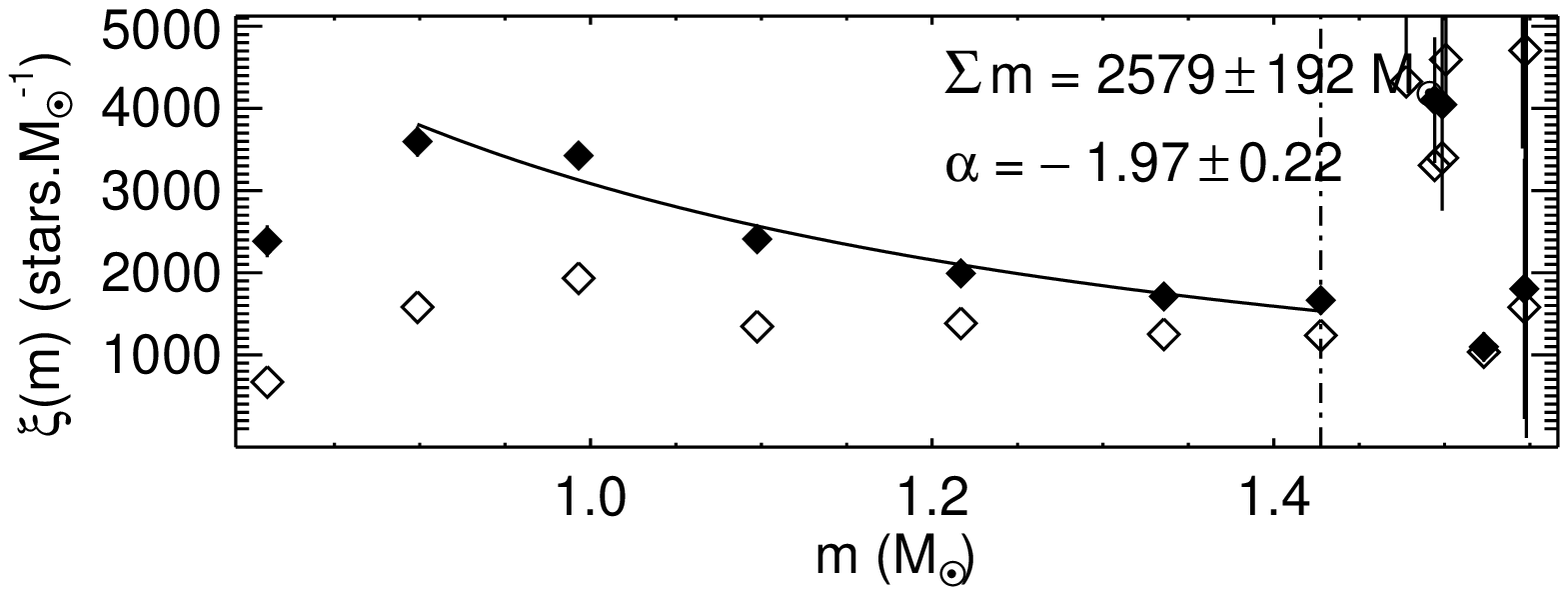}
\caption{Top panel: $M_I$ luminosity function for Kron 37 built with the observed (filled histogram)
and completeness corrected (open histogram) samples. Bottom panel: resulting mass function of Kron
37 from observed (open symbols) and completeness corrected (filled symbols) samples. The turn-off
magnitude and mass are indicated by a vertical dashed line and the best fitting power-law is
represented by the solid line. Total observed mass and resulting MF slope are also indicated.}
\label{fig:mf}
\end{figure}

Fig.~\ref{fig:mf} shows the luminosity function, the resulting mass function and the fit of 
Eq.~\ref{eq:mf} for Kron 37. Figs.~\ref{fig:MFs_SMC} and \ref{fig:MFs_LMC} show the resulting 
LF and MF for the remaining samples clusters. We typically reach stellar masses as low as 0.8
$M_{\odot}$ under good AO performance, and about 1.0 $M_{\odot}$ otherwise. This limit is deeper
than that reached by large surveys in the crowded regions of star clusters (e.g. MCPS will reach
$\sim$2.5 $M_\odot$ at 50\% completeness level for a typical main sequence star in the SMC). We 
note that the spatially resolved completeness correction employed is crucial in probing the 
low-mass regime.

Whenever it could be assumed that a cluster stellar content follows the IMF, i.e. it presents a
(high mass) MF slope that is compatible with the expected value of $\alpha=-2.30 \pm 0.36$ given by 
\citet{Kroupa:2013}, its total mass was estimated by integrating this analytical IMF down to the
theoretical mass limit of 0.08 M$_\odot$. Uncertainties on the IMF analytical parameters and the
normalization constant $A$, derived in the MF fit, were properly propagated into the total
integrated mass (M$_\mathrm{int}$), shown in Table~\ref{tab:results}.

Since clusters K37, NGC796, SL61 and SL897 presented sizes ($r_t$) outside or very close to the 
image boundaries, their mass functions were estimated using only stars inside their inner region 
(within half $r_t$). Their total observed masses were later corrected to their full 
spatial extent based on integrations of their King profiles. 
Given the way the stars are distributed in each cluster, the correction factors amounted 
to 1.01-1.35, being higher for less concentrated clusters like SL61 and almost negligible to 
the concentrated ones like NGC796.

This was also reflected on the MF slope of these two clusters, which were found slightly flatter 
than the IMF, indicating a deficit of low mass content in their inner region. This could be 
interpreted as a sign of mass segregation or preferential loss of the low mass content, depending on
whether these stars are found in the periphery of these clusters or not. Both hypotheses have 
implications regarding the clusters dynamical evolution and the external tidal field acting on them. 

Similarly, AM3 and OHSC3 presented MF slopes significantly flatter than expected by the IMF. Since 
their full extent was sampled by the images, it is possible to assert that severe depletion of 
their lower mass content took place. Their low mass budget and advanced ages makes them specially 
susceptible to stellar evaporation and tidal stripping effects. The remaining clusters showed no
such signs of depletion of their stellar content.

In most cases the total integrated mass is 2-4 times the observable mass of the cluster. This can 
be explained by the shape of the IMF which peaks around 0.5 M$_\odot$, below the minimum observed
mass of $\sim$0.8--1.0 M$_\odot$, implying that most of the cluster mass lies in the less massive
stellar content, unseen by our observations. The errors of the integrated masses are larger than
those of the observed masses because they include (and are dominated by) the uncertainty in the
exponents of the adopted IMF \citep{Kroupa:2013} in this lower mass regime.



\section{First Results}
\label{sec:results}

Tables \ref{tab:structpar} and \ref{tab:results} summarize the parameters determined for a sample of 9 clusters from the present data set. These were chosen to represent the large variety of cluster types found, in terms of richness, ages, metal content and density. In this section, we discuss our results in comparison with those provided in the literature. Many clusters had their ages previously derived from integrated photometry and ours are the first estimates based on stellar isochrone fitting. Similarly, distances and/or metallicities were often assumed constant in previous photometric studies, making our values the first set of simultaneously derived, self-consistent parameters. In addition, determinations of most of the clusters' mass budgets and mass distributions were done for the first time in this work. Particularly, we derived for the first time the considered astrophysical parameters for HW20 and KMHK228. We discuss below the results for each cluster and compare them with the available literature.

\subsubsection*{OHSC\,3 (LMC)}
From integrated spectroscopy, \cite{Dutra:2001} obtained an age of 1-2\,Gyr for OHSC\,3, in agreement with our determination, and reddening $E(B-V)$=0.12 from \cite{Schlegel:1998} dust maps, a little over our estimate from isochrone fitting. 

\subsubsection*{SL\,576 (LMC)}
\cite{Bica:1996} derived for SL\,576 an age in the range 200-400\,Myr from the measured integrated colours ($U-B$)=0.08 and ($B-V$)=0.38 and their calibration with \cite{Searle:1980} SWB type. Our analysis gave an age consistent with a much older cluster (0.97\,Gyr). Integrated colours may be affected by stochastic effects from bright field stars superimposed on the cluster direction, specifically in this case a non-member blue star would contribute to lower the cluster integrated colours, and so mimicking a younger cluster. On the other hand, in our photometry this issue was accounted for with the decontamination procedure where any outsider is excluded before the isochrone fitting.

\subsubsection*{SL\,61 (LMC)}
Among the LMC clusters in our sample, SL\,61 (=LW\,79) is the most studied. \cite{Geisler:1997} determined an age of 1.8\,Gyr by measuring the magnitude difference between main sequence turnoff and red clump and using a calibration of this parameter with age. Its integrated colours, ($U-B$)=0.27 and ($B-V$)=0.59, place SL\,61 in the age range 0.6-2.0\,Gyr \citep{Girardi:1995,Bica:1996}. By adopting ($m-M$)$_\circ$ = 18.31 and $E(B-V)$ = 0.08 from independent measurements, \cite{Mateo:1988} performed isochrone fits to the clusters' cleaned CMD built from $BVR$ photometry \citep{Mateo:1987}, obtaining [Fe/H]=0.0 and an age of 1.8\,Gyr or 1.5\,Gyr depending on the stellar models used, with or without overshooting, respectively. \cite{Grocholski:2007} redetermined an age of 1.5\,Gyr based on the cluster photometry by \cite{Mateo:1987} and updated isochrones.  Using the red clump $K$ magnitude, they obtained a distance of $49.9\pm 2.1$\,kpc, and considering \cite{Burstein:1982} extinction maps, a reddening of $E(B-V)=0.11$ was adopted. From a calibration of the Ca\,II triplet with metallicity, \cite{Grocholski:2006} derived [Fe/H]$=-0.35\pm0.04$ from 8 stars and \cite{Olszewski:1991}, using the same technique, obtained [Fe/H]=-0.50 based on a single cluster star. In general, our results are in agreement with those of the literature, which are compatible among themselves. Regarding the cluster age, our value (2.08\,Gyr) is consistent with literature upper estimates given the uncertainties quoted in Table~\ref{tab:results}. Since our deep photometry resolves stars some magnitudes below the turnoff, we are confident of the age derived, because the CMD region most sensitive to age was assessed and thus a reliable isochrone match was possible.  Our derived metallicity is intermediate between those determined from Ca\,II triplet spectra. The same conclusion can be drawn for the reddening and distance derived.

\subsubsection*{SL\,897 (LMC)}
Integrated photometry of SL\,897 (=LW\,483) yielded colours 
($U-B$)=0.24 and ($B-V$)=0.56, that are compatible with an intermediate-age 
(400-800\,Myr) cluster \citep{Bica:1996}.
\cite{Piatti:2016} investigated the cluster by means of $gi$ photometry using the 8-m Gemini-S telescope obtaining a deep, high quality CMD. Isochrone fits to a cleaned CMD determined an age of $1.25\pm0.15$\,Gyr by adopting initial values of metallicity ([Fe/H]=-0.4), reddening ($E(B-V)$=0.075) and distance modulus  (($m-M$)$_\circ=18.49\pm0.09$) from previous observational constraints. Recalling that in our analysis all parameters were free in the search for the best solution, we found similar age, metallicity and reddening (see Table~\ref{tab:results}). As for the  distance, our study places the cluster closer than the LMC average, the value used by \cite{Piatti:2016}. 

This is also the only cluster in our LMC sample that had its structural properties previously investigated, allowing a direct comparison with our results. \cite{Piatti:2016} derived $r_c=2.7\pm0.5$\,pc and $r_t=36.4\pm2.4$\,pc from star counts. While our determined core radius is similar ($r_c=2.6\pm0.4$\,pc), our tidal radius ($r_t=19.2\pm2.2$\,pc) is considerably smaller, but comparable to their value for the cluster radius ($r_{cls}=21.8\pm1.2$\,pc). Besides the distance difference, we identified two possible reasons for this discrepancy: (i) while \cite{Piatti:2016} RDP extends to $\sim 160\,\arcsec$, ours is restricted to $\sim 80\,\arcsec$ and (ii) their photometry being slightly deeper, it may catch lower mass stars which occupy cluster peripheral regions as a consequence of evaporation and mass segregation. We postpone a detailed analysis of this issue for a forthcoming paper dealing with structural parameters of VISCACHA clusters.

\subsubsection*{KMHK\,228 (LMC)}
For KMHK\,228 we provide astrophysical parameters for the first time.

\subsubsection*{AM\,3 (SMC)}
This is one of the three clusters discovered by \cite{madore+79} who indicated it as the possible westernmost cluster of the SMC. It is also in the west halo group classified by \cite{Dias:2014}. The reddening was derived only by \cite{Dias:2014} as E(B-V)=0.08$\pm$0.05 which agrees very well with our derived value of E(B-V)=0.06$^{+0.01}_{-0.02}$. Distance was only derived by \cite{Dias:2014} as 63.1$^{+1.8}_{-1.7}$ kpc in good agreement with our result of 64.8$^{+2.1}_{-2.0}$ kpc. The age of AM\,3 was derived by \cite{Dias:2014} as 4.9$^{+2.1}_{-1.5}$ Gyr, and also by \cite{piatti+15wash,piatti+11,dacosta99} as 4.5$\pm$0.7 Gyr, 6.0$\pm$1.0 Gyr, and 5-6 Gyr respectively, but the last three fixed distance and reddening values to derive the age. Nevertheless all age estimates agree with ours of 5.5$^{+0.5}_{-0.7}$ Gyr. Metallicity was only derived from photometry so far: [Fe/H] = -0.75$\pm$0.40, -0.8$^{+0.2}_{-0.6}$, -1.25$\pm$0.25, -1.0 by \cite{piatti+15wash,Dias:2014,piatti+11,dacosta99} respectively, and now we derived [Fe/H] = -1.36$^{+0.31}_{-0.25}$. 
This rather large uncertainty in metallicity is owing to the low number of RGB stars to properly trace its slope. We are carrying out a spectroscopic follow-up to better constrain the AM3 metallicity. 

The structural parameters were only derived by \cite{Dias:2014}: $r_c=18.1\pm1.1\arcsec$ and $r_t=62\pm6\arcsec$. The tidal radius agrees with our value of $r_t=54\pm8\arcsec$ and with the estimated size of 0.9$\arcmin$ from the Bica catalogue \citep{bica+95}. The core radius is larger than that derived by us, $r_c=5.6\pm0.8\arcsec$. The difference comes from the unresolved stars in the centre of the cluster using SOI photometry by \cite{Dias:2014}, who derived only the RDP and were limited by some bright stars in the inner region. We could resolve the central stars using AO with SAMI and we confirmed the core radius using the SBP.
\cite{dacosta99} estimated M$_{\rm V}=-3.5\pm0.5$ mag as the total luminosity of AM\,3, which corresponds to M$\sim2.5\times10^3$M$_{\odot}$.  We refrained from calculating a total integrated mass for AM3, given that its MF slope showed heavy depletion of its lower mass stellar content. This behavior implies a smaller contribution from the unseen low mass content, meaning that its integrated mass would be closer to the observed mass budget. 


\subsubsection*{HW\,20 (SMC)}
This cluster belongs to the wing/bridge group in the classification of \cite{Dias:2014}. We derive accurate age, metallicity, distance, and reddening for the first time and found 1.10$^{+0.08}_{-0.14}$ Gyr, [Fe/H] = -0.55$^{+0.13}_{-0.10}$, E(B-V) = 0.07$^{+0.02}_{-0.01}$, d = 62.2$^{+2.5}_{-1.2}$ kpc. The only previous estimatives of age and metallicity were done by \cite{rafelski+05} fitting integrated colours to two models and different metallicities. The combination with smaller error bars is using STARBURST: [Fe/H] $\approx$ -1.3 and age 5.7$^{+0.8}_{-4.3}$ Gyr, which is very different from our determinations. Another combination agrees better with our results but with larger error bars using GALEV: [Fe/H] $\approx$ -0.7, age 1.2$^{+9.1}_{-0.5}$ Gyr.

The structural parameters were derived before by \cite{hill+06}: $r_c = 3.05$ pc and the 90\% light radius as 18.28 pc. The core radius agrees well with our determination of $r_c = 3.26\pm0.61$ pc, but their 90\% radius is significantly larger than the tidal radius derived here: $r_t = 11.2\pm3.3$ pc. The size estimated in the Bica catalogue \citep{bica+95} of 0.75$\arcmin$ agrees better with our tidal radius of 37$\pm$11$\arcsec$. \cite{hill+06} used photometry from the MCPS that is limited to $V<21$ mag while we included also fainter stars down to $V<24$ mag. Figs. \ref{fig:SBPs_SMC_king} and \ref{fig:RDPs_SMC} show that the sky background is high, and that a tidal radius much larger than 11-12$\arcsec$ would not fit the profile. It is possible that the fitting by \cite{hill+06} was limited by a poor determination of the sky background based only on bright stars in a crowded region. \cite{rafelski+05,hill+06} derived $M_{\rm V}$ = 14.97 and 16.2, which corresponds to $M \sim$ 4.3$\times10^3$M$_{\odot}$ and $\sim$ 1.2$\times10^3 M_{\odot}$. Our mass determination is within this range: $M = 2.06\pm 0.43\times10^3 M_{\odot}$.

\subsubsection*{K\,37 (SMC)}
This is also a wing/bridge cluster in the classification of \cite{Dias:2014}. SIMBAD classifies it as an open Galactic cluster, but based on its position and distance, it is probably an SMC cluster. Accurate age was derived only by \cite{piatti+11} as 2.0$\pm$0.3 Gyr based on the magnitude difference between MSTO and RC. \cite{glatt+10} estimated $\sim$1.0 Gyr with error bars larger than 1-2 Gyr based on MCPS photometry that is limited to clusters younger than 1 Gyr. \cite{rafelski+05} derived ages based on integrated colours, and the combination of model, metallicity, and age with smaller error bars led to an age of 1.13$^{+0.05}_{-0.10}$ Gyr for a metallicity of [Fe/H] $\sim$ -0.7. Accurate spectroscopic metallicity was derived by \cite{parisi+15} as [Fe/H] = -0.79$\pm$0.11 based on CaII triplet lines. \cite{piatti+11} derived [Fe/H] = -0.90$\pm$0.25 based on the RGB slope. Although both values agree with ours [Fe/H] = -0.81$^{+0.13}_{-0.14}$ within uncertainties, we call attention to the fact, that the very good agreement with the spectroscopic value gives strength to the VISCACHA metallicities whenever the cluster has enough RGB stars.

The structure parameters from previous works do not agree very well. \cite{hill+06} and \cite{kontizas+85} derived $r_c = 3.36_{-0.92}^{+2.14}$ pc and $r_c = 1.3$ pc, respectively, and our result of $r_c = 3.42\pm0.47$ pc agrees well with the most recent value. The same authors derived $r_{90} = 11.07_{-3.29}^{2.2}$ pc and $r_t = 40.3$ pc and none of them are close to our derived value of $r_t = 25.1\pm5.2$ pc. As the case of HW\,20, our photometry is deeper and our images have better spatial resolution, therefore  we are not biased by bright stars only as it may be the case of the previous works. In fact, our $r_t = 83\pm 17\arcsec$ agrees with the cluster size by \cite{piatti+11} and \cite{bica+95} of r=70$\pm$10$\arcsec$ and 1.0$\arcmin$, respectively,  but not with \cite{glatt+10} who derived r=0.5$\arcmin$. The difference is probably because of their shallow MCPS photometry. All previous integrated magnitudes agree between M$_{\rm V}$=14.1-14.2 \citep{hill+06,rafelski+05,bica+86,gascoigne66}, which means 9-10$\times10^3$M$_{\odot}$, in good agreement with our determination of M=9.2$\pm2.0\times10^3$M$_{\odot}$.

\subsubsection*{NGC\,796 (SMC)}
This is another wing/bridge cluster based on the classification of \cite{Dias:2014}. It is possibly the youngest cluster in the Magellanic Bridge, the only one with an IRAS counterpart, defined by Herbig Ae/Be and OB stars \citep{nishiyama+07}. Accurate age was derived by \cite{kalari+18} who observed the cluster in the very same night as we did using SAMI@SOAR, but using griH$\alpha$ filters. They derived 20$^{+12}_{-5}$ Myr assuming a metallicity of [Fe/H] $< -0.7$. \cite{Bica:2015} derived 42$^{+24}_{-15}$ Myr, which agrees with our determination of 0.04$^{+0.01}_{-0.02}$ Gyr and with the estimates of a young age based on integrated spectroscopy ranging from 3-50Myr \citep{santos+95,ahumada+02}. The older age derived by \cite{piatti+07} of 110$^{+50}_{-20}$ Myr (assuming $d=$56.8\,kpc, $E(B-V)=0.03$, [Fe/H] = -0.7 to -0.4) was explained by \cite{Bica:2015}: their CMD did not include some saturated stars. Metallicity was only derived by \cite{Bica:2015} as [Fe/H] = -0.3$^{+0.2}_{-0.3}$ which agrees very well with our value of [Fe/H] = -0.31$^{+0.09}_{-0.12}$. Reddening is very similar: 0.03 derived by \cite{ahumada+02}, \cite{Bica:2015} and \cite{kalari+18} in agreement with ours of 0.02$^{+0.01}_{-0.01}$. The distance derived by \cite{kalari+18} of 59$\pm$0.8 kpc agrees very well with ours (60.3$^{+2.7}_{-2.4}$ kpc), and the much closer distance of 40.6$\pm$1.1 kpc derived by \cite{Bica:2015} was considered very unlikely by \cite{kalari+18} based on spectroscopic parallax.

The structural parameters were derived by \cite{kontizas+86} and \cite{kalari+18}: $(r_c,r_t) = (0.2, 36.5)$ pc and $ (1.4\pm0.3, 13.9\pm1.2)$ pc, respectively. These values do not agree with each other and our determinations lie in between: 
$(r_c,r_t) = (0.94\pm0.15, 28.4\pm2.9)$ pc. The photometric quality obtained by \cite{kalari+18} is very similar to ours, but they used rings of similar density instead of circles around the cluster centre as we did, and they found anomalies in their fit, possibly because of this choice. Another difference is that they fit \cite{Elson+87} profiles and we fit King profiles. \cite{kalari+18} found an MF slope of $\alpha = -1.99\pm0.2$, similar to the value we found $\alpha = -2.31\pm0.17$. Their derived integrated mass of 990$\pm$220 $M_{\odot}$ considered only stars more massive than 0.5 $M_\odot$, and used their derived MF slope, which is slightly flatter than ours, for integration. In our experience, the stellar content less massive than 0.5 $M_\odot$ usually accounts for roughly half the cluster's integrated mass budget when it can be assumed to follow the IMF. Correcting for this and for the difference in the MF slopes, their reported mass becomes compatible with ours. The integrated magnitude by \cite{gordon+83} of $M_{\rm V} = -0.97\pm0.03$ mag, meaning $M \sim$ 200 M$_{\odot}$, should be taken with caution as the bright stellar content of this young cluster introduces a lot of stochasticity in the integrated magnitudes. Finally the derived mass by \cite{kontizas+86} of $4\times10^3$ $M_{\odot}$ agrees with our determination of $(3.6\pm0.7)\times10^3$ $M_{\odot}$.

\section{Conclusions and Perspectives}
\label{sec:conclusions}

We presented the VISCACHA survey, an observationally homogeneous optical photometric database 
of star clusters in the Magellanic Clouds, most of them located in their outskirts and having low 
surface brightness and for this reason largely neglected in the literature.
Images of high quality (sub-arcsecond) and depth were collected with adaptive optics at the 4-m 
SOAR telescope. Our goals are: (i) to investigate Magellanic Cloud regions as yet unexplored with
such comprehensive, detailed view, in order to establish a more complete chemical enrichment and 
dynamical evolutionary scenario for the Clouds, since their peripheral clusters are the best witnesses 
of the ongoing gravitational interaction among the Clouds and the MW; (ii) to assess relations 
between cluster structural parameters and astrophysical ones, aiming at studying evolutionary effects 
on the clusters' structure associated with the tidal field (location in the galaxy); (iii) to map
the outer cluster population of the Clouds and identify chemical enrichment episodes linked to 
major interaction epochs; (iv) to evaluate the cluster distribution of both galaxies with the 
purpose of establishing the 3D structures of the SMC and the LMC.

In this first paper, the methods used to explore the cluster properties and their 
connections with the Clouds were detailed. 
We have shown that the careful image processing, PSF extraction and calibration methods employed, 
delivered high quality photometric data, unmatched by previous studies. Furthermore, a detailed 
spatially resolved completeness treatment allied with a robust analysis methodology proved crucial 
in deriving corrections to the most commonly used techniques in cluster analysis, 
such as the ones used to determine density profiles, CMDs and luminosity and mass functions. 
A reliable and 
homogeneously derived compilation of astrophysical parameters was provided for a 
sample of 9 clusters. Enlargement of this sample will allow us to better understand 
the galactic environment at the Magellanic Clouds periphery and to address our longer term goals.

In future work we intend to present a more detailed 
analysis of the whole cluster sample on each topic described in this paper, and present more general results concerning both Clouds. 
Then, we shall study the mass function and possible mass segregation, as well as constrain the 
star formation and tidal history in both Clouds.   

\section*{Acknowledgements}

We thank the anonymous referee for the suggestions and critics which helped to improve this manuscript. It is a pleasure to thank the SOAR staff for the efficiency and pleasant times at the telescope, and thus contributing to the accomplishment of VISCACHA. 
F.F.S.M. acknowledge FAPESP funding through the fellowship n$^o$ 2018/05535-3. J.A.H.J. thanks the Brazilian institution CNPq for financial support through postdoctoral fellowship (project 150237/2017-0) and Chilean institution CONICYT, Programa de Astronom\'ia, Fondo ALMA-CONICYT 2017, C\'odigo de proyecto 31170038. A.P.V. acknowledges FAPESP for the postdoctoral fellowship no. 2017/15893-1. This study was financed in part by the Coordena\c c\~ao de Aperfei\c coamento de Pessoal de N\'ivel Superior - Brasil (CAPES) - Finance Code 001. The authors also acknowledge support from the Brazilian Institutions CNPq, FAPESP and FAPEMIG.



\bibliographystyle{mnras}
\bibliography{bibliography} 




\appendix

\section{Structural analysis charts for studied clusters}
\label{sec:app1}

\begin{figure}
\centering
\includegraphics[width=0.24\textwidth]{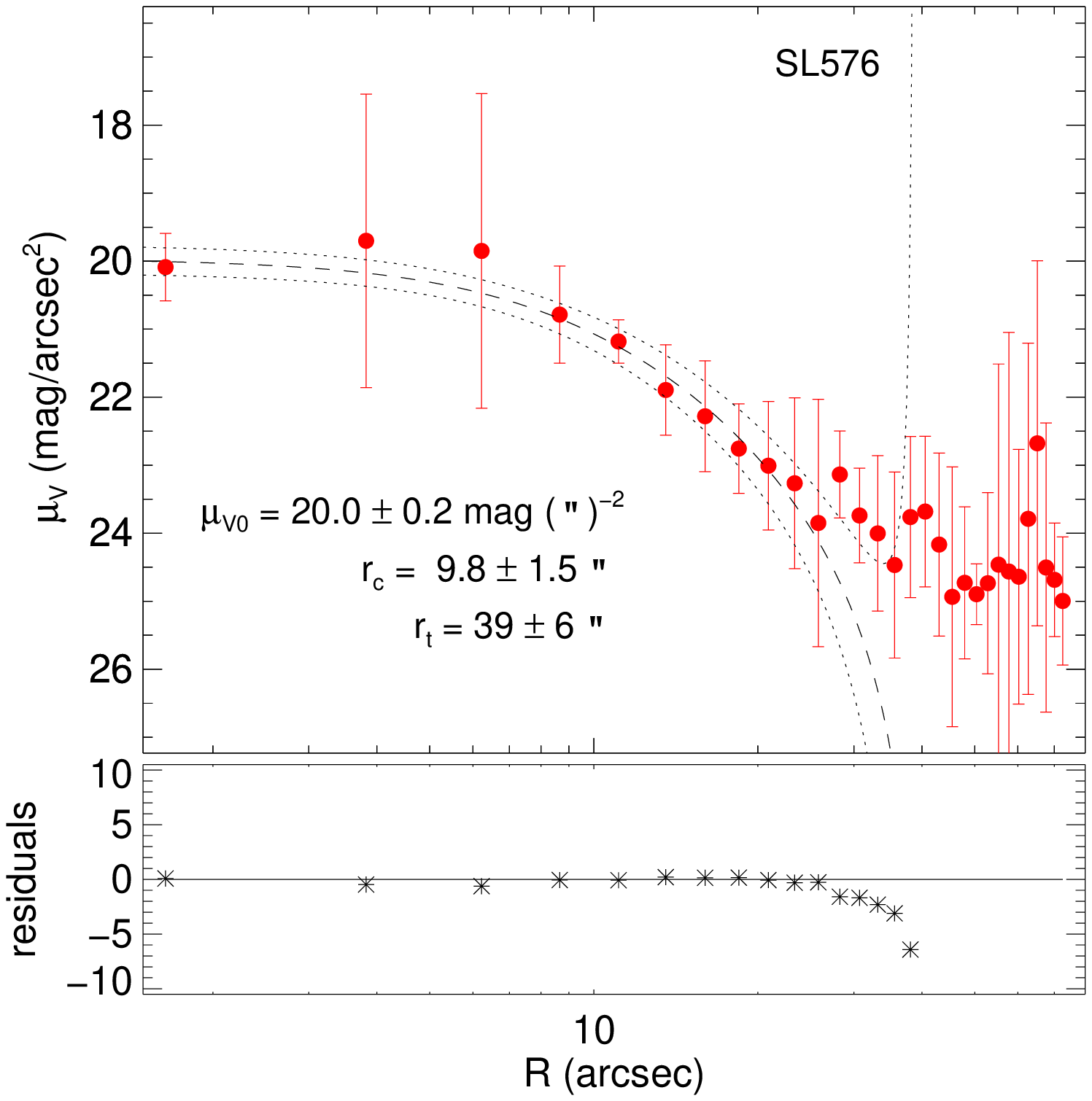}
\includegraphics[width=0.24\textwidth]{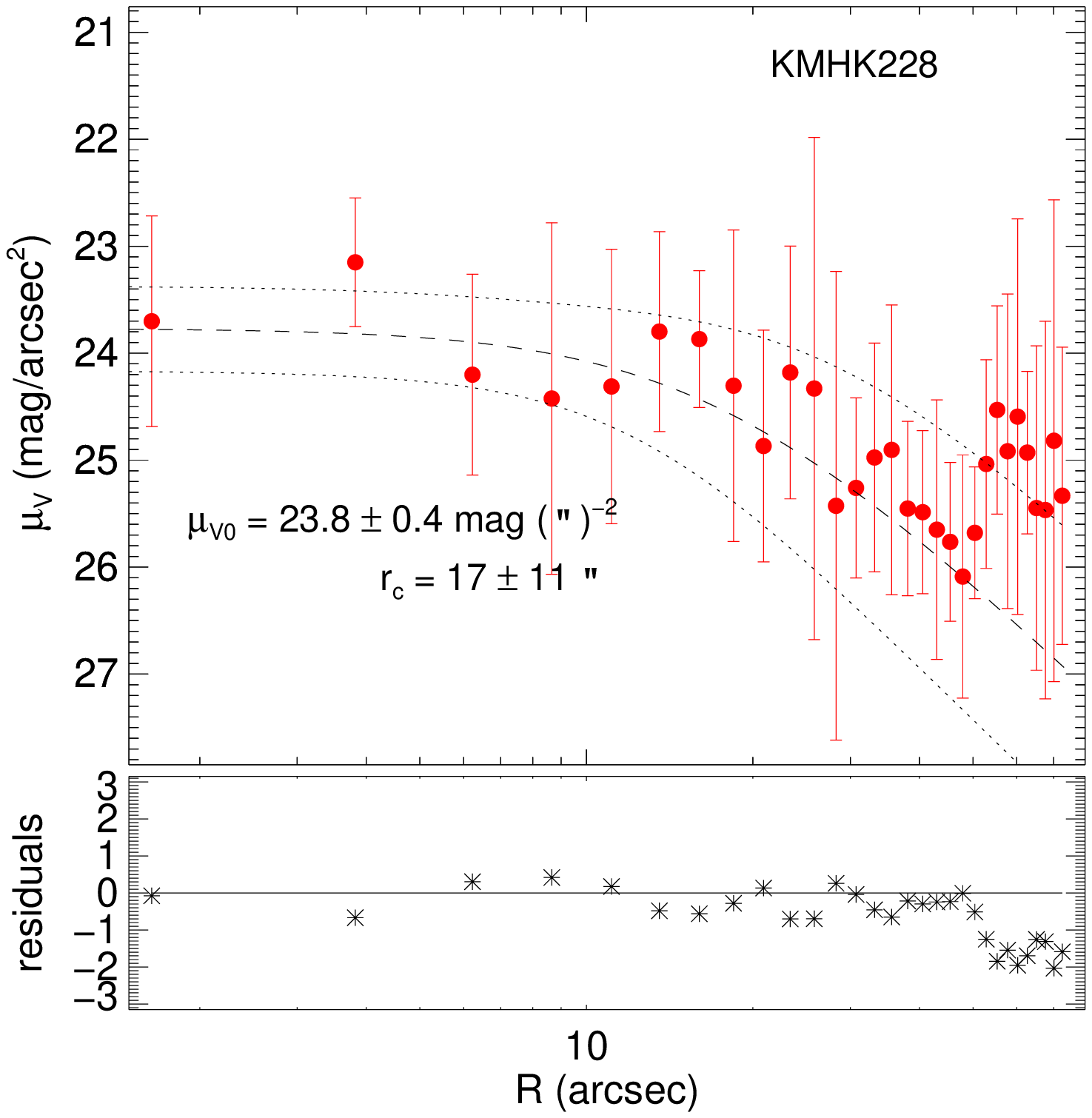}

\includegraphics[width=0.24\textwidth]{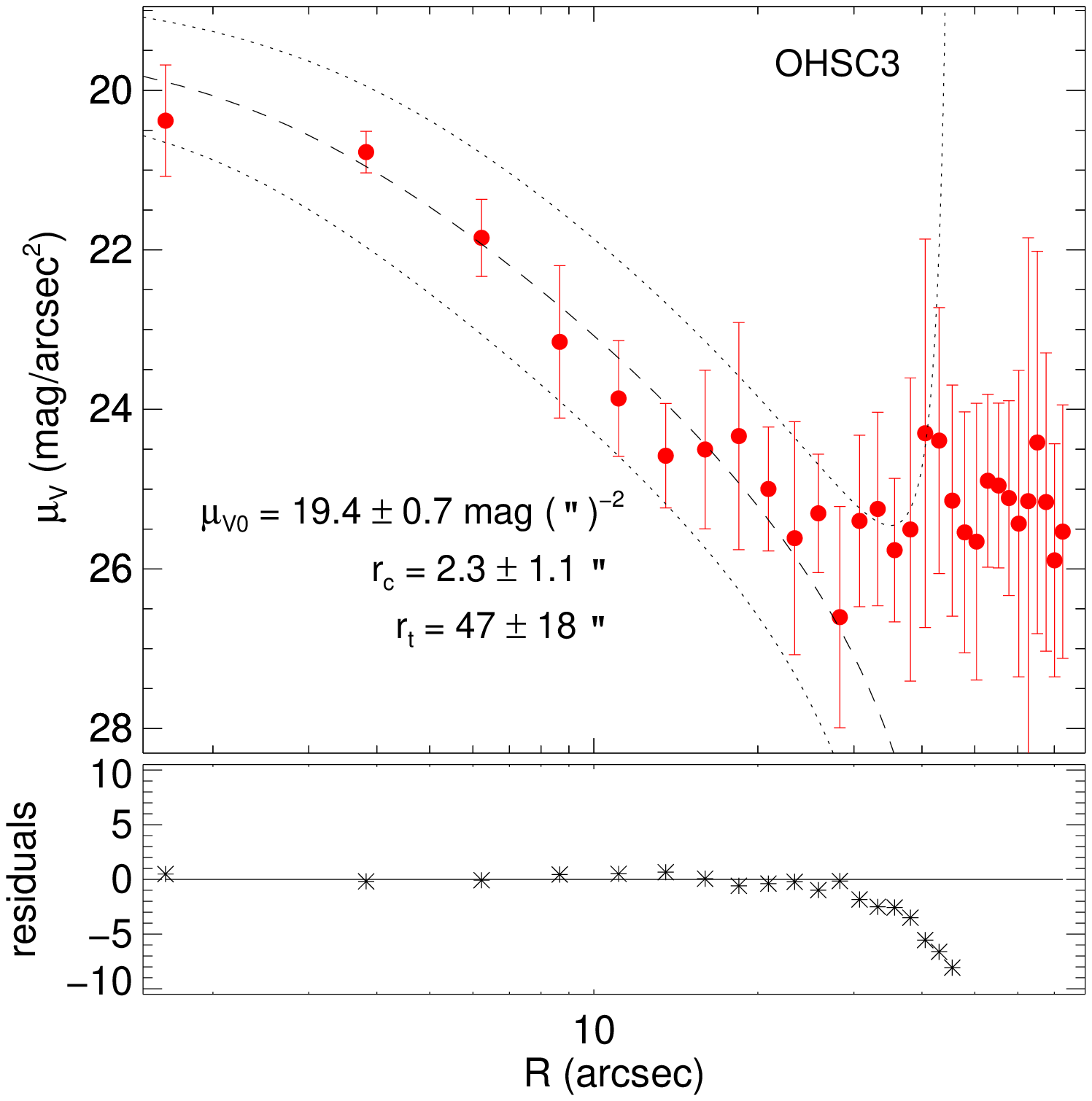}
\includegraphics[width=0.24\textwidth]{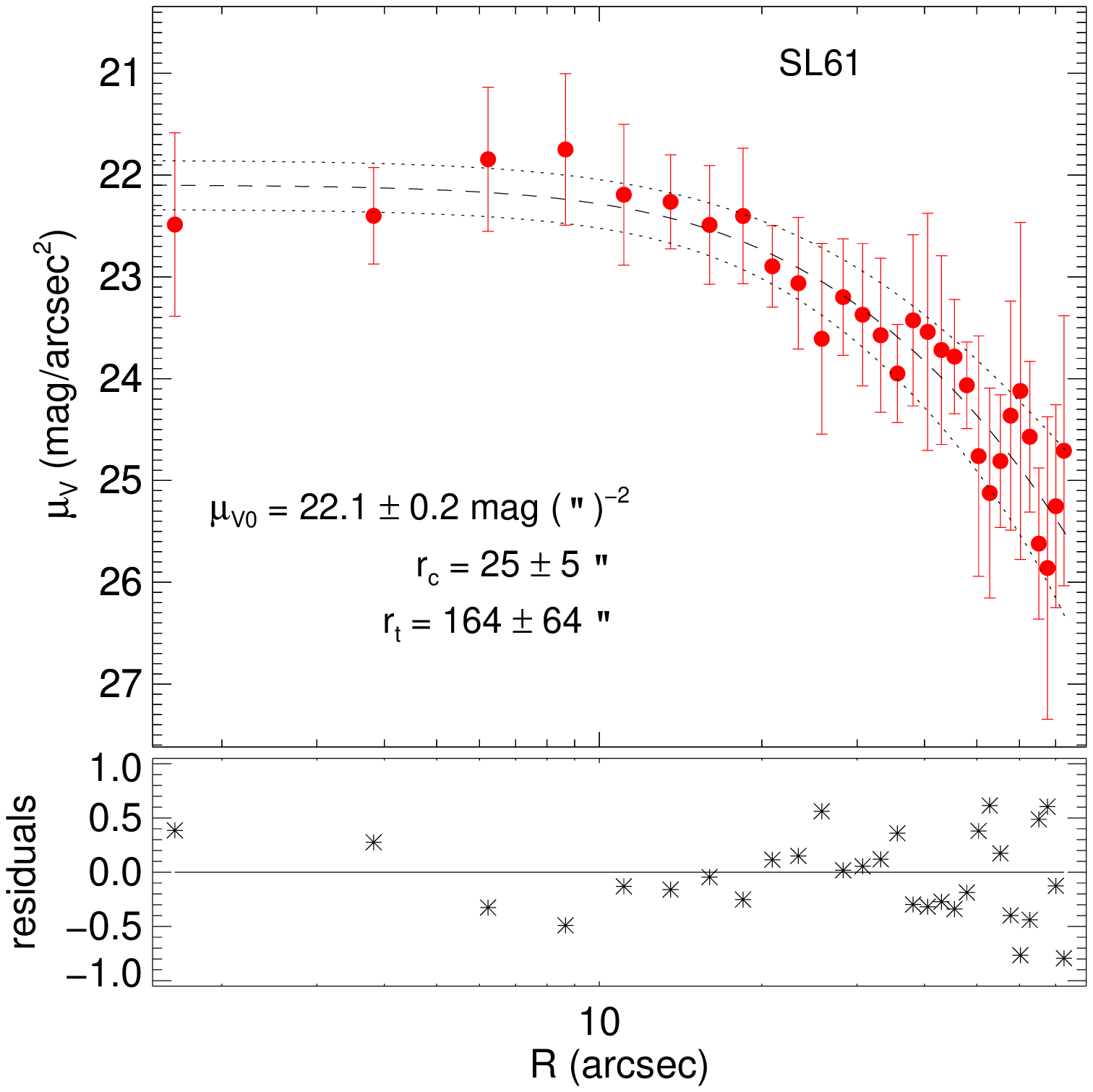}
\caption{King model fits (dashed lines) with 1-sigma uncertainty (dotted lines) to SBPs (red dots and error bars) of clusters SL576, KHMK228, OHSC3 
and SL61 (from top left to bottom right). The lower panel in each plot shows the residuals. The resulting parameters are indicated.}
\label{fig:SBPs_LMC_king}
\end{figure}

\begin{figure}
\centering
\includegraphics[width=0.24\textwidth]{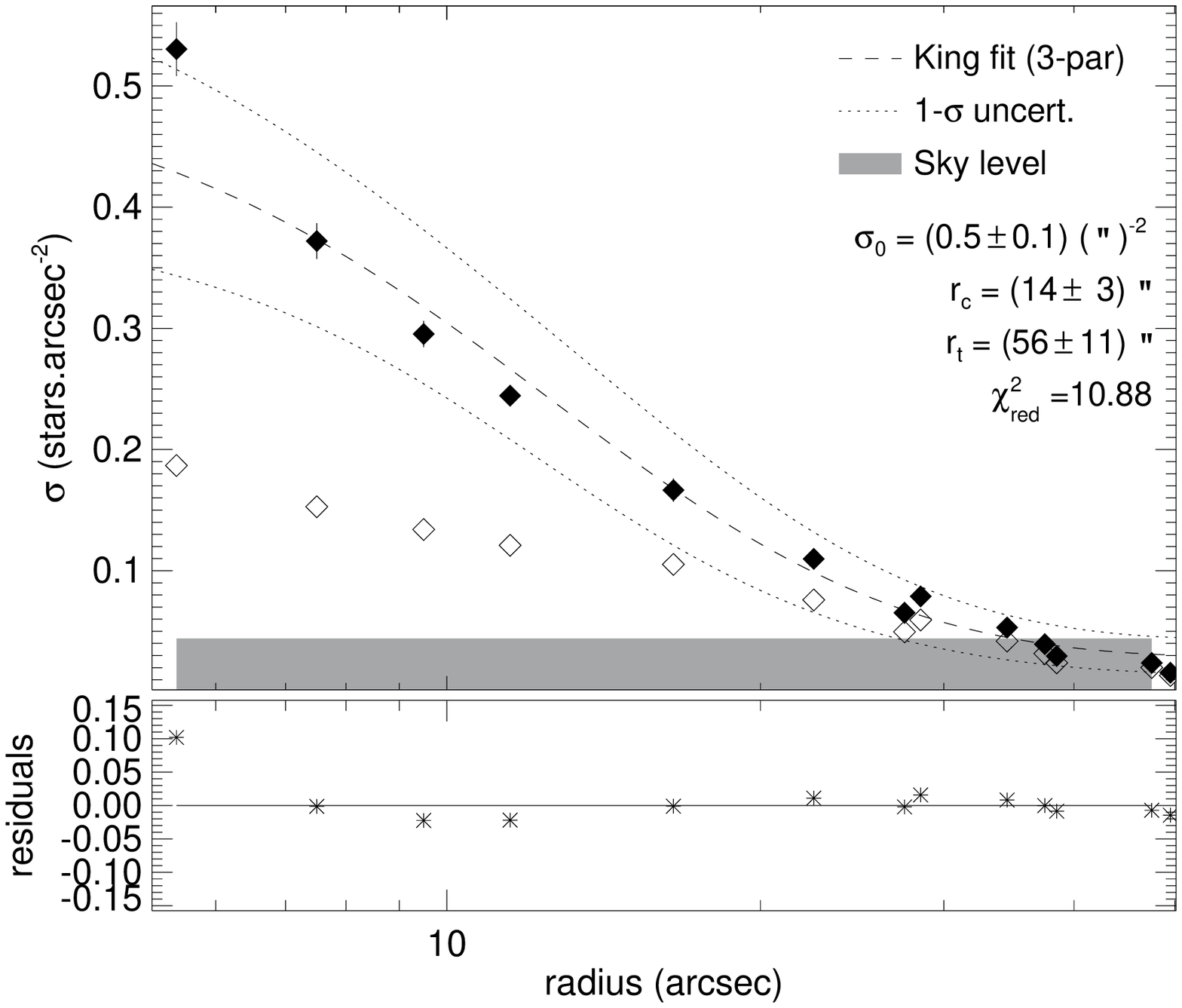}
\includegraphics[width=0.24\textwidth]{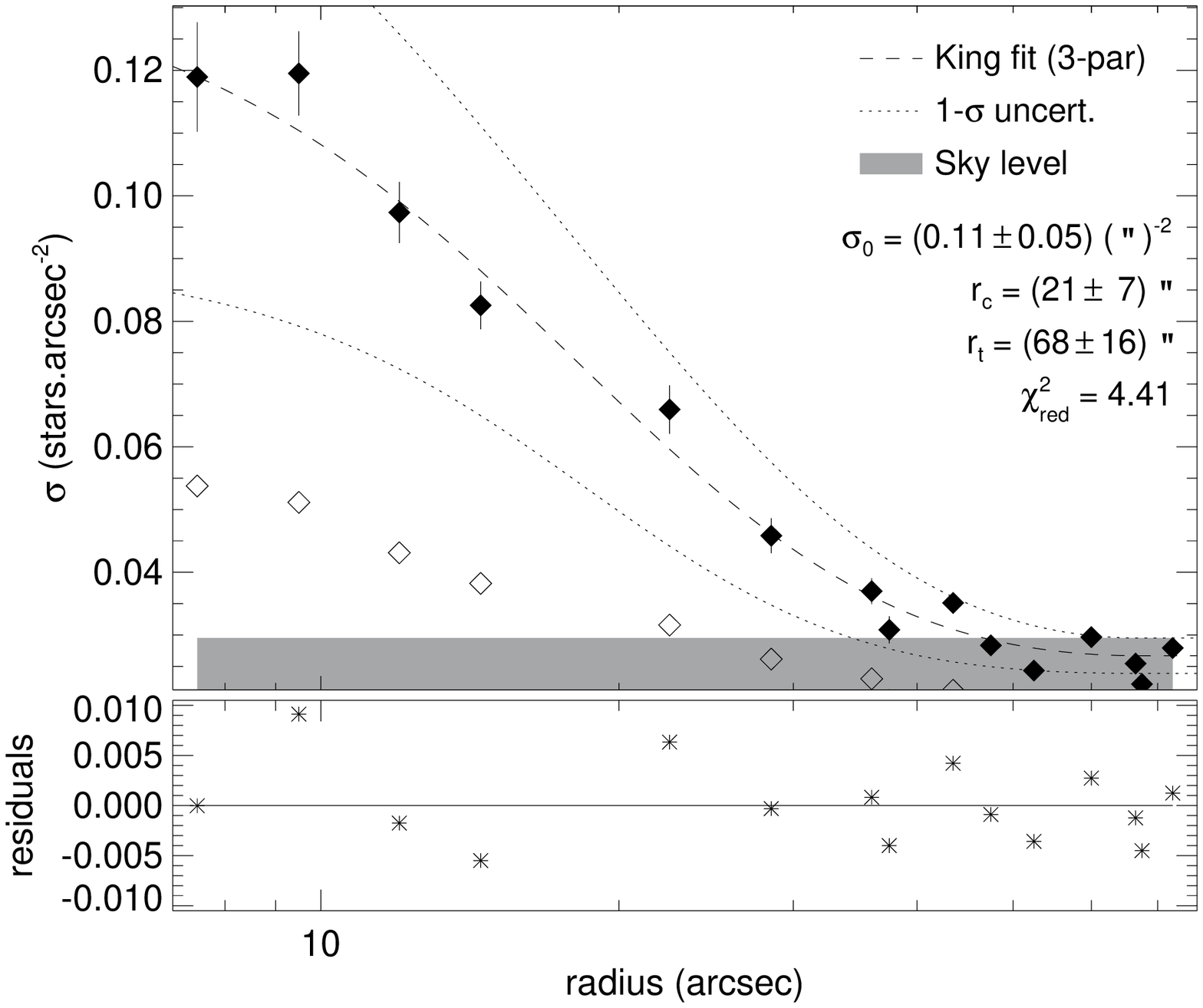}

\includegraphics[width=0.24\textwidth]{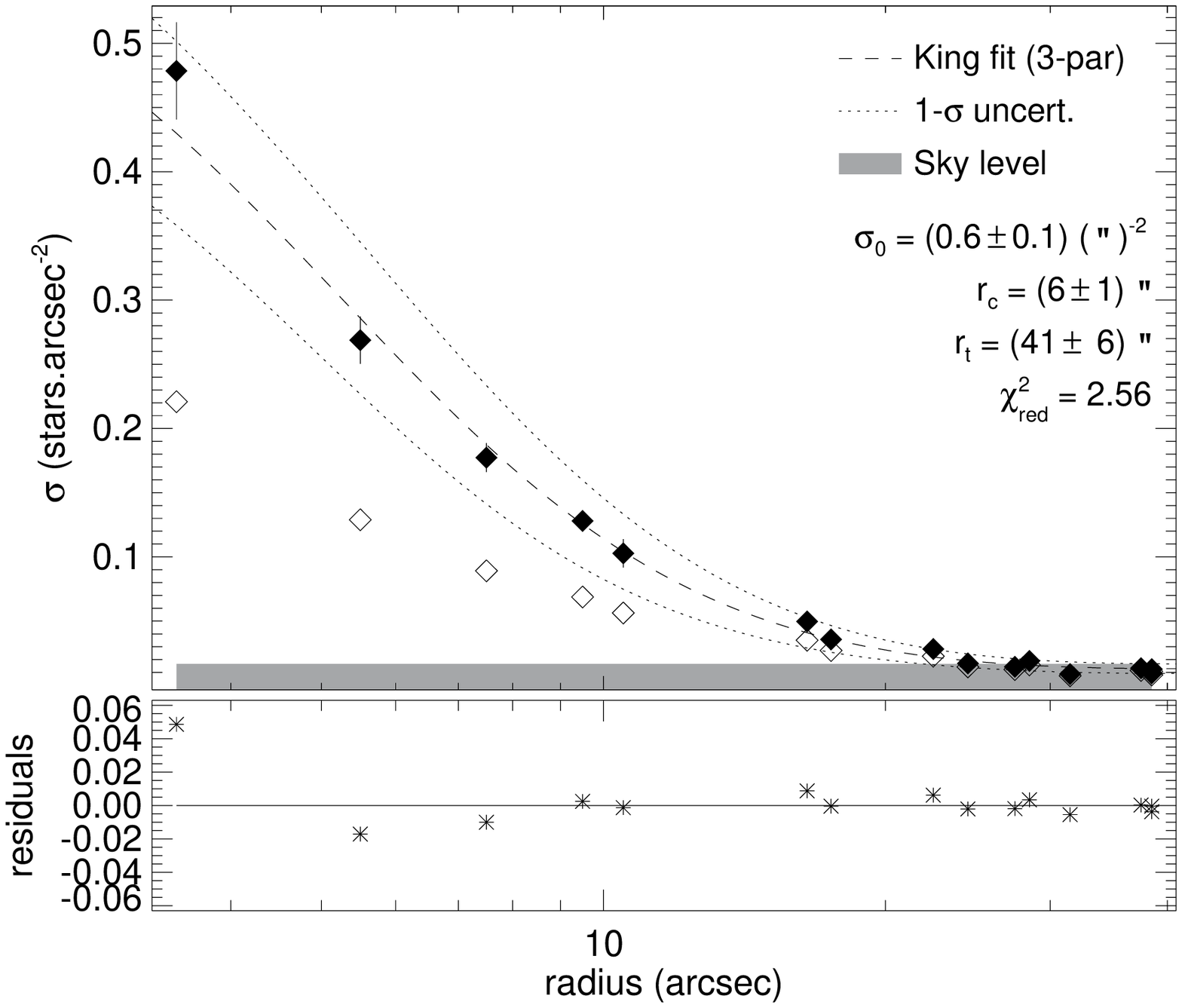}
\includegraphics[width=0.24\textwidth]{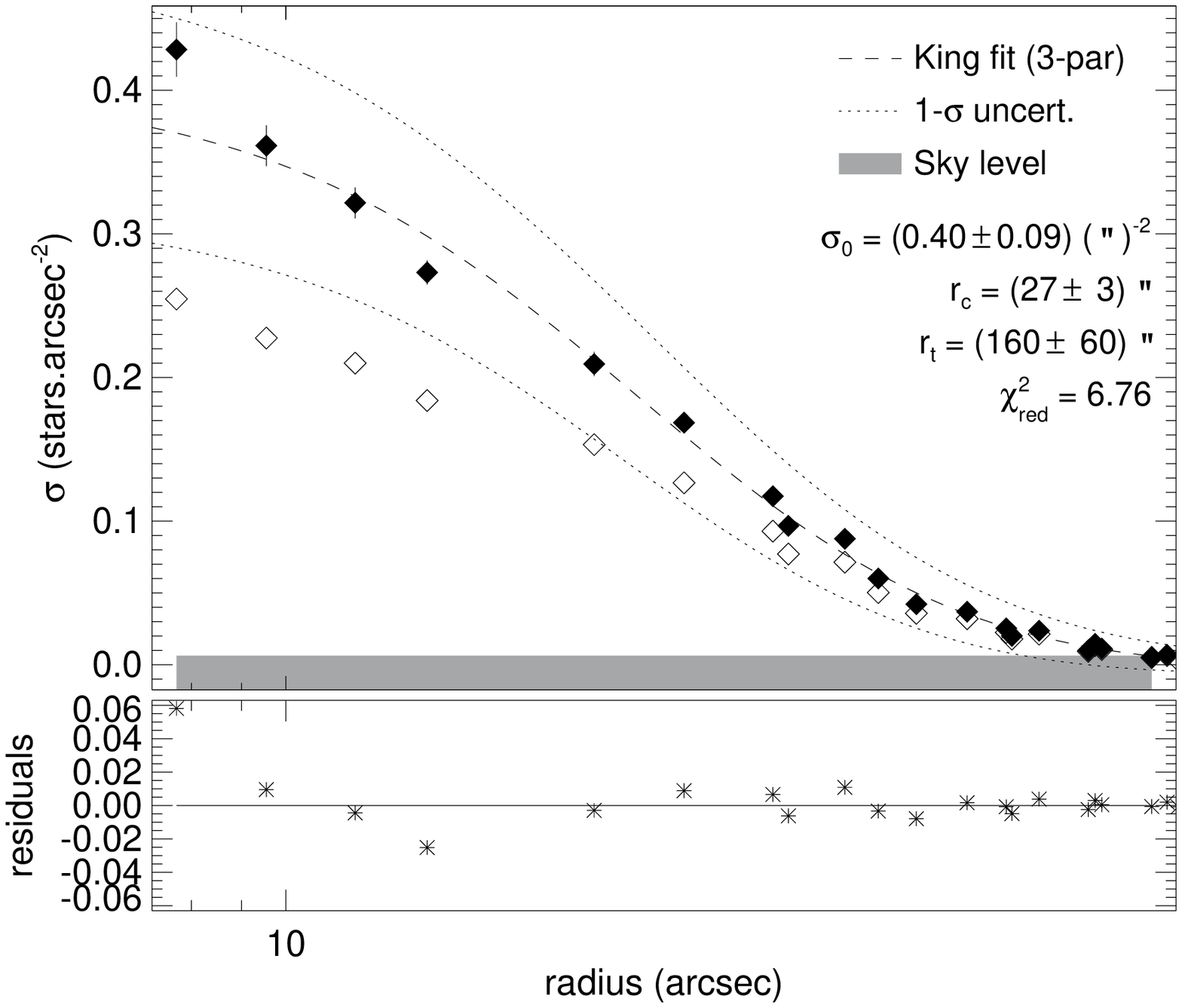}
\caption{King model fits (dashed lines) with 1-sigma uncertainty (dotted lines) to the original (open symbols) and completeness corrected RDPs (filled symbols) of clusters SL576, KMHK228, OHSC3, SL61 (from top left to bottom right). The lower panel in each plot shows the residuals. The resulting parameters are indicated.}
\label{fig:RDPs_LMC}
\end{figure}

\begin{figure}
\centering
\includegraphics[width=0.24\textwidth]{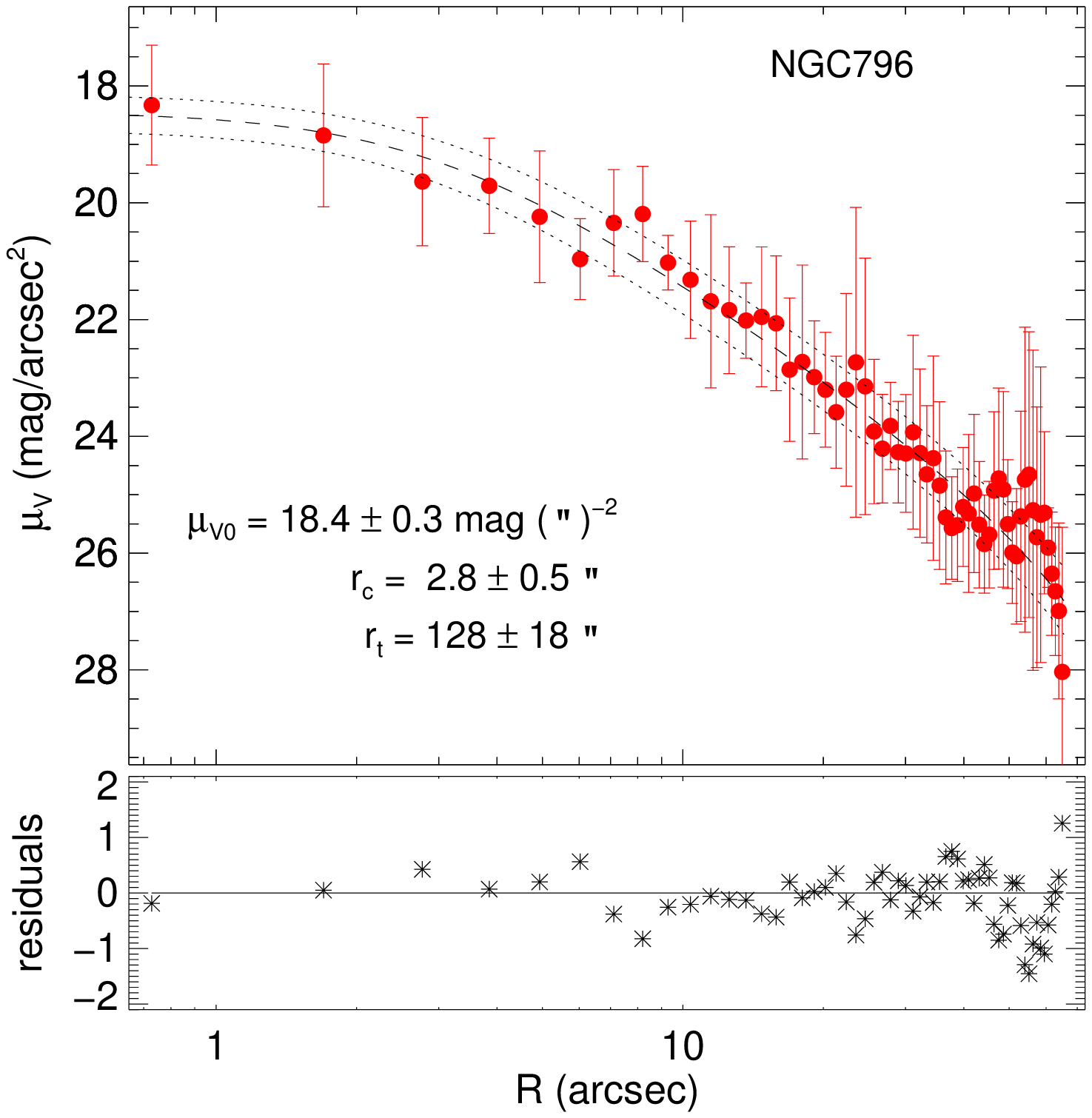}
\includegraphics[width=0.24\textwidth]{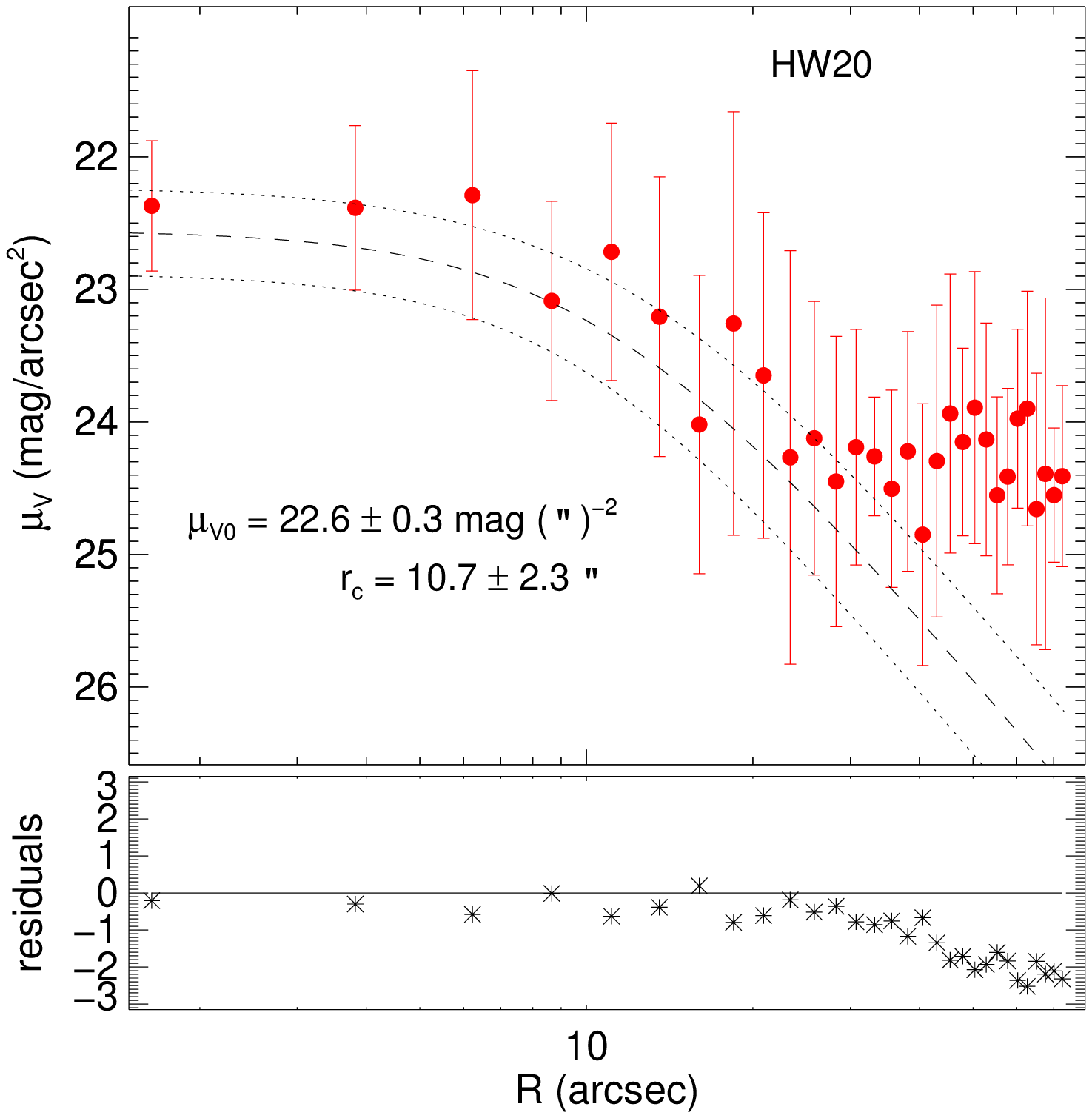}

\includegraphics[width=0.24\textwidth]{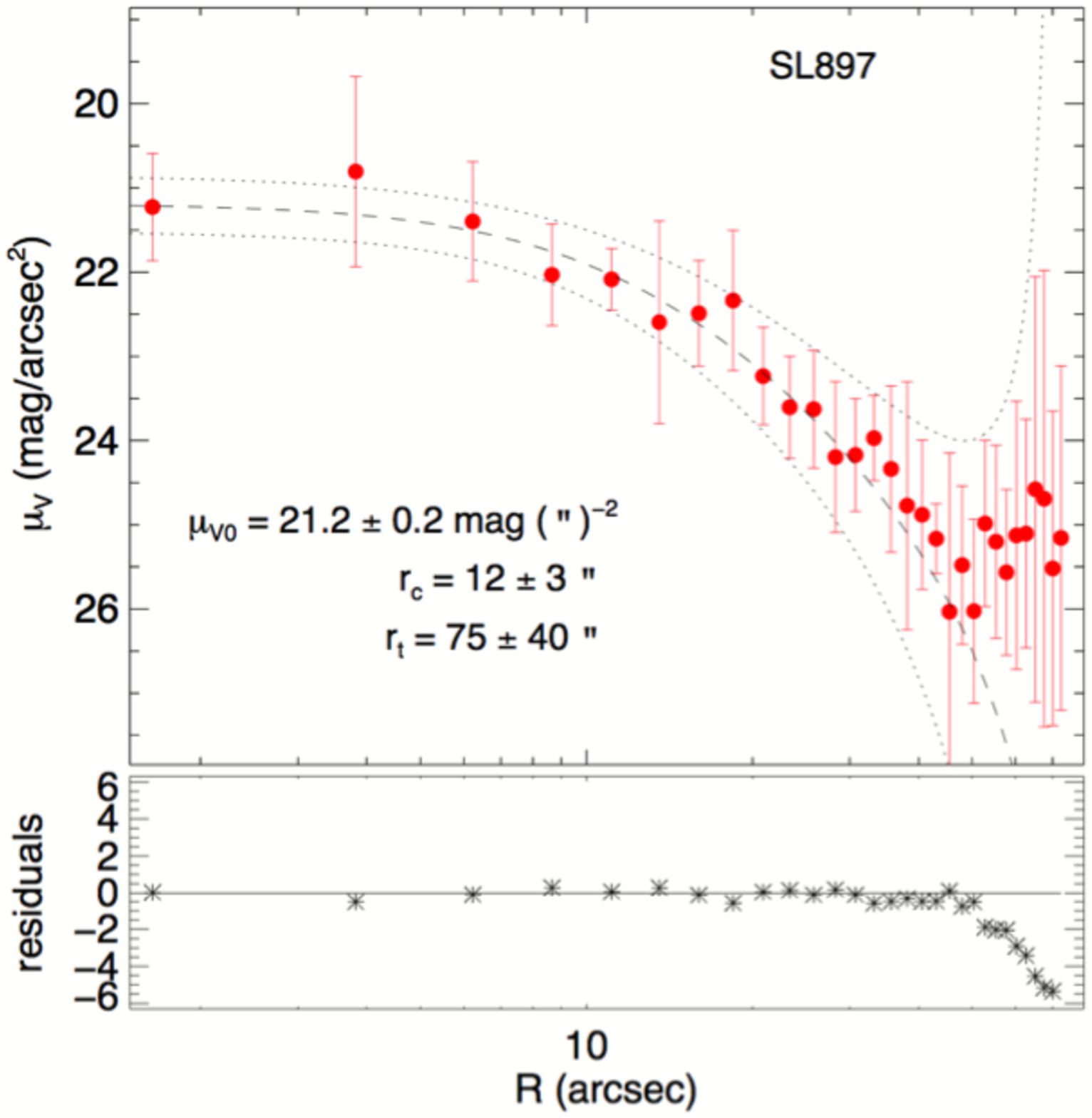}
\includegraphics[width=0.24\textwidth]{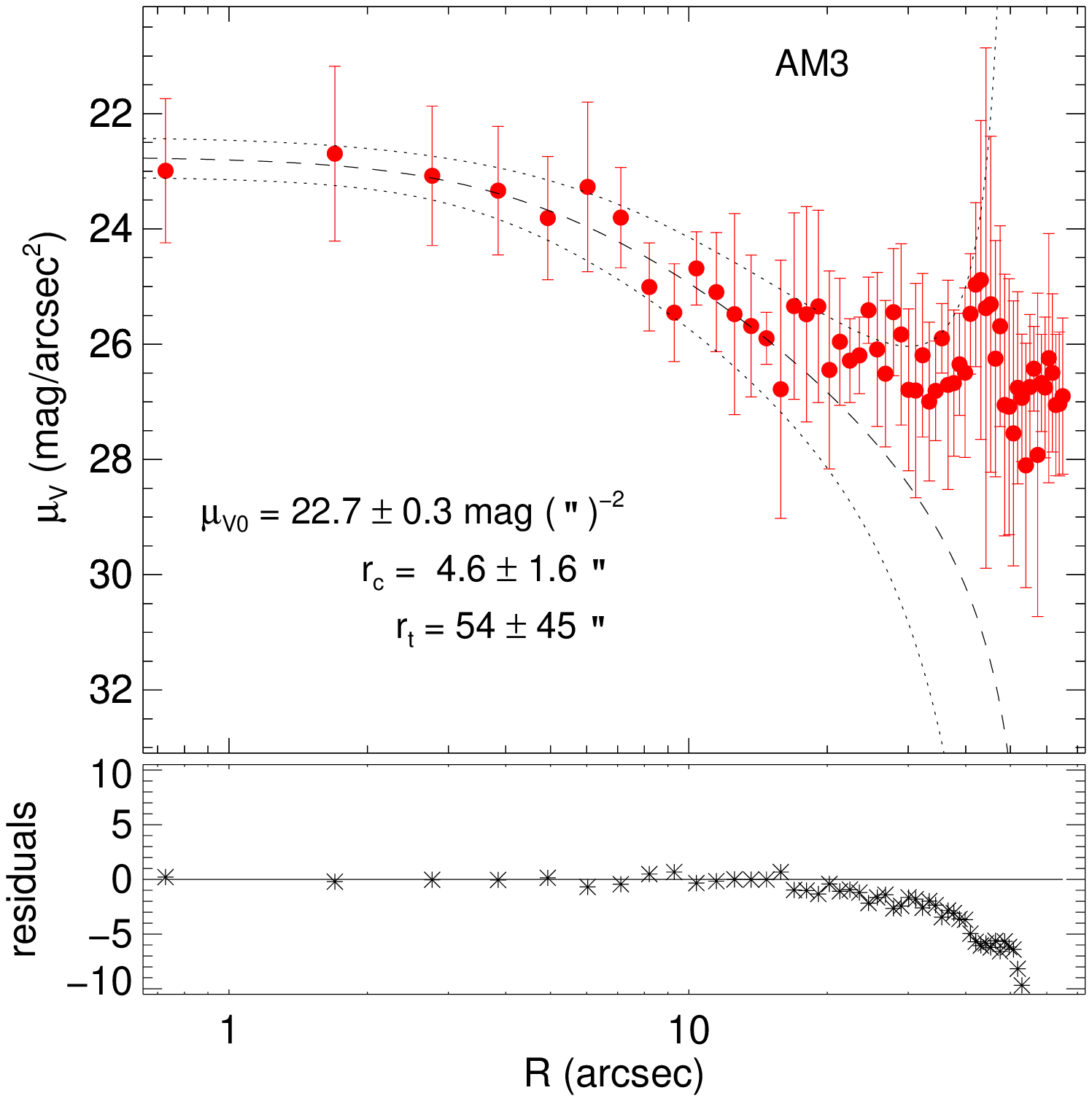}
\caption{King model fits to SBPs of clusters NGC796, HW20, SL897 and AM3 (from top left to bottom right). Details as in Fig.~\ref{fig:SBPs_LMC_king}.} 
\label{fig:SBPs_SMC_king}
\end{figure}

\begin{figure}
\centering
\includegraphics[width=0.24\textwidth]{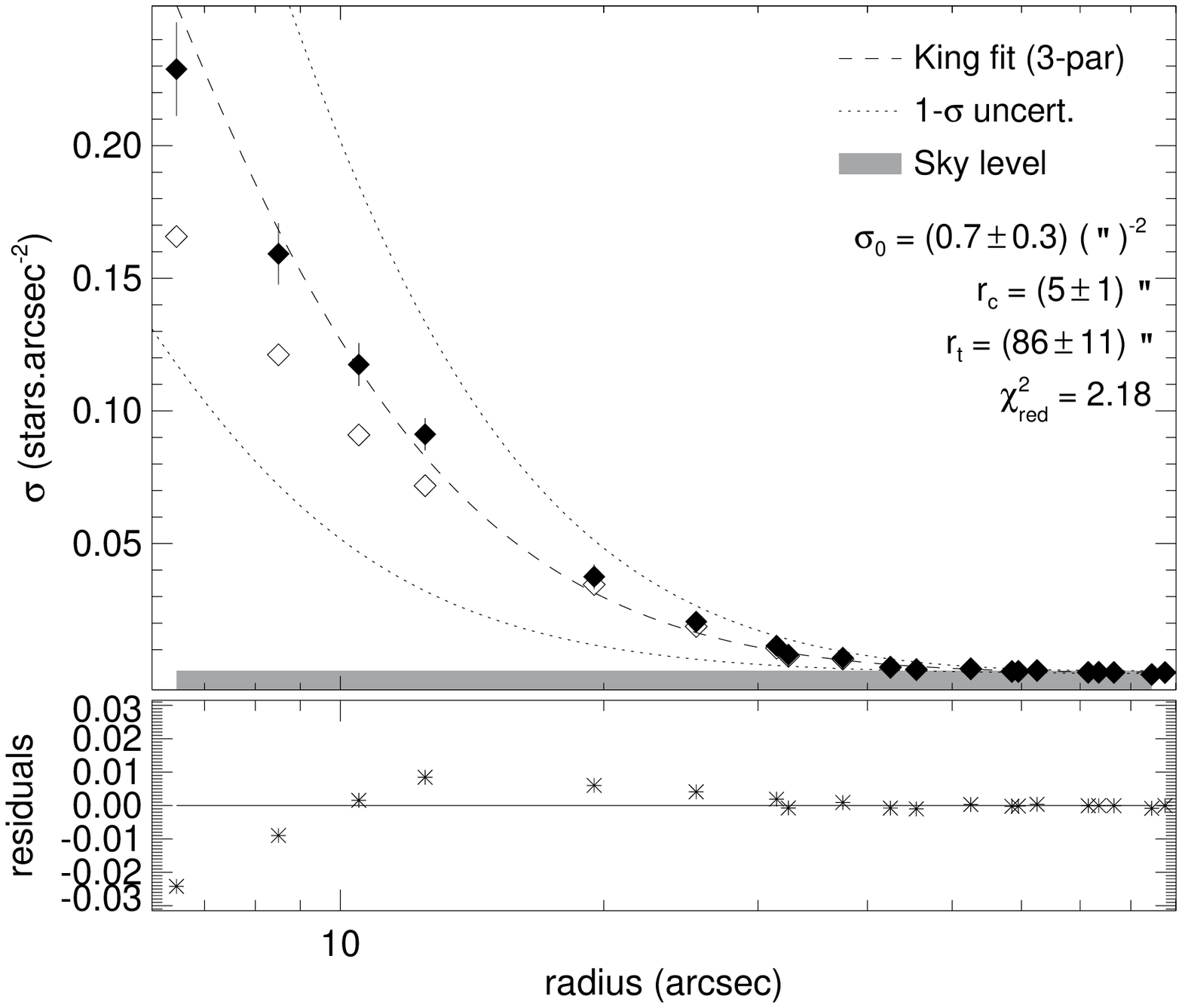}
\includegraphics[width=0.24\textwidth]{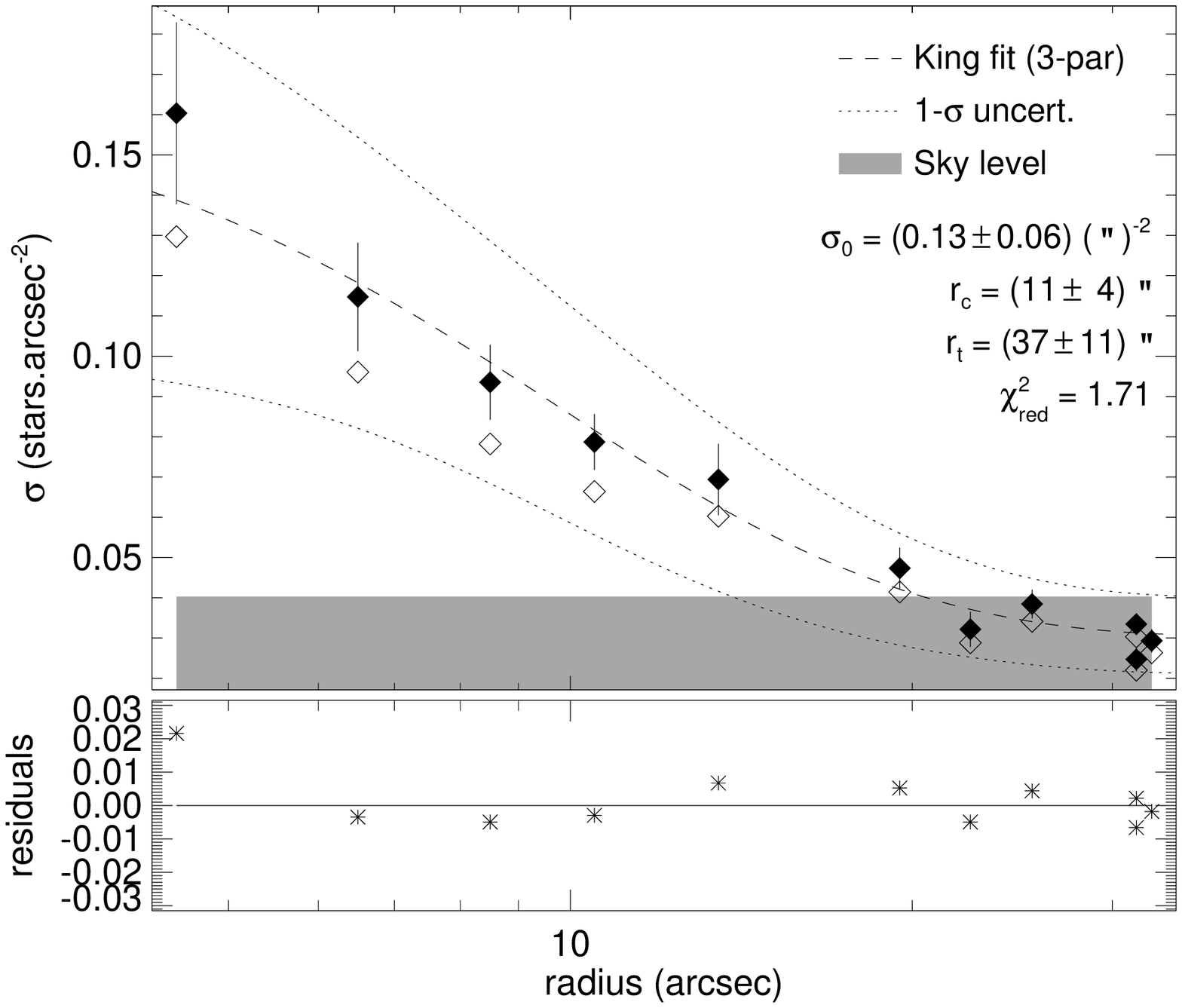}

\includegraphics[width=0.24\textwidth]{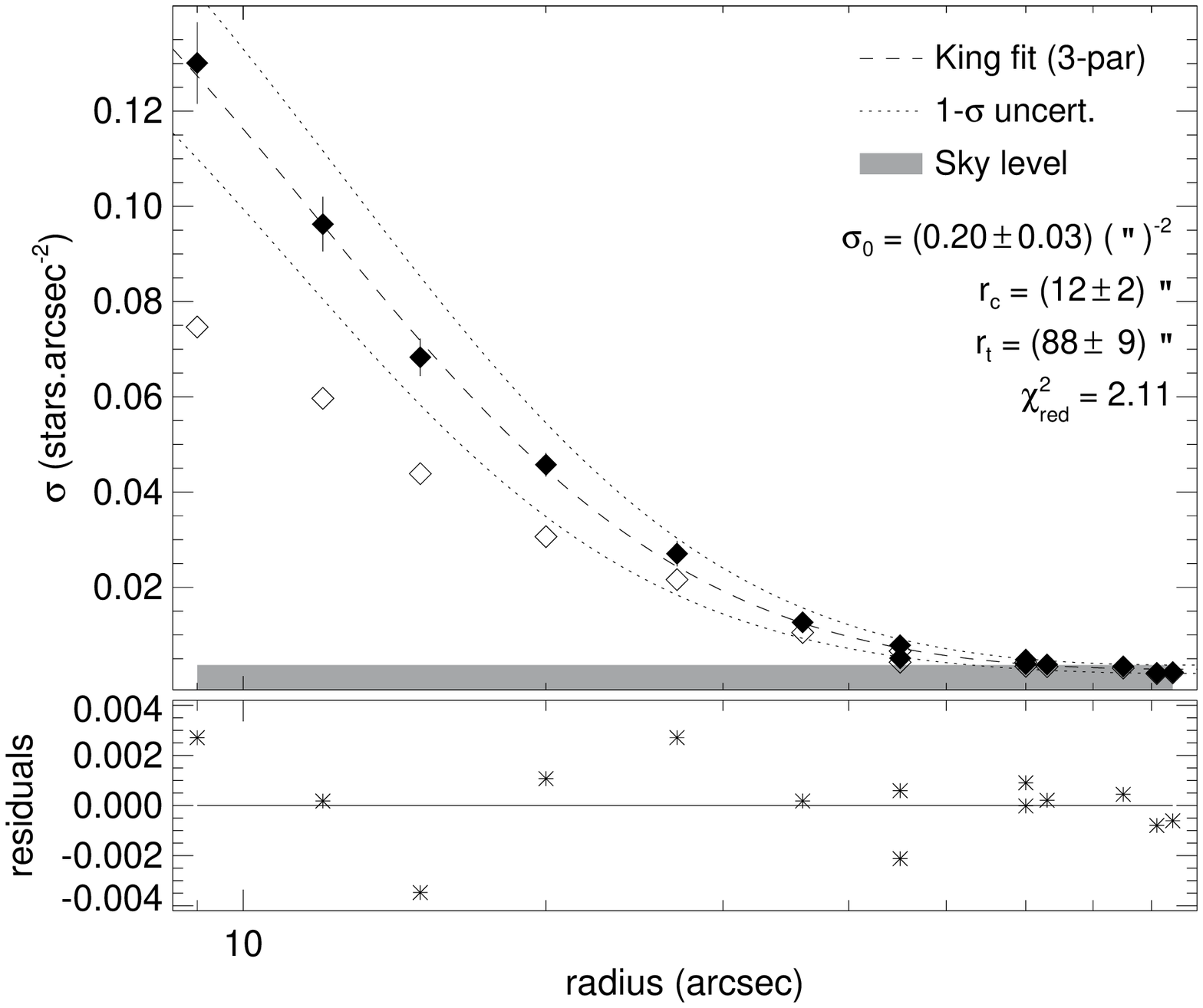}
\includegraphics[width=0.24\textwidth]{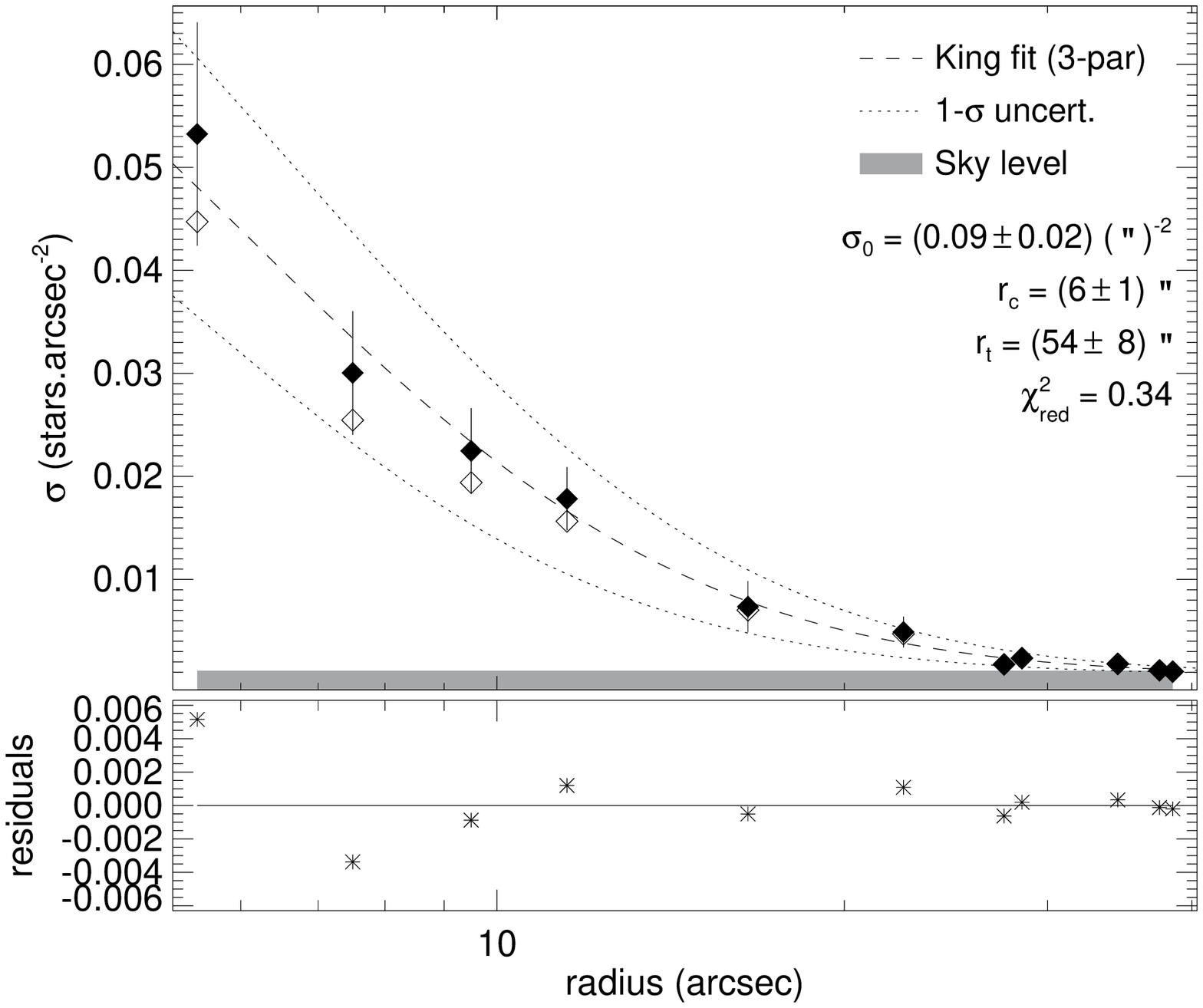}
\caption{King model fits to RDPs of clusters NGC796, HW20, SL897 and AM3 (from top left to bottom right). Details as in Fig.~\ref{fig:RDPs_LMC}.}
\label{fig:RDPs_SMC}
\end{figure}

This appendix compiles figures resulting from the structural analysis of the studied clusters, 
as described in Sect.~\ref{sec:analysis_rdp}. Figs.~\ref{fig:SBPs_LMC_king} and \ref{fig:SBPs_SMC_king} 
shows the fits of the King function (Eq.~\ref{eq:king_sbp}) over the surface brightness profiles 
(SBPs) of studied clusters. Figs.~\ref{fig:RDPs_LMC} and \ref{fig:RDPs_SMC} shows the fits of the 
King function (Eq.~\ref{eq:king_rdp}) over the radial density profiles (RDPs) of the studied clusters. 

\section{Isochrone fits charts}
\label{sec:app2}

\begin{figure}
\includegraphics[width=0.3\textwidth]{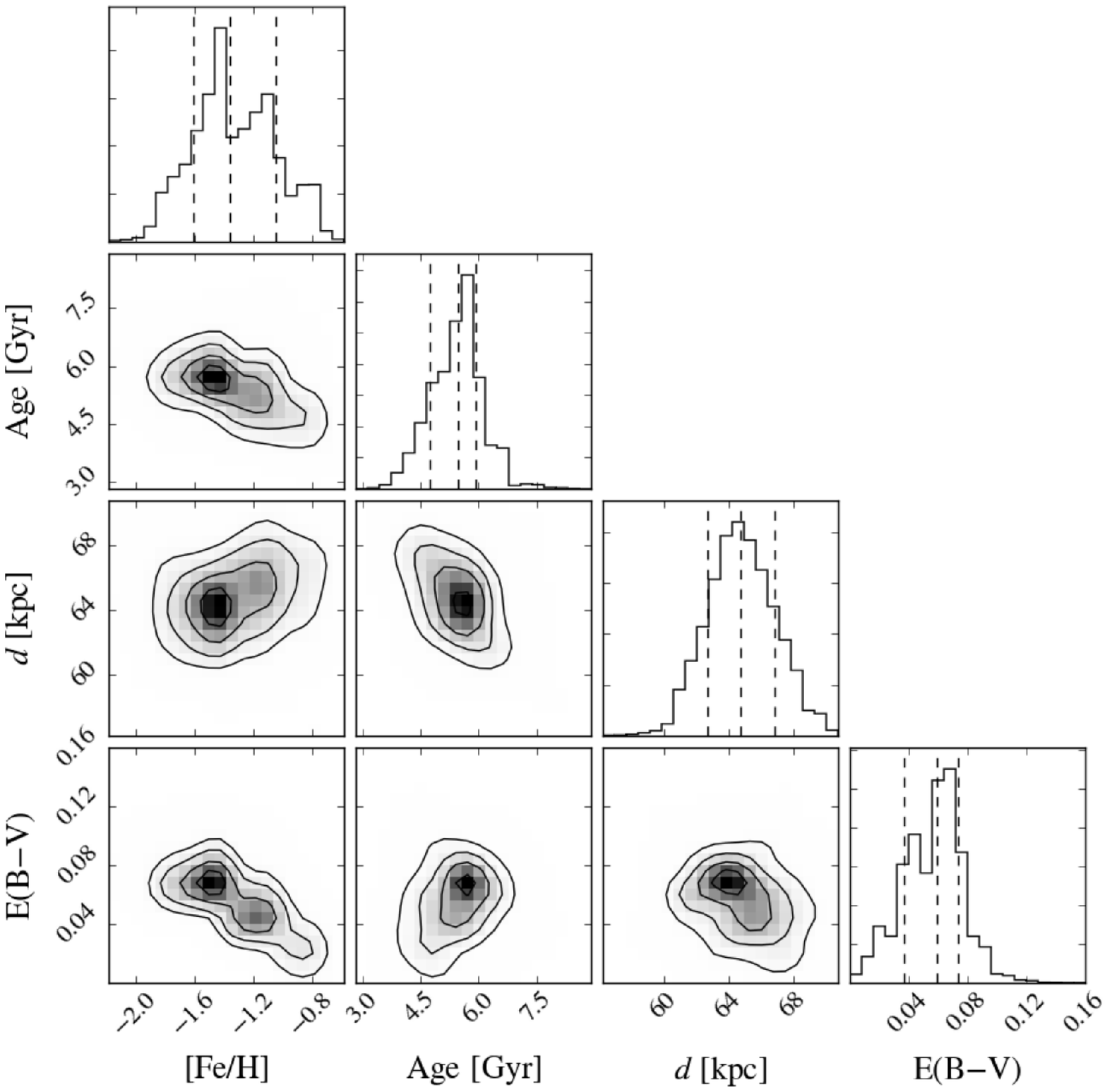}\hspace{-1.35cm}
\includegraphics[width=0.245\textwidth,trim=0 -7.0cm 0 0]{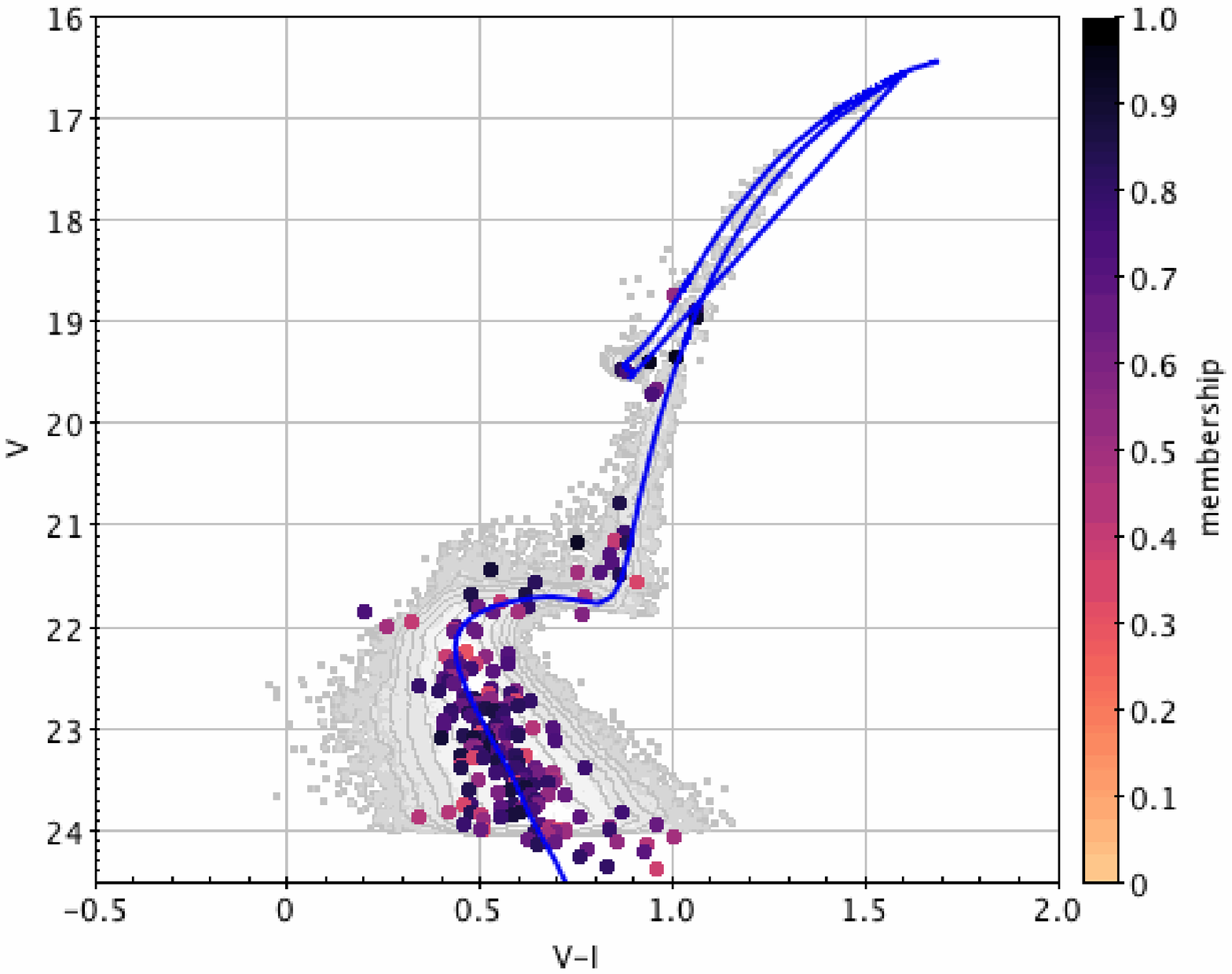}
\includegraphics[width=0.3\textwidth]{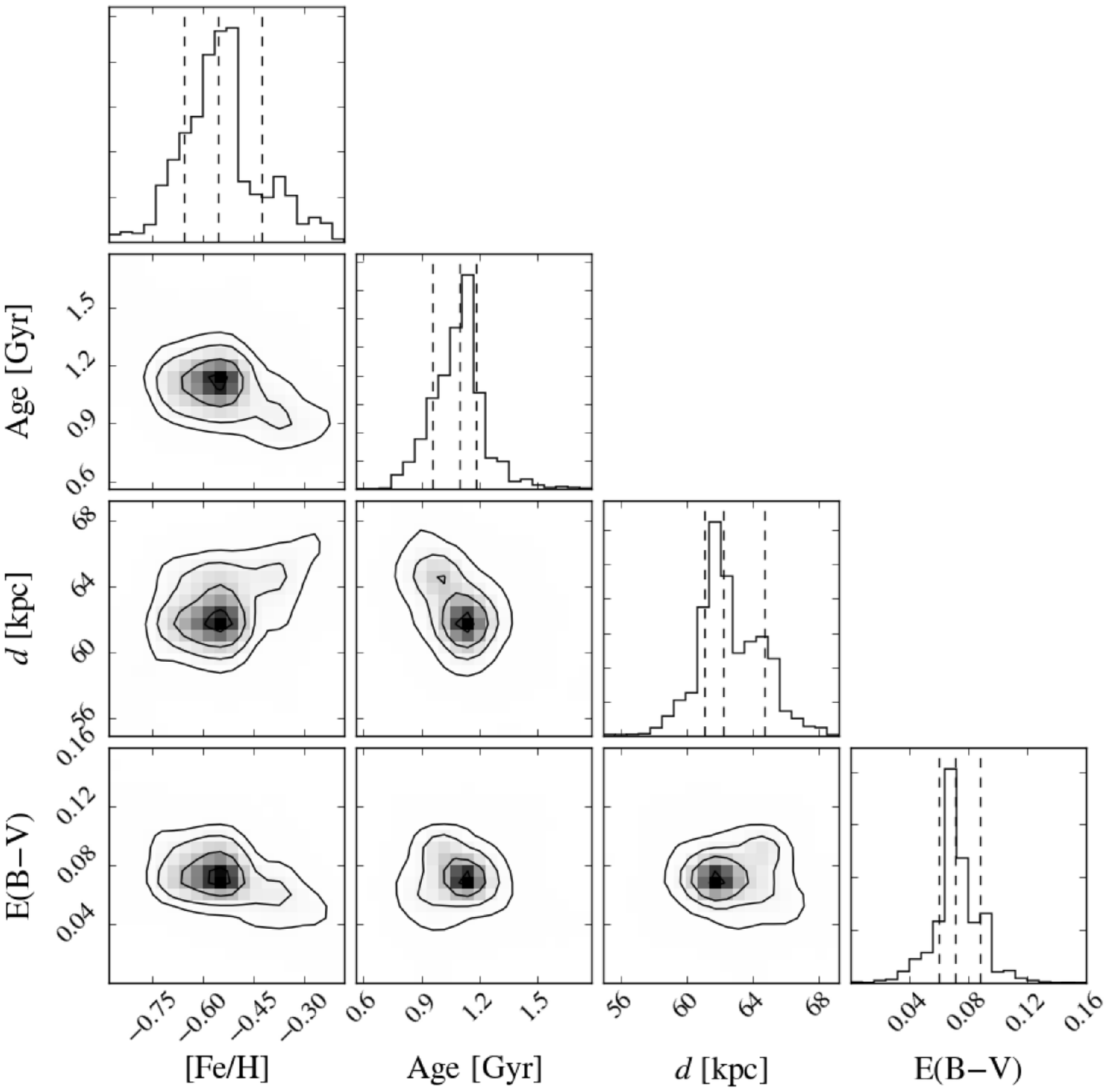}\hspace{-1.35cm}
\includegraphics[width=0.245\textwidth,trim=0 -7.0cm 0 0]{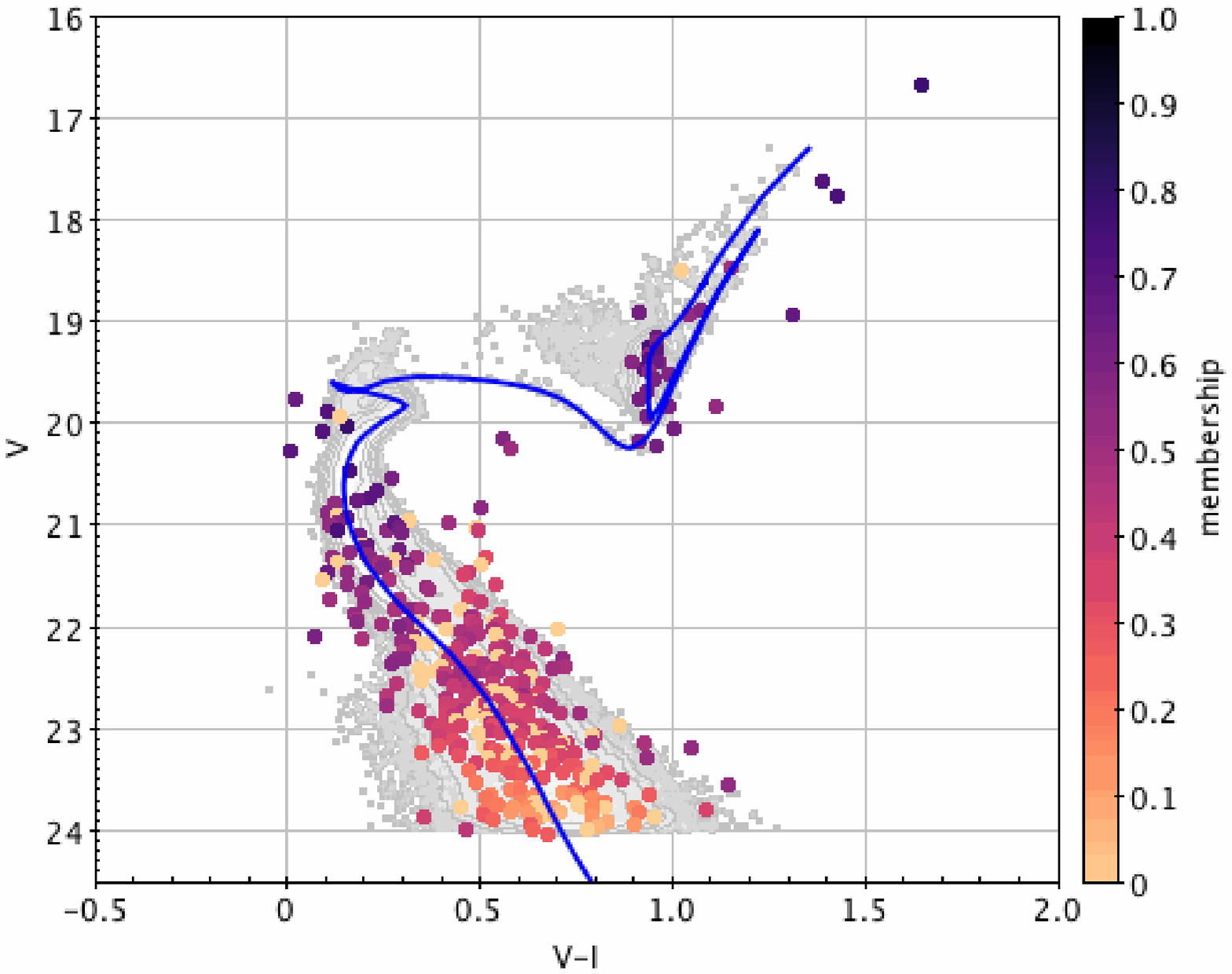}
\includegraphics[width=0.3\textwidth]{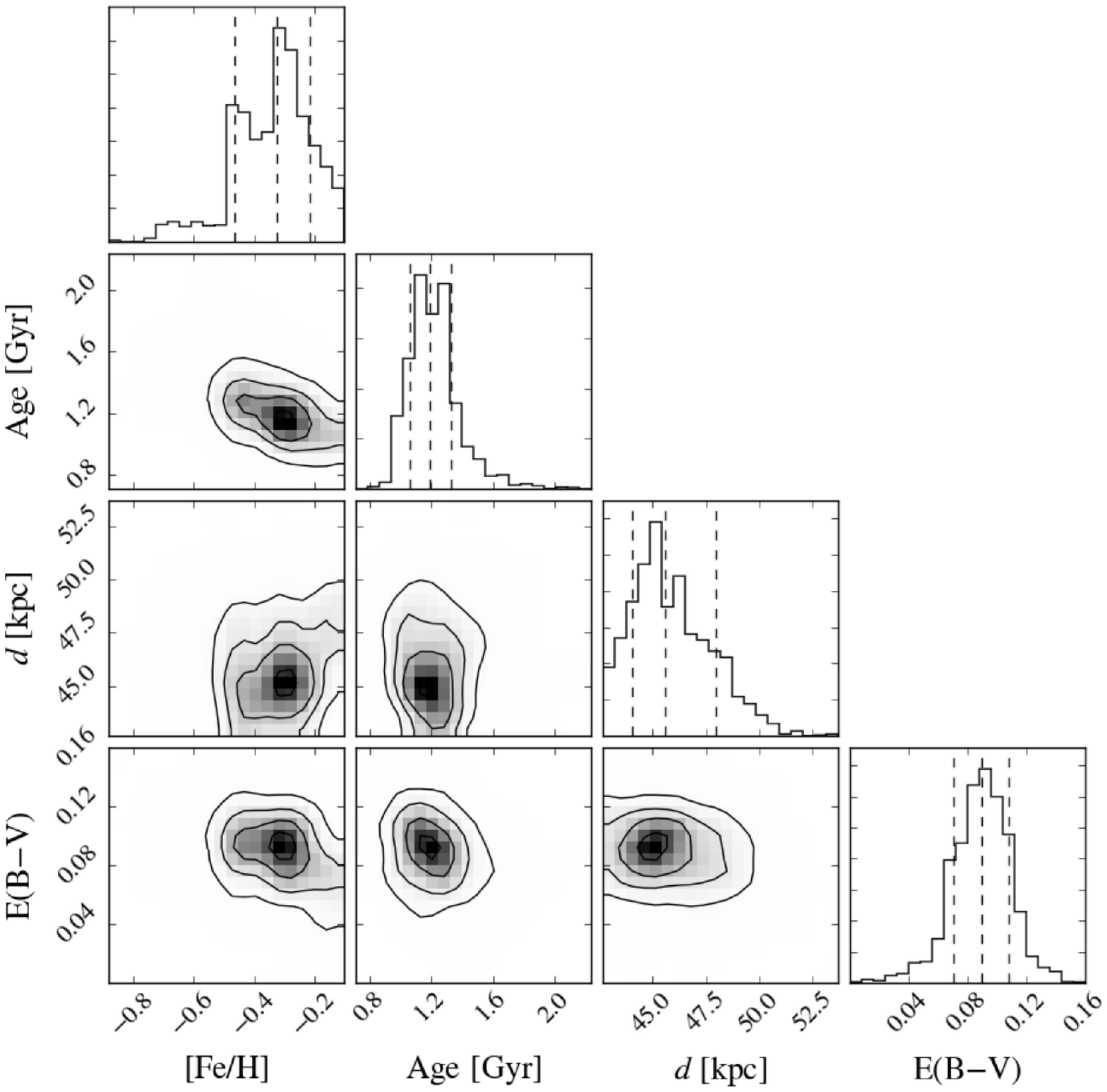}\hspace{-1.35cm}
\includegraphics[width=0.245\textwidth,trim=0 -7.0cm 0 0]{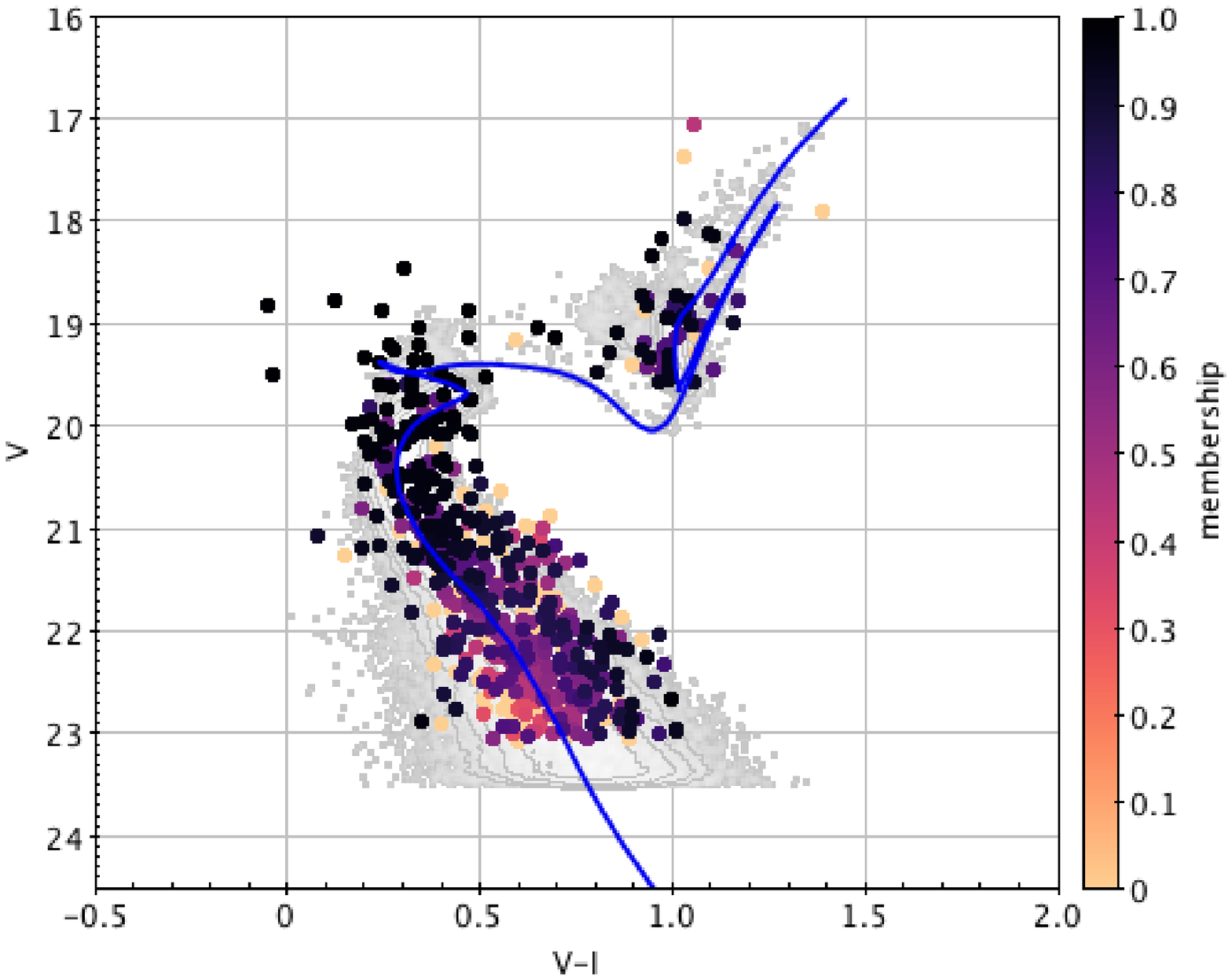}
\includegraphics[width=0.3\textwidth]{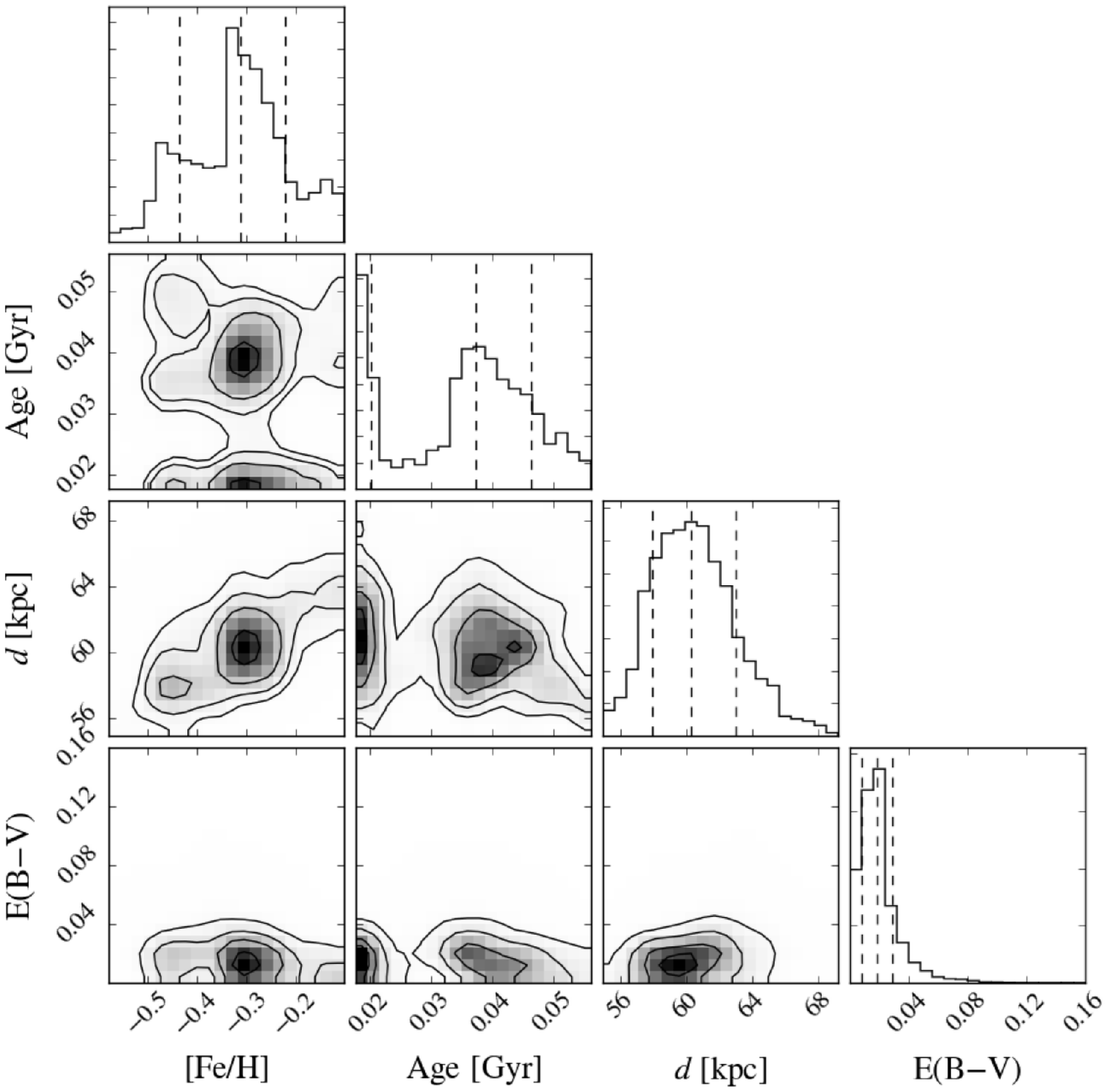}\hspace{-1.35cm}
\includegraphics[width=0.245\textwidth,trim=0 -7.0cm 0 0]{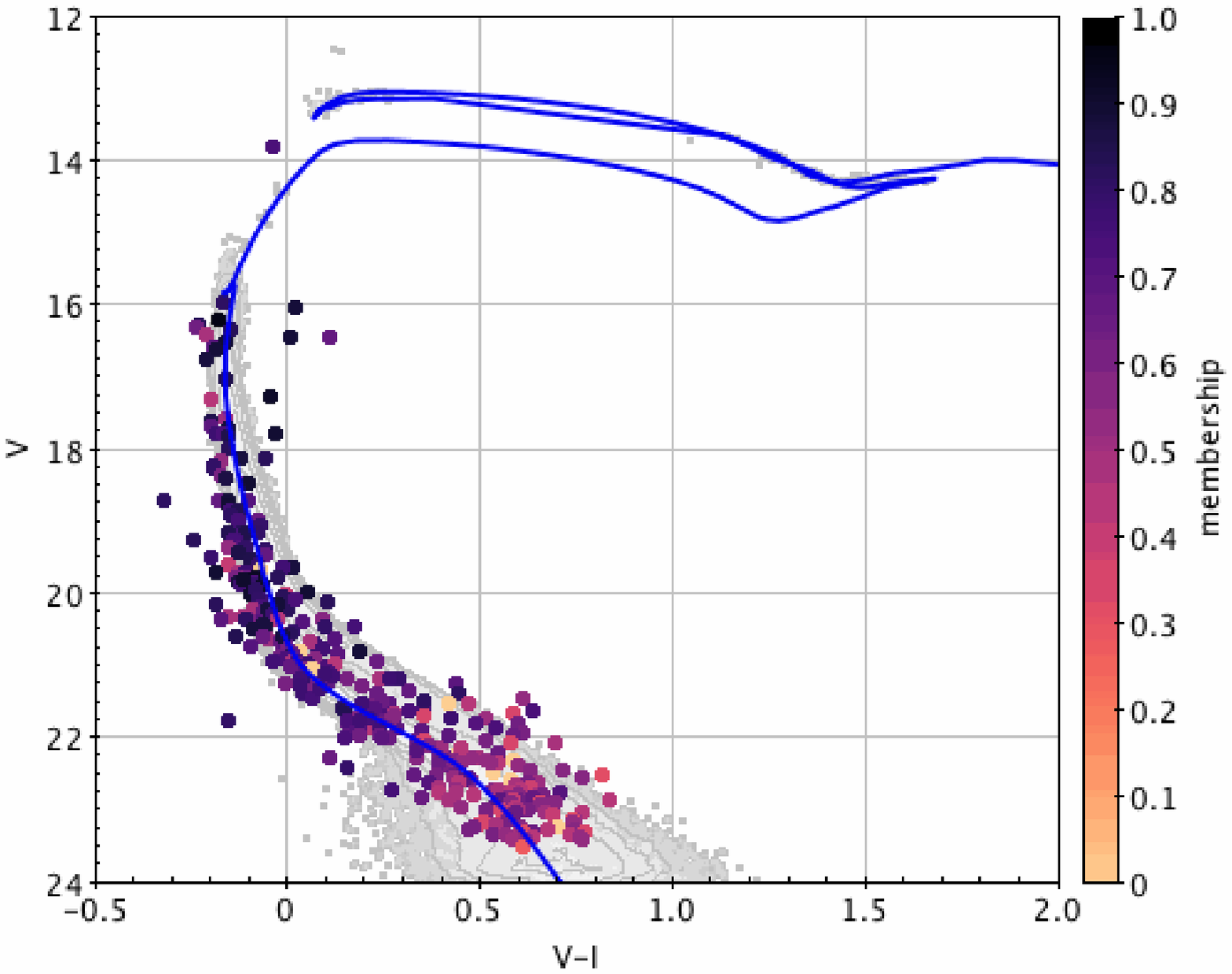}
\caption{Left: corner plots showing the posterior distribution of the astrophysical parameters derived
from MCMC simulations. Right: decontaminated CMDs showing the best model isochrones (solid lines) and the 
synthetic populations used in the MCMC procedure (gray dots). From top to bottom: AM3, HW20, SL897, NGC796.}
\label{fig:iso_SMC}
\end{figure}

\begin{figure}
\includegraphics[width=0.3\textwidth]{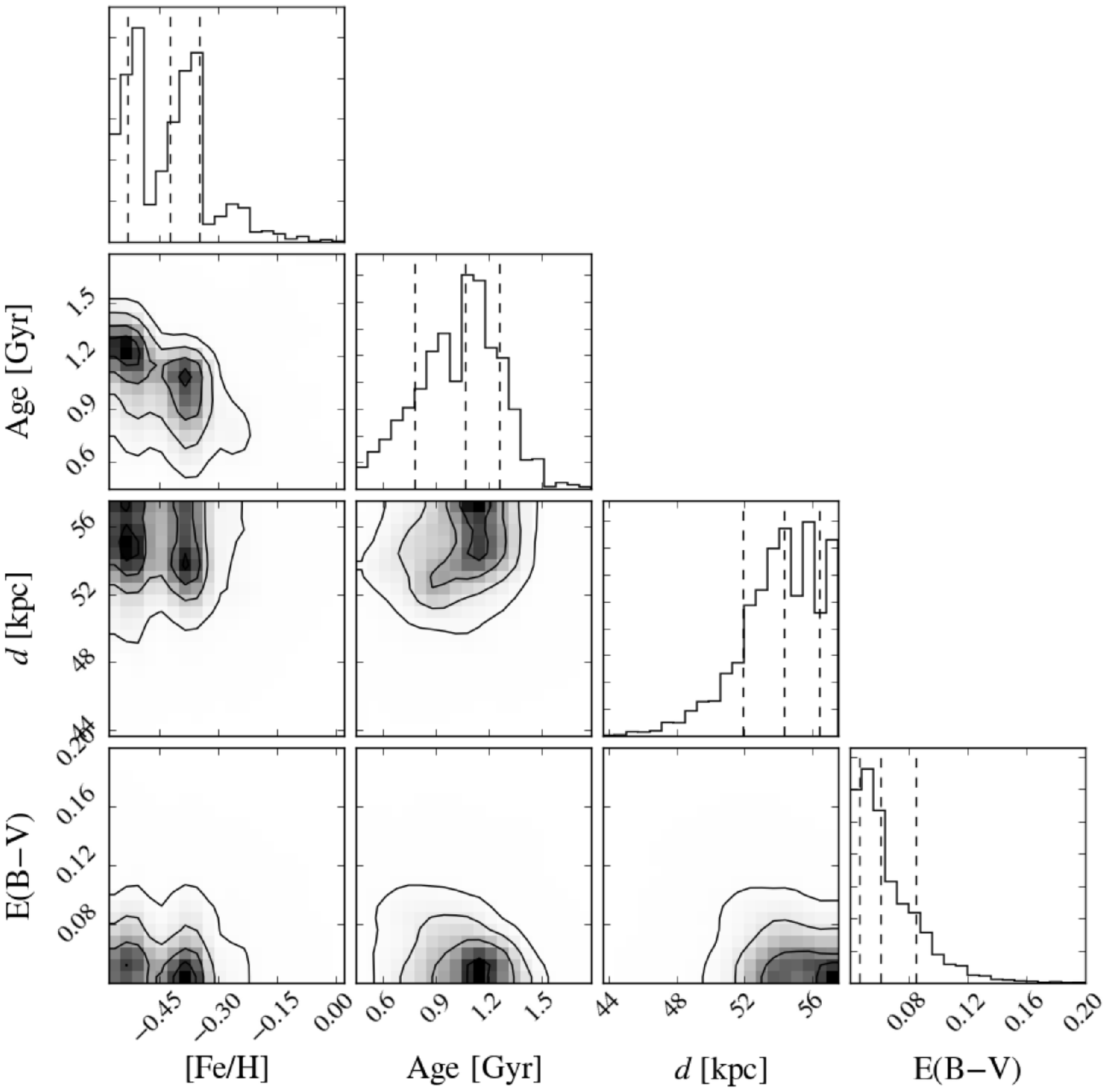}\hspace{-1.35cm}
\includegraphics[width=0.245\textwidth,trim=0 -7.0cm 0 0]{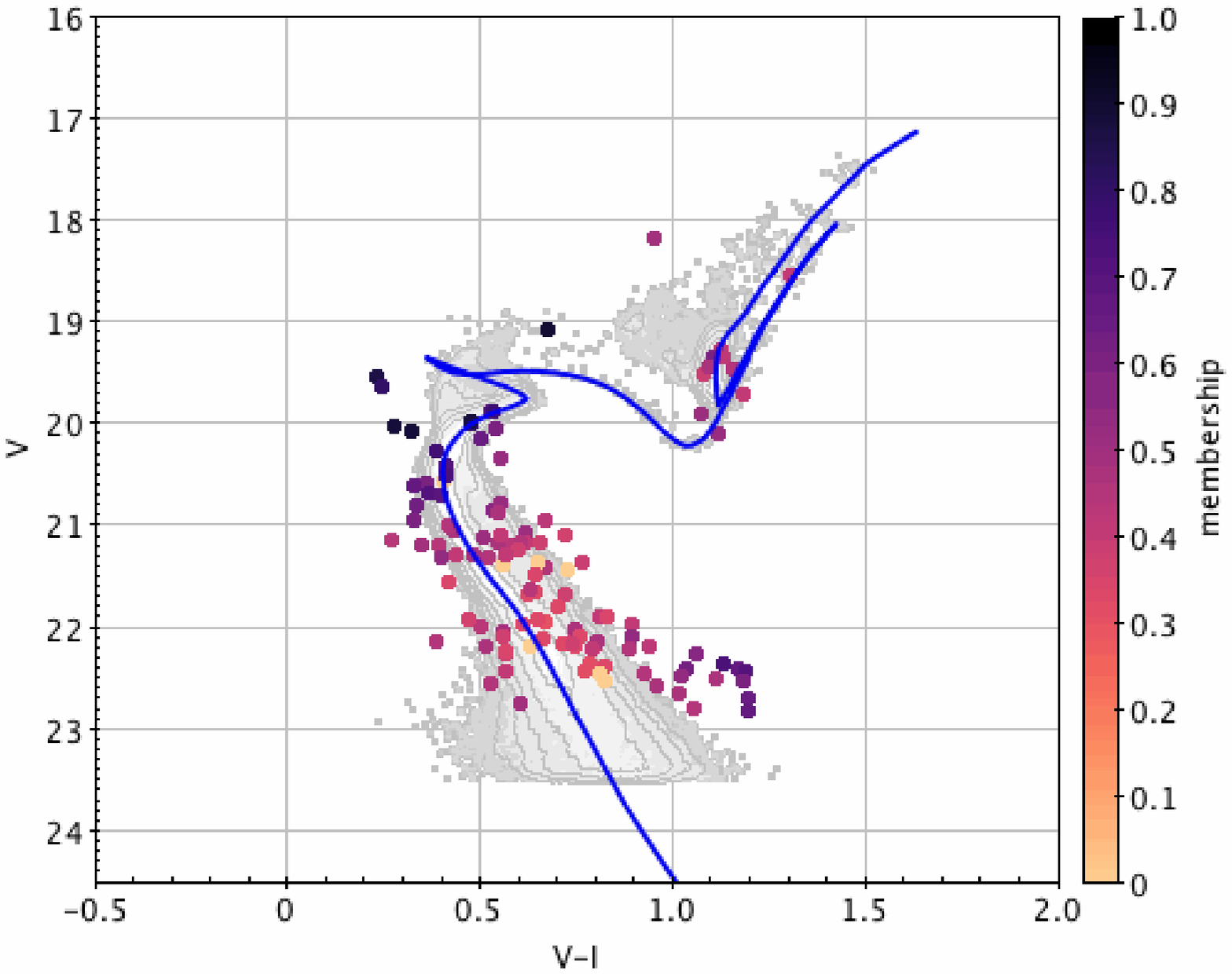}
\includegraphics[width=0.3\textwidth]{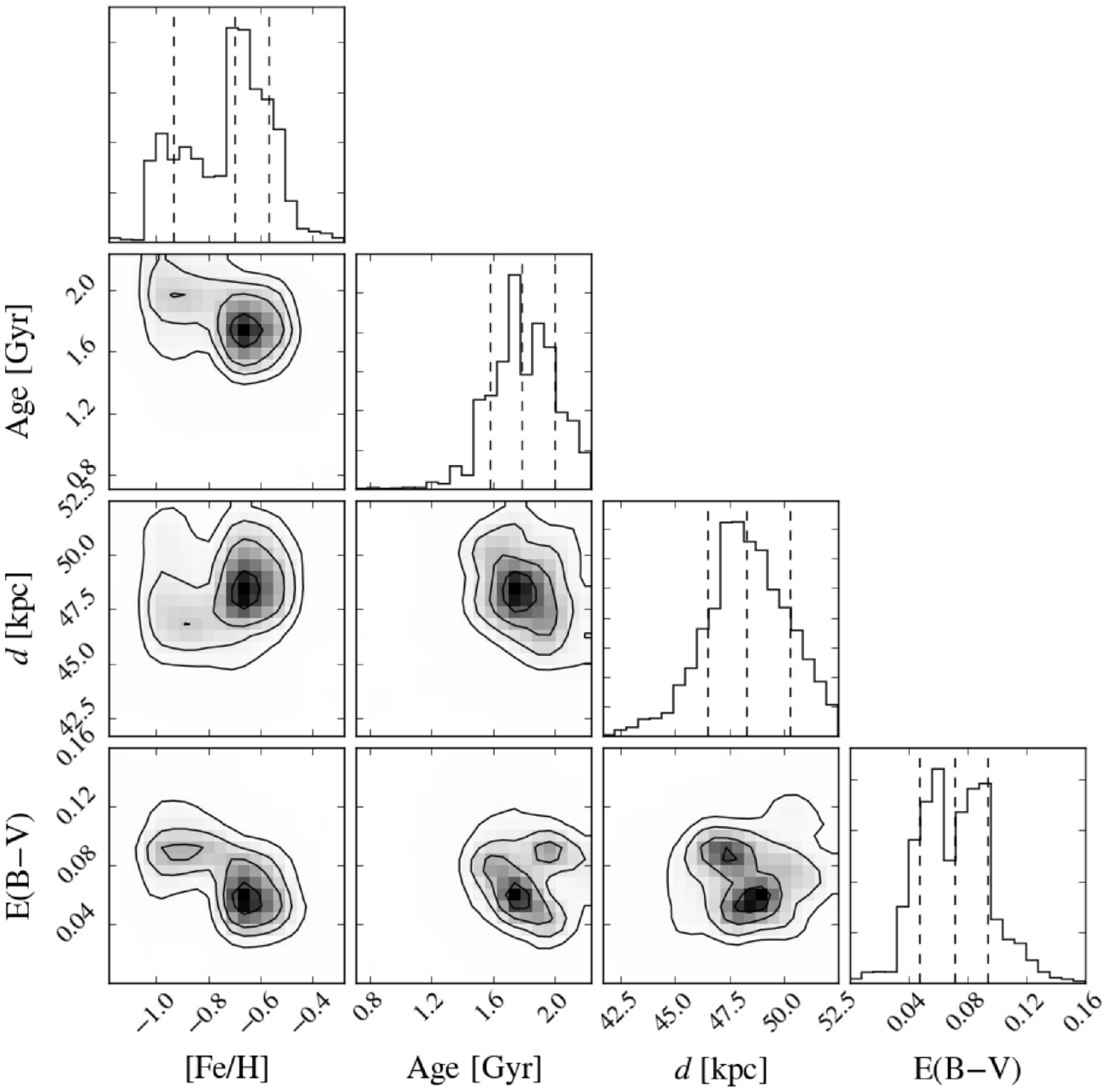}\hspace{-1.35cm}
\includegraphics[width=0.245\textwidth,trim=0 -7.0cm 0 0]{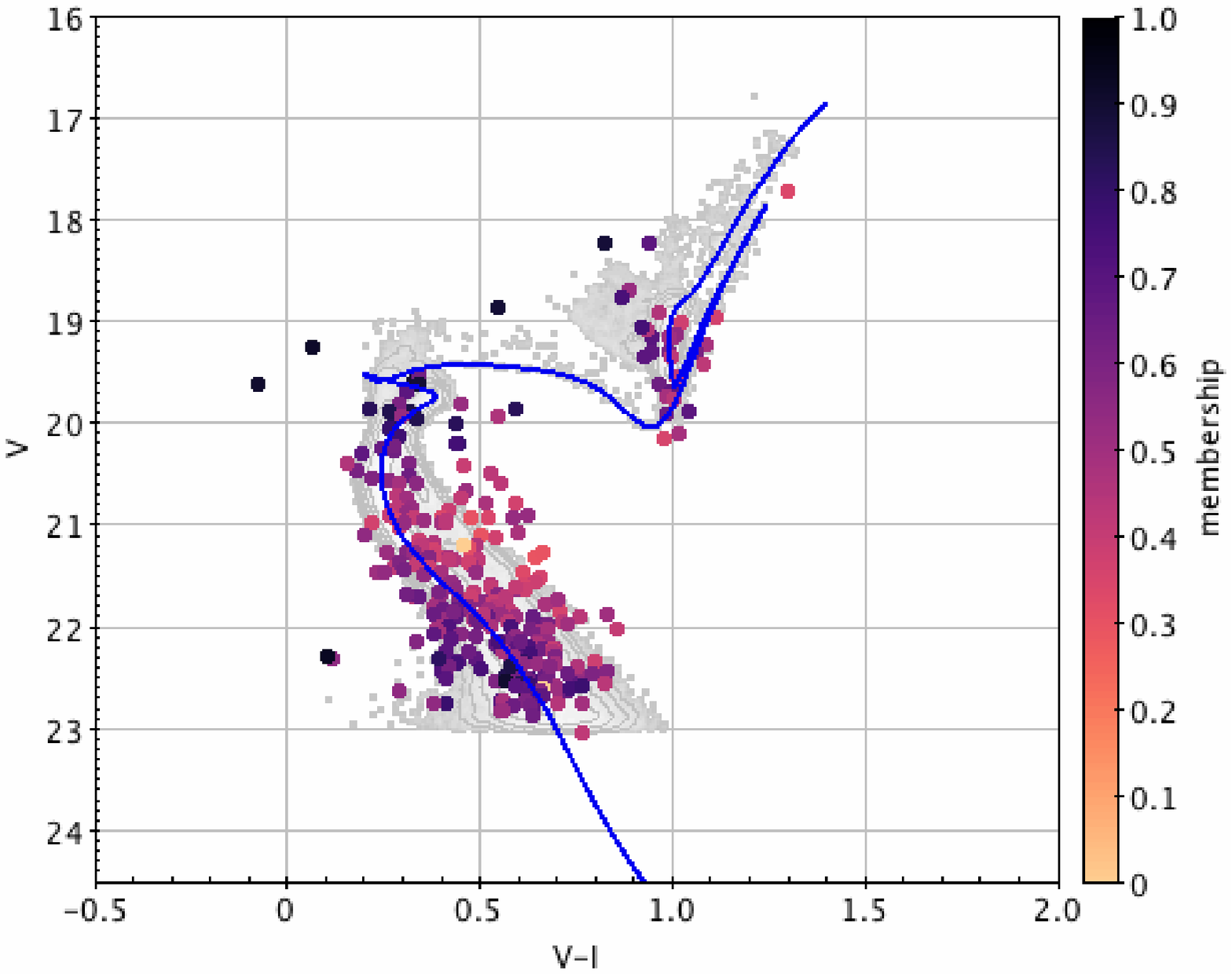}
\includegraphics[width=0.3\textwidth]{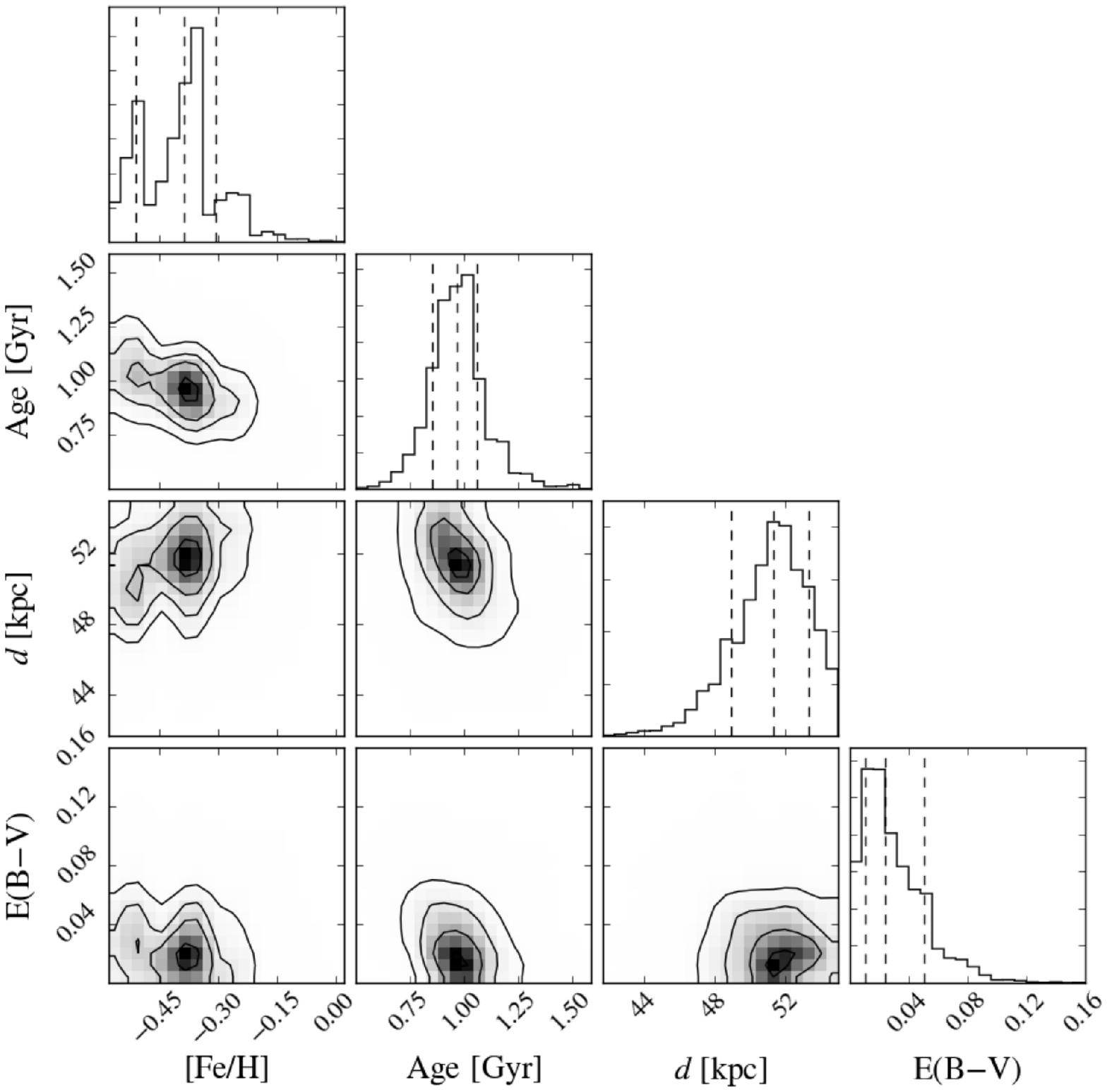}\hspace{-1.35cm}
\includegraphics[width=0.245\textwidth,trim=0 -7.0cm 0 0]{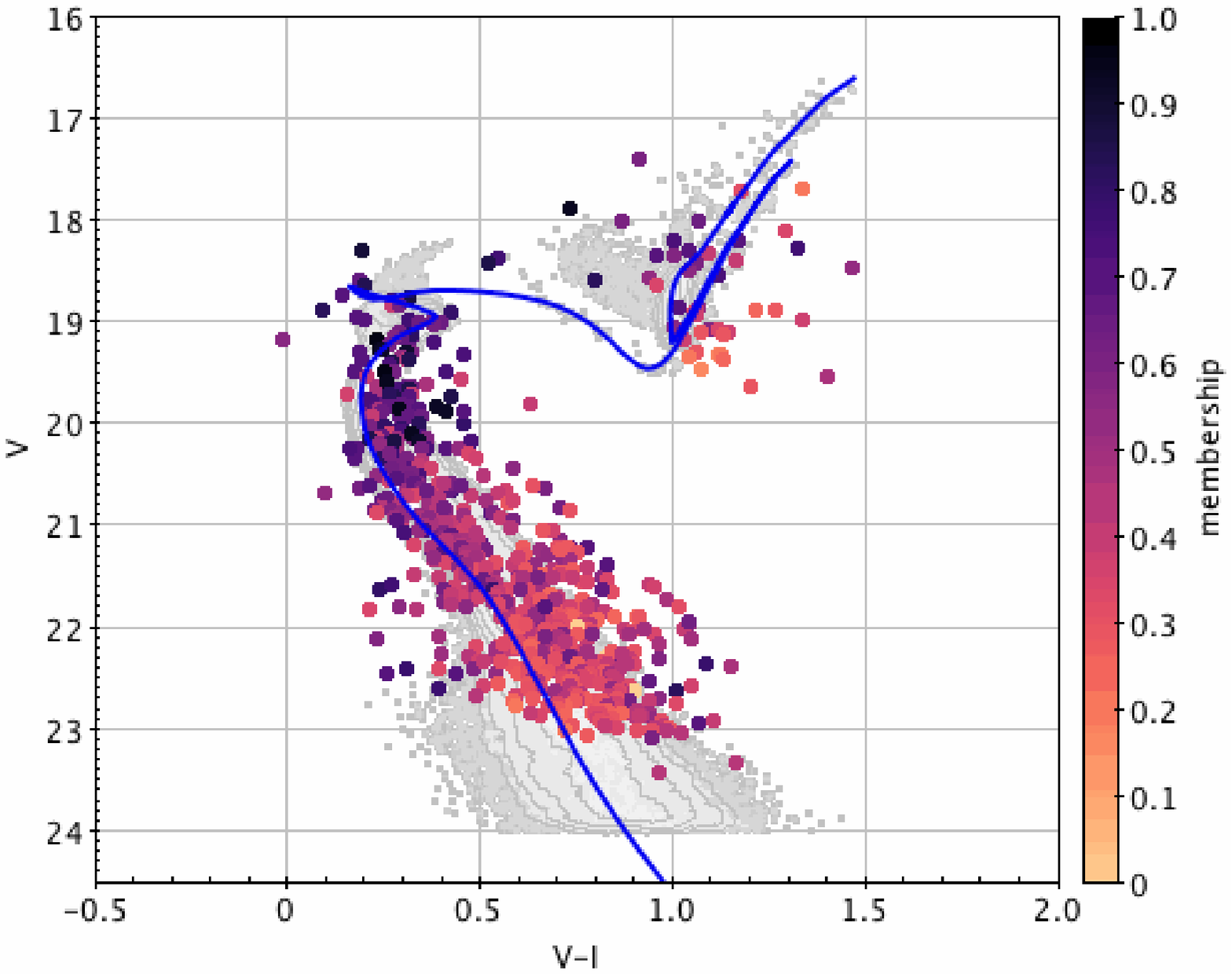}
\includegraphics[width=0.3\textwidth]{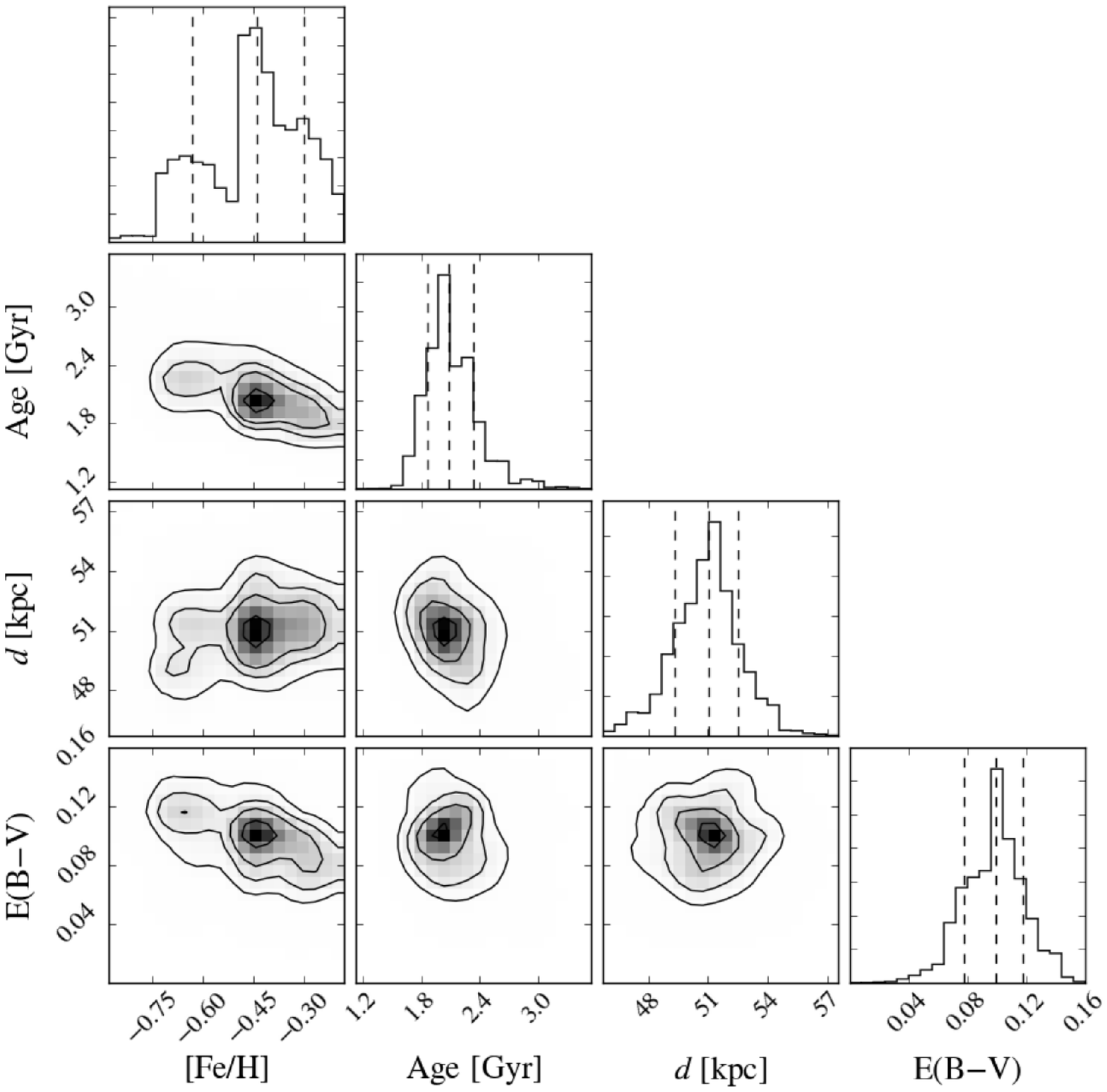}\hspace{-1.35cm}
\includegraphics[width=0.245\textwidth,trim=0 -7.0cm 0 0]{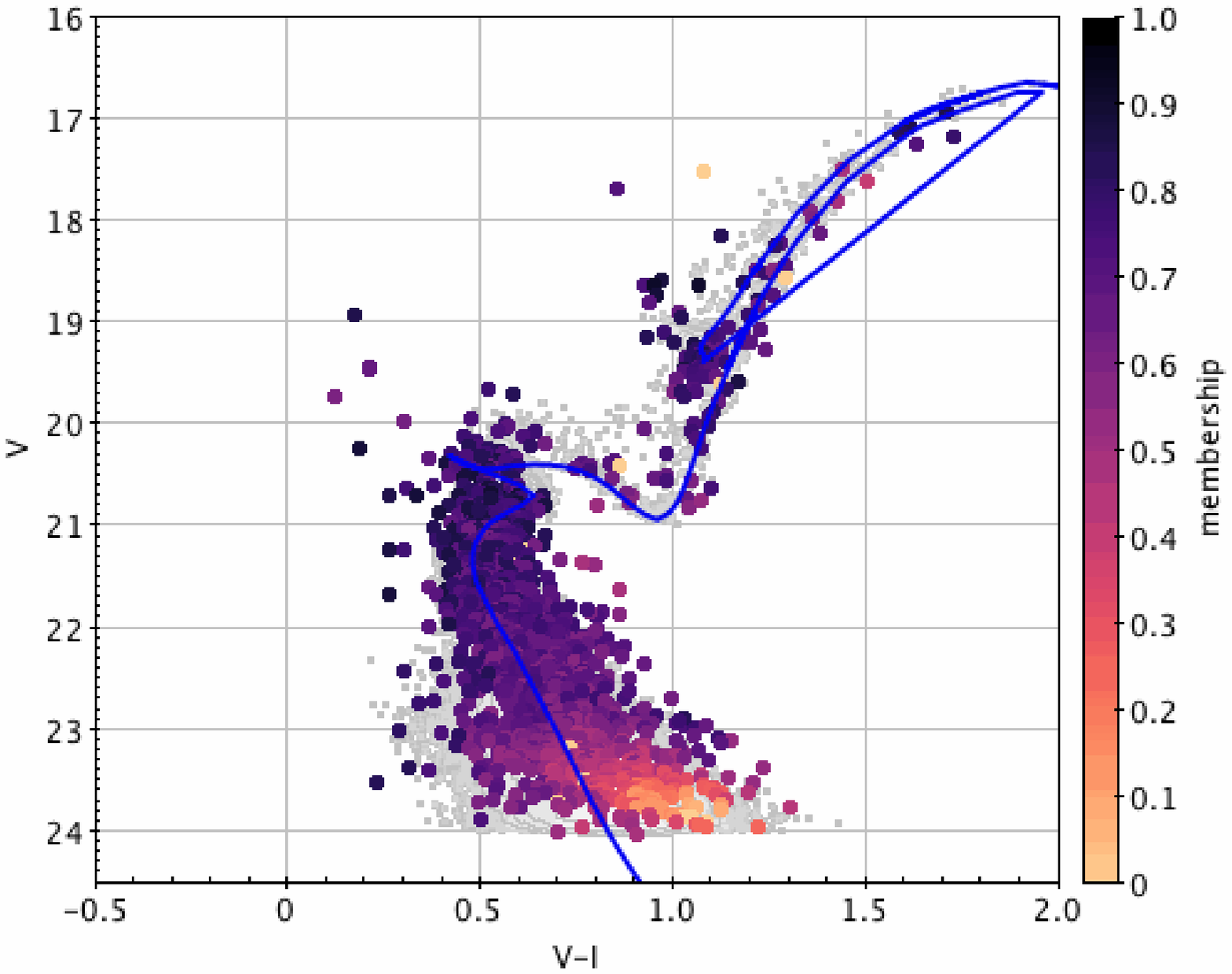}
\caption{Same as Fig.~\ref{fig:iso_SMC}, but for clusters (from top to bottom): 
KMHK228, OHSC3, SL576, SL61}
\label{fig:iso_LMC}
\end{figure}

This appendix compiles figures resulting from the isochrone fits of the studied clusters,
using a MCMC approach, as described in Sect.~\ref{sec:analysis_iso}. Figs.~\ref{fig:iso_SMC} and
\ref{fig:iso_LMC} shows the posterior distribution of the MCMC parameters used to infer the best 
model isochrones, and their representations over the clusters CMD. 

\section{Mass function fitting charts}
\label{sec:app3}

\begin{figure}
\centering
\includegraphics[width=0.22\textwidth]{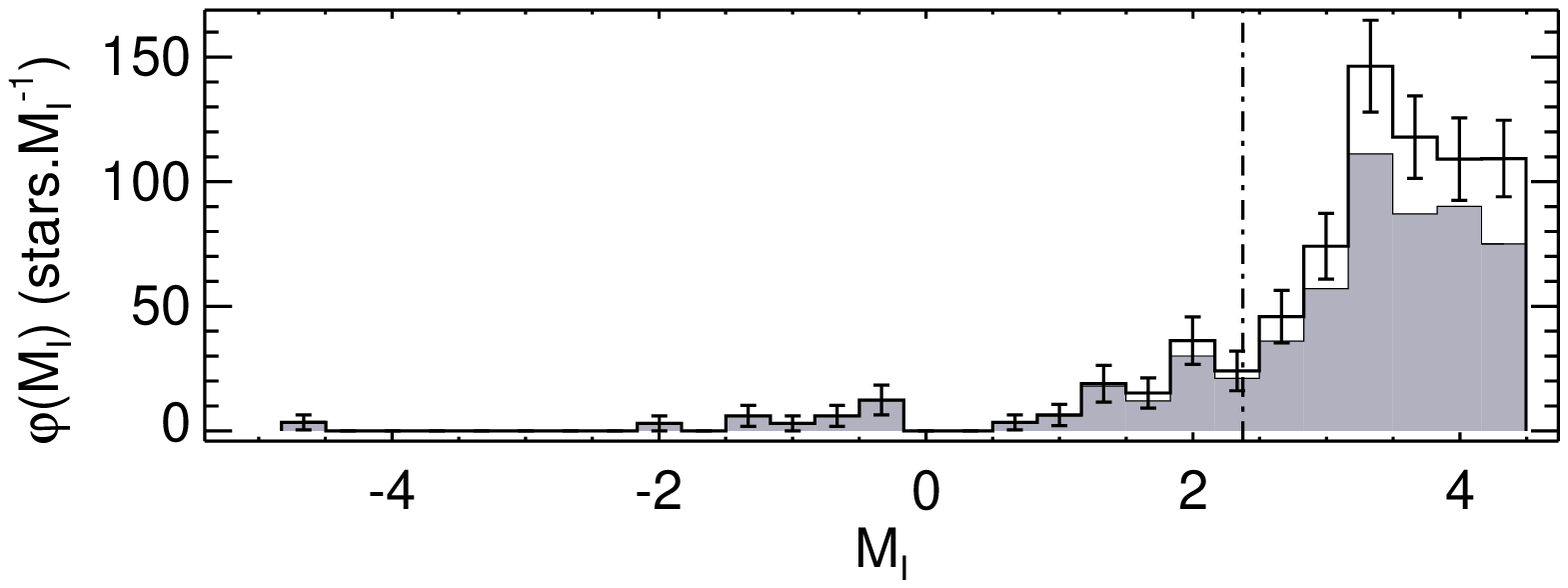} \hskip 0.5cm \includegraphics[width=0.22\textwidth]{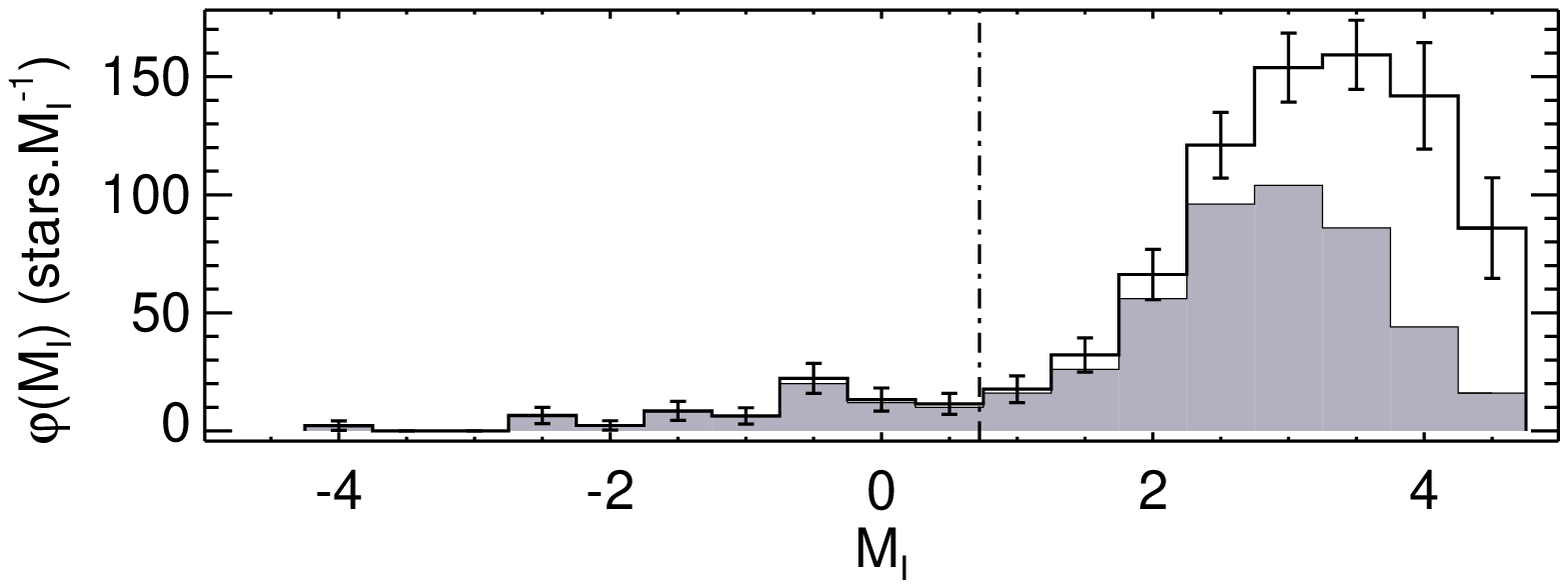} \\
\includegraphics[width=0.22\textwidth]{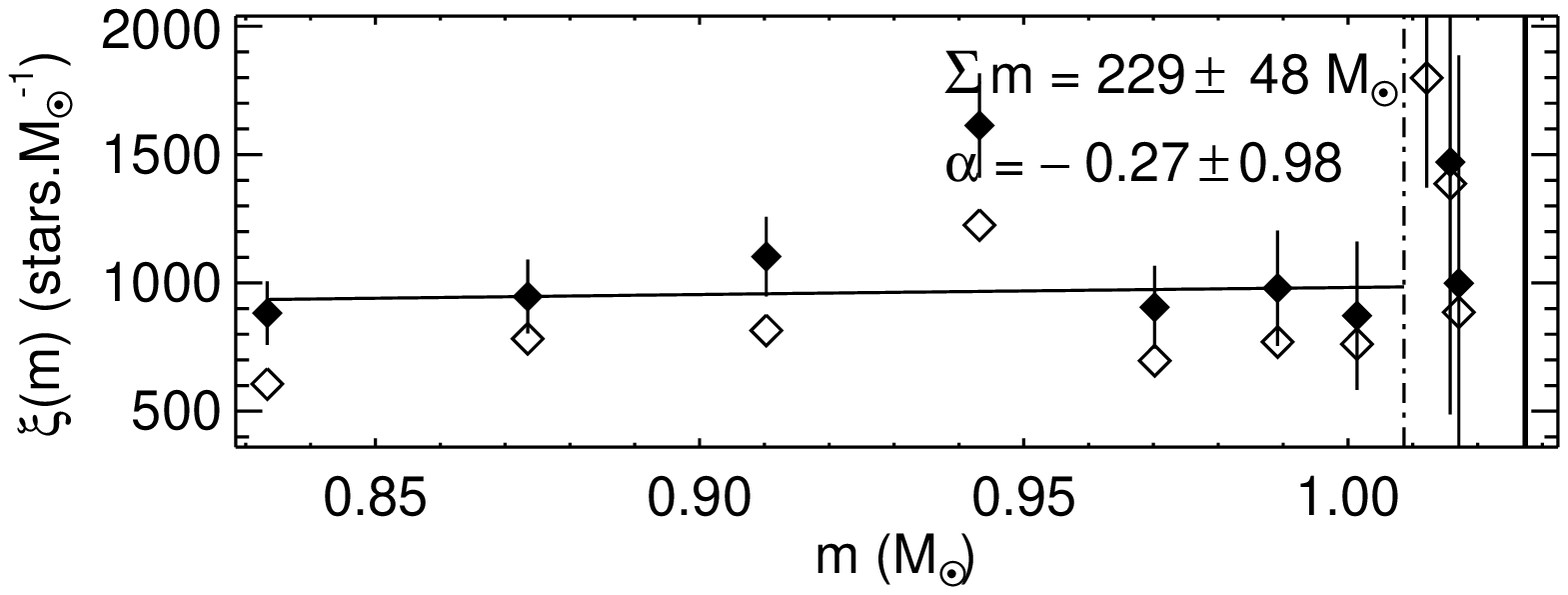} \hskip 0.5cm \includegraphics[width=0.22\textwidth]{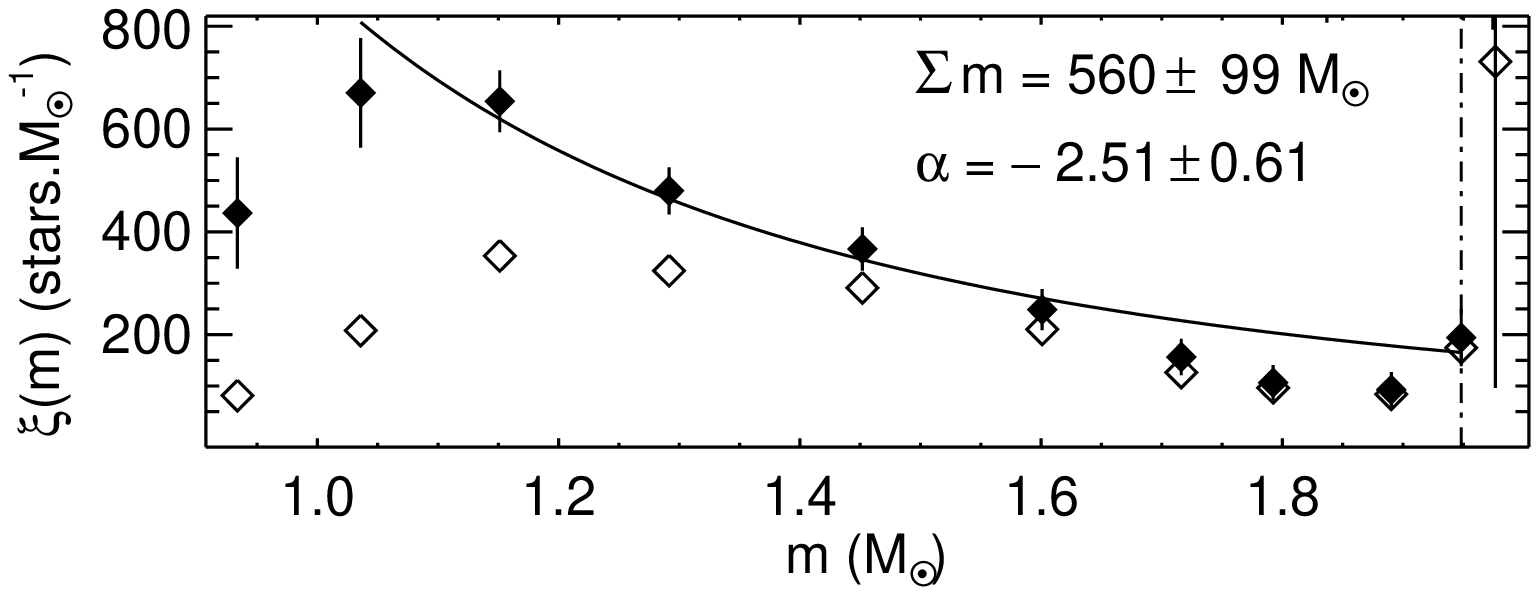} \\
\vskip 0.3cm
\includegraphics[width=0.22\textwidth]{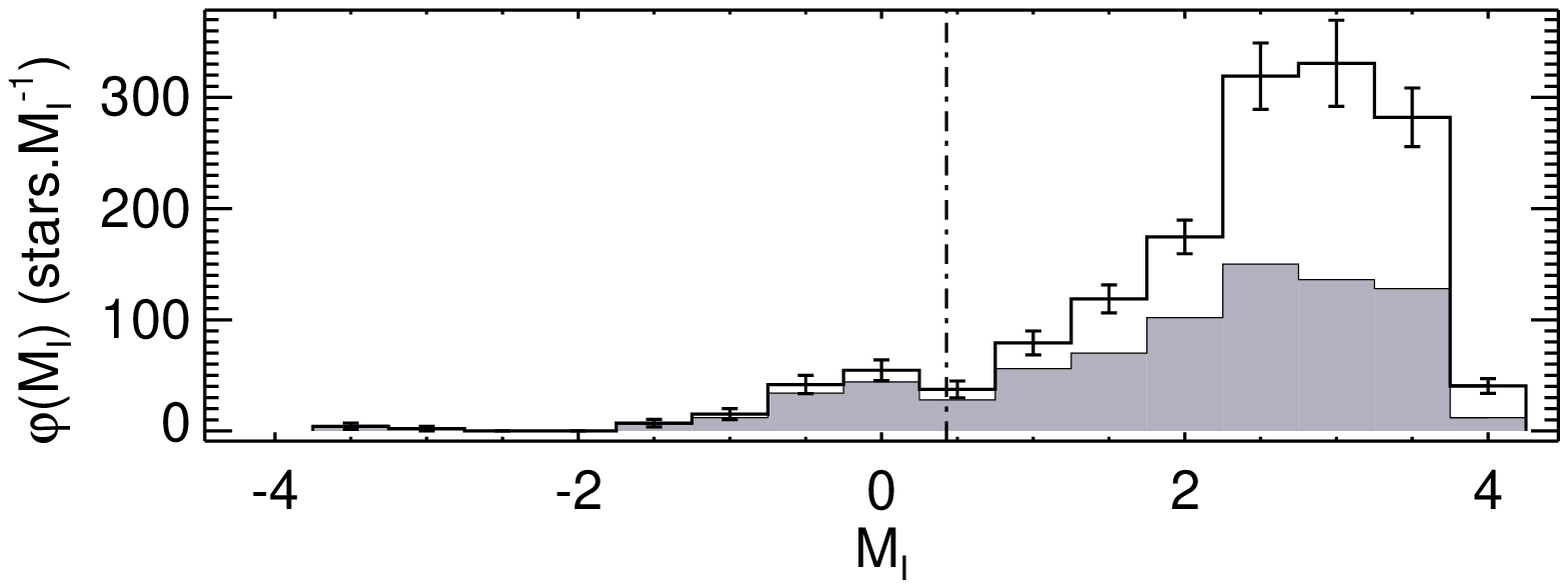} \hskip 0.5cm \includegraphics[width=0.22\textwidth]{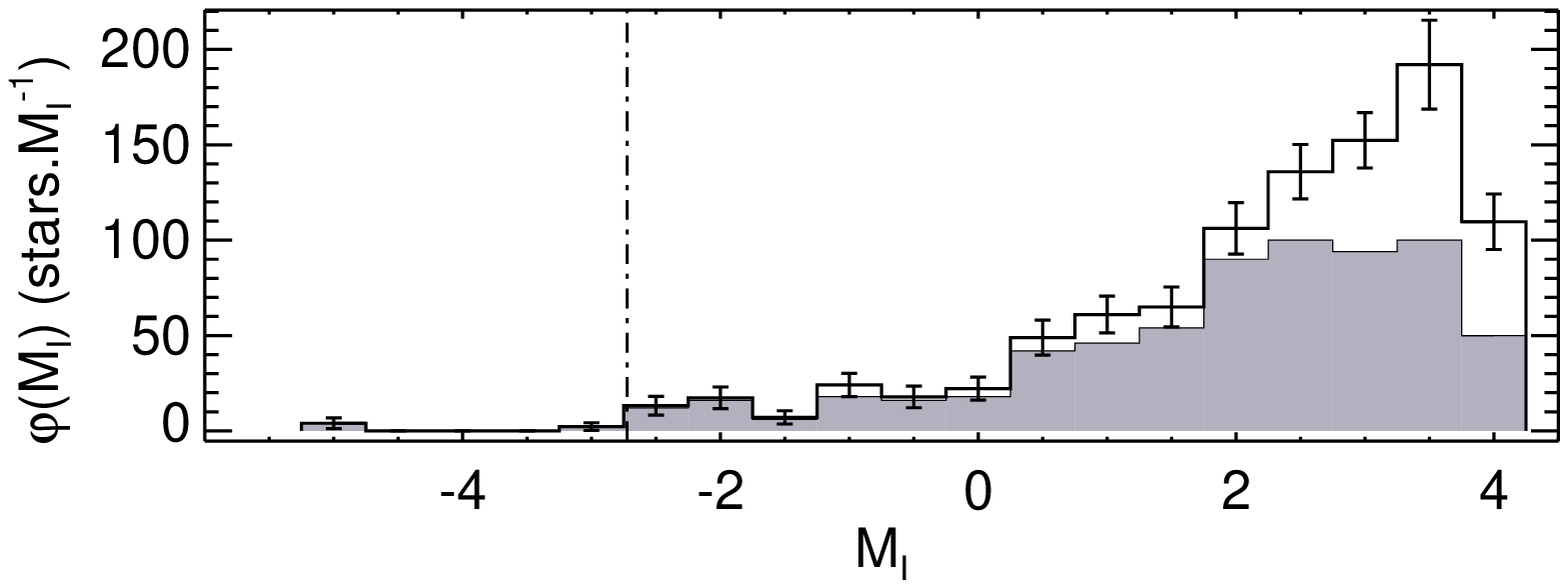} \\
\includegraphics[width=0.22\textwidth]{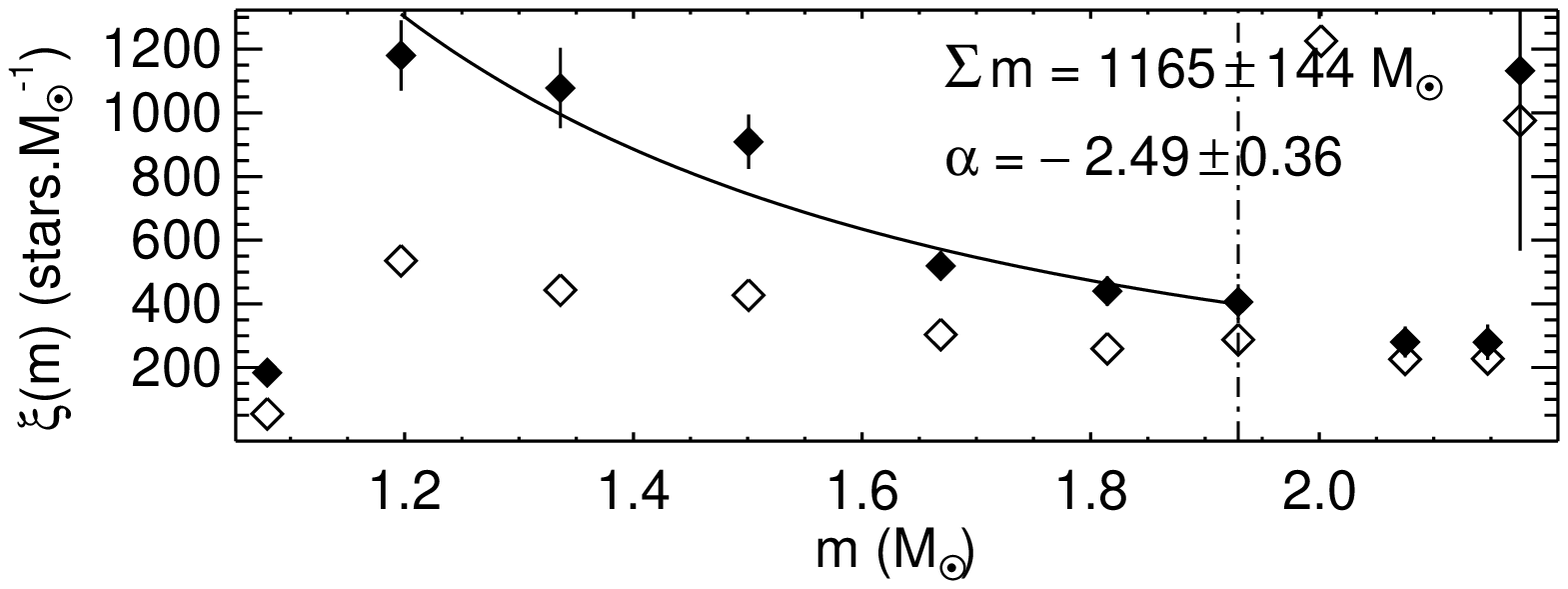} \hskip 0.5cm \includegraphics[width=0.22\textwidth]{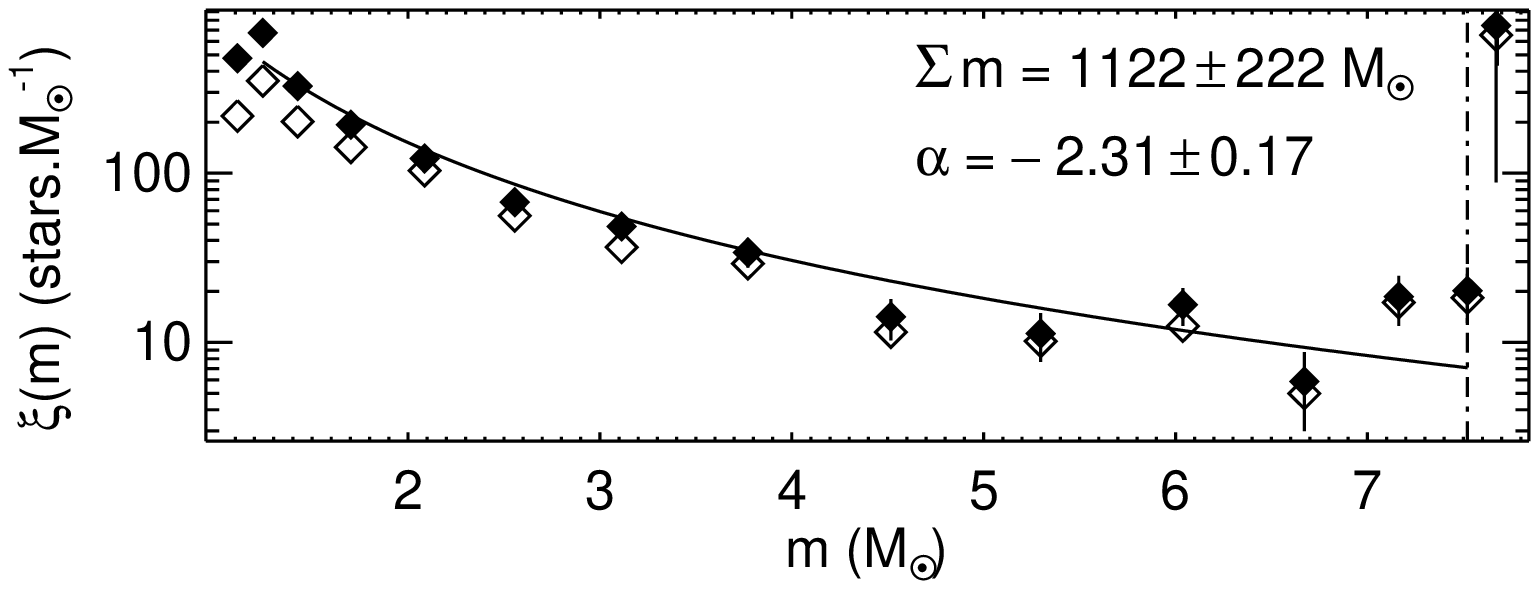} \\
\caption{Observed and completeness corrected LFs (filled and open histograms, respectively) and MFs 
(open and filled symbols, respectively). From top left to bottom right: 
LFs (top panels) and MFs (bottom panels) of AM3, HW20, SL897 and NGC796. The vertical dashed lines 
correspond to the turn-offs and the solid lines represent the MF fits.}
\label{fig:MFs_SMC}
\end{figure}

\begin{figure}
\centering
\includegraphics[width=0.22\textwidth]{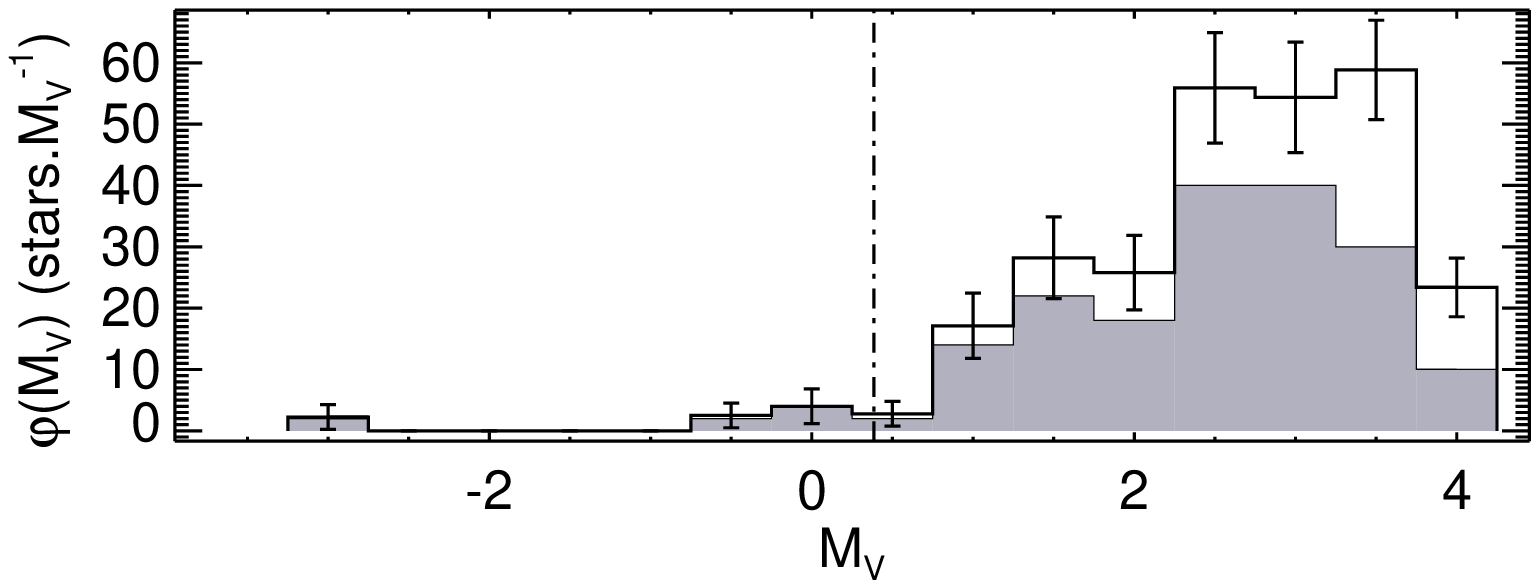} \hskip 0.5cm \includegraphics[width=0.22\textwidth]{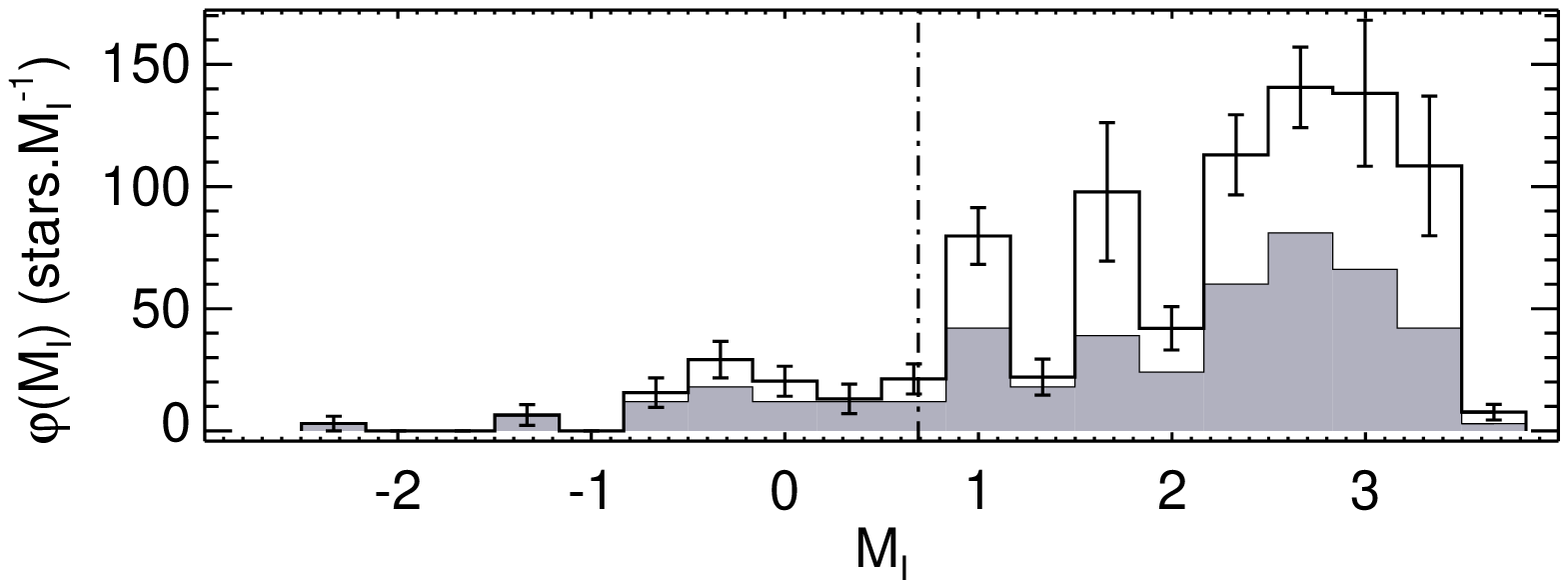} \\
\includegraphics[width=0.22\textwidth]{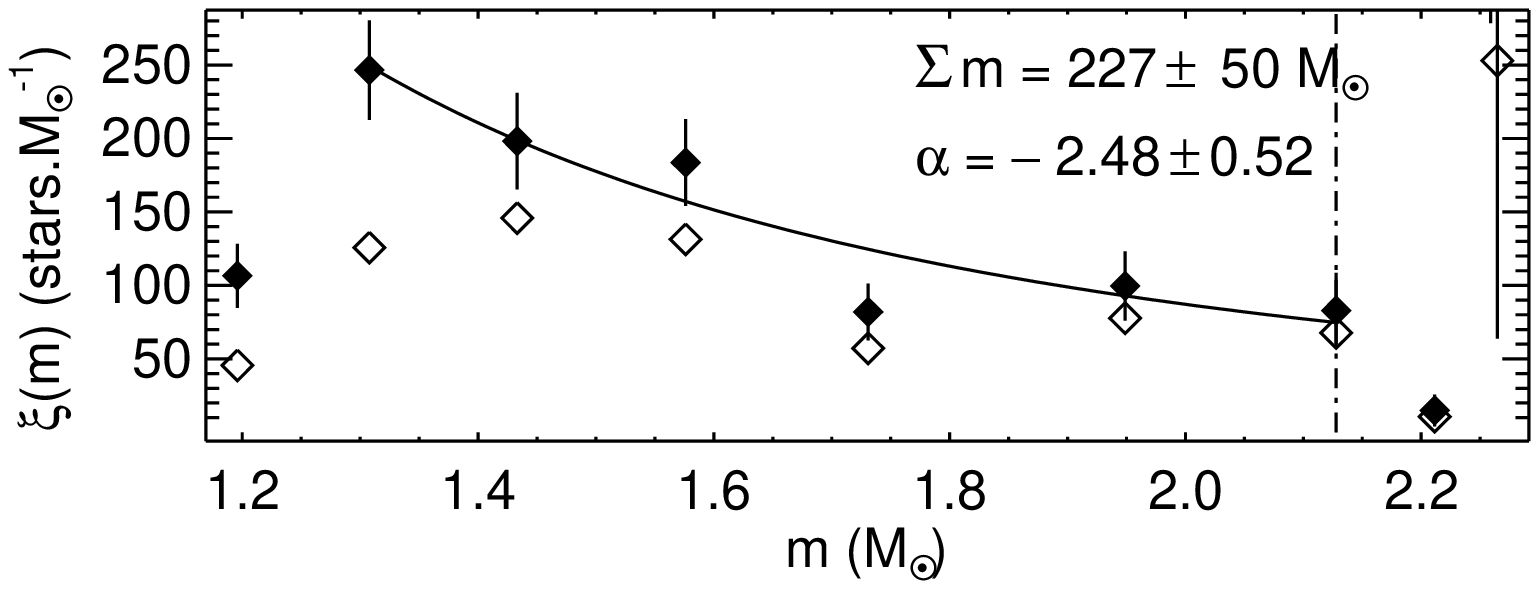} \hskip 0.5cm \includegraphics[width=0.22\textwidth]{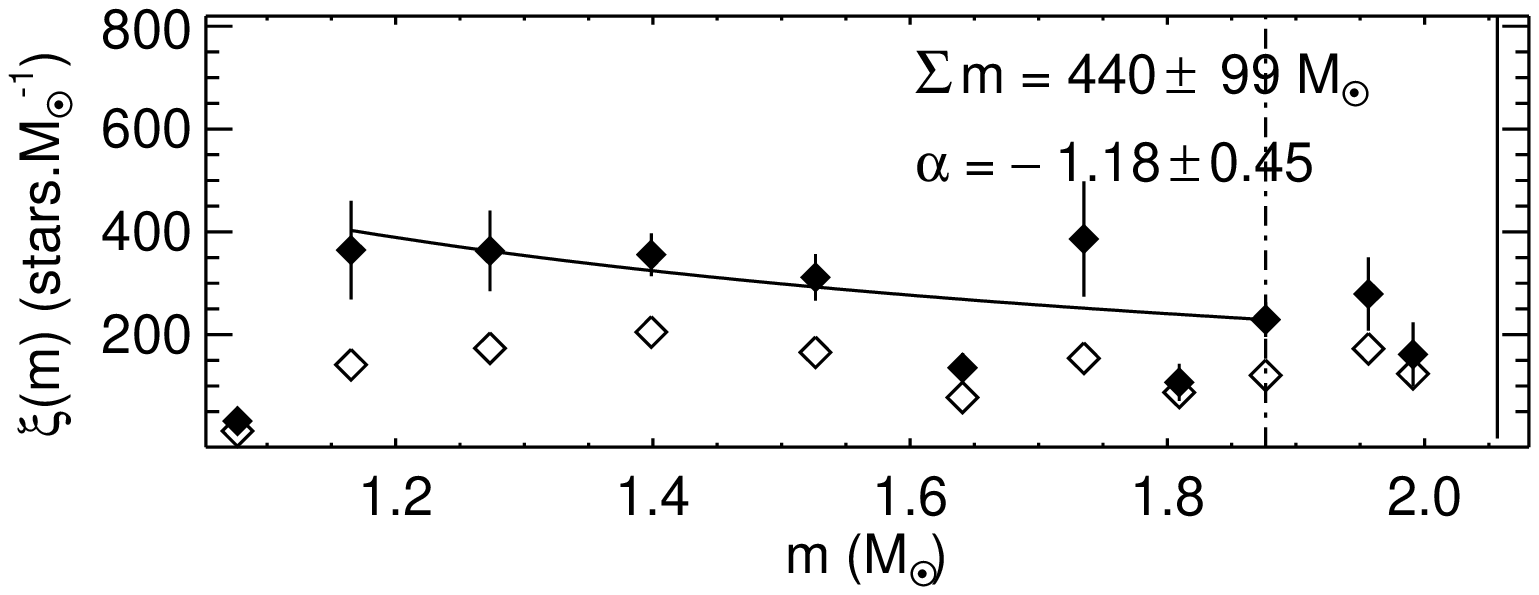} \\
\vskip 0.3cm
\includegraphics[width=0.22\textwidth]{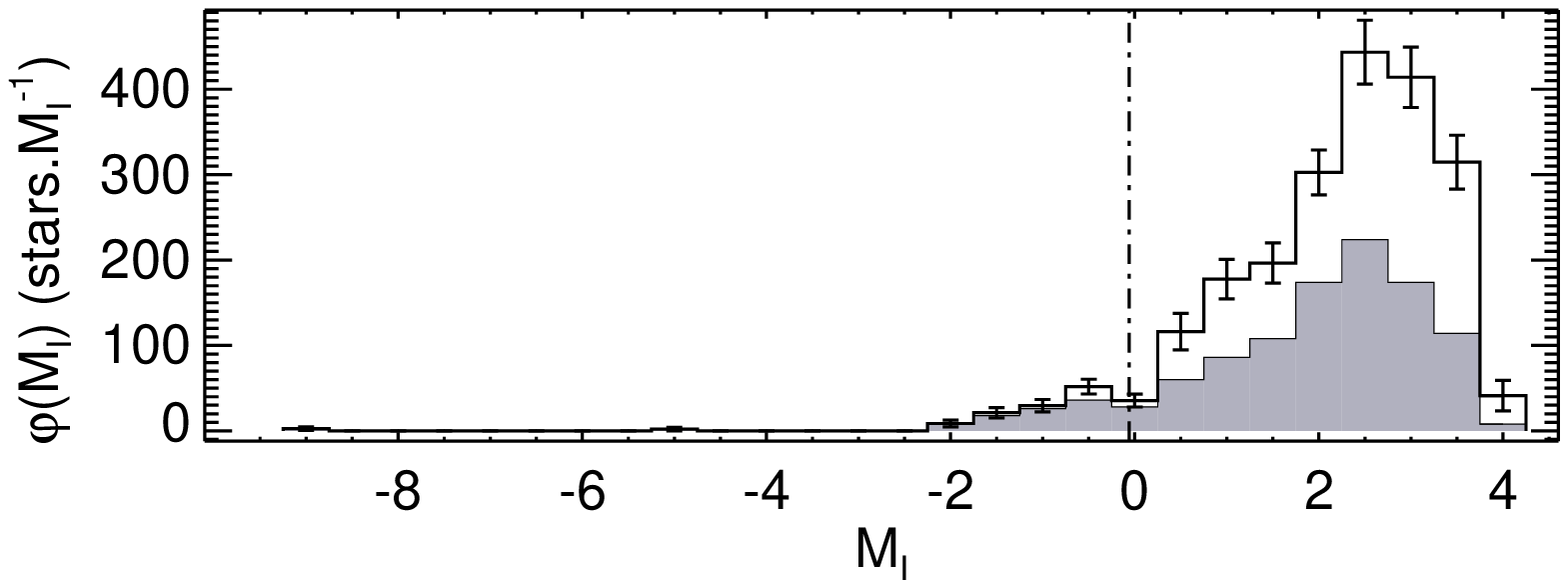} \hskip 0.5cm \includegraphics[width=0.22\textwidth]{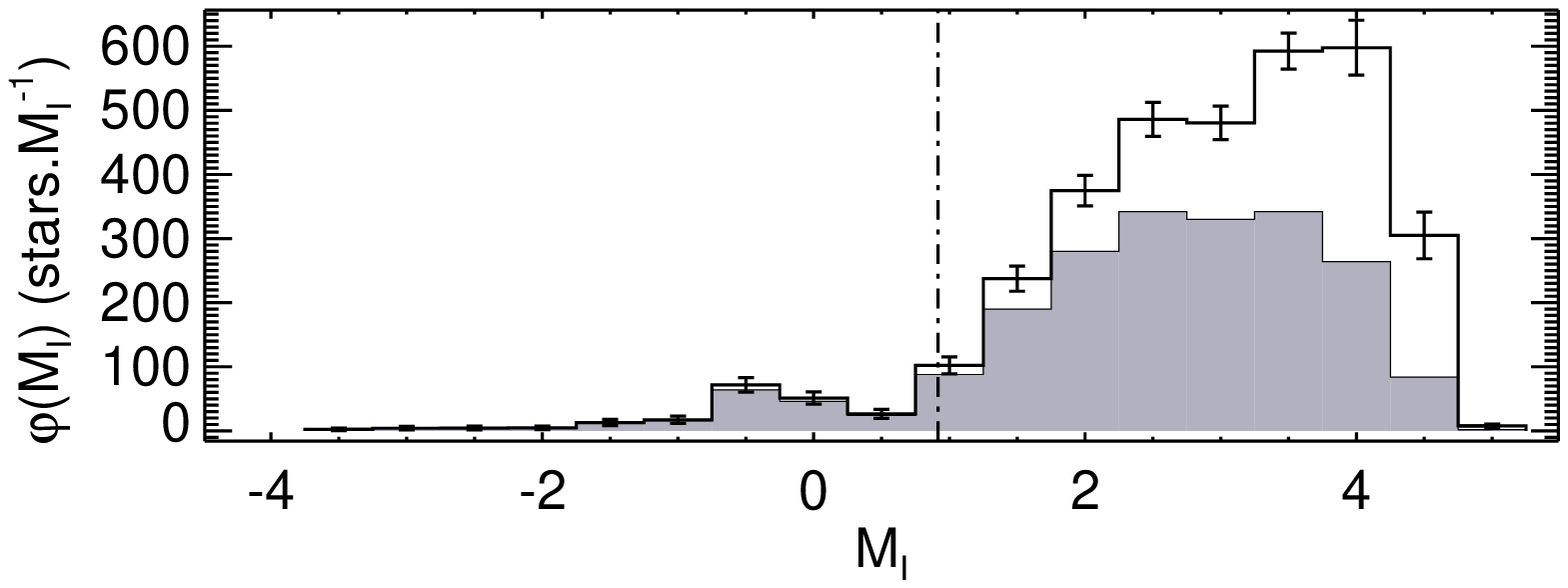} \\
\includegraphics[width=0.22\textwidth]{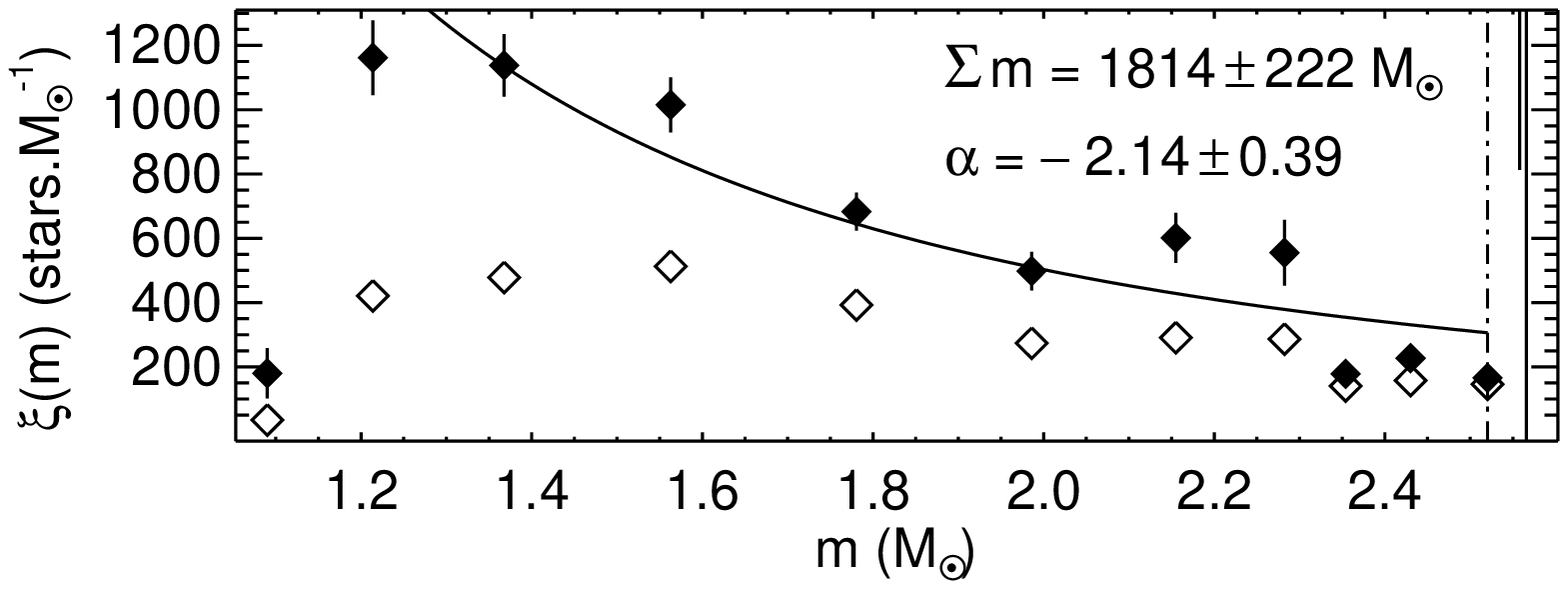} \hskip 0.5cm \includegraphics[width=0.22\textwidth]{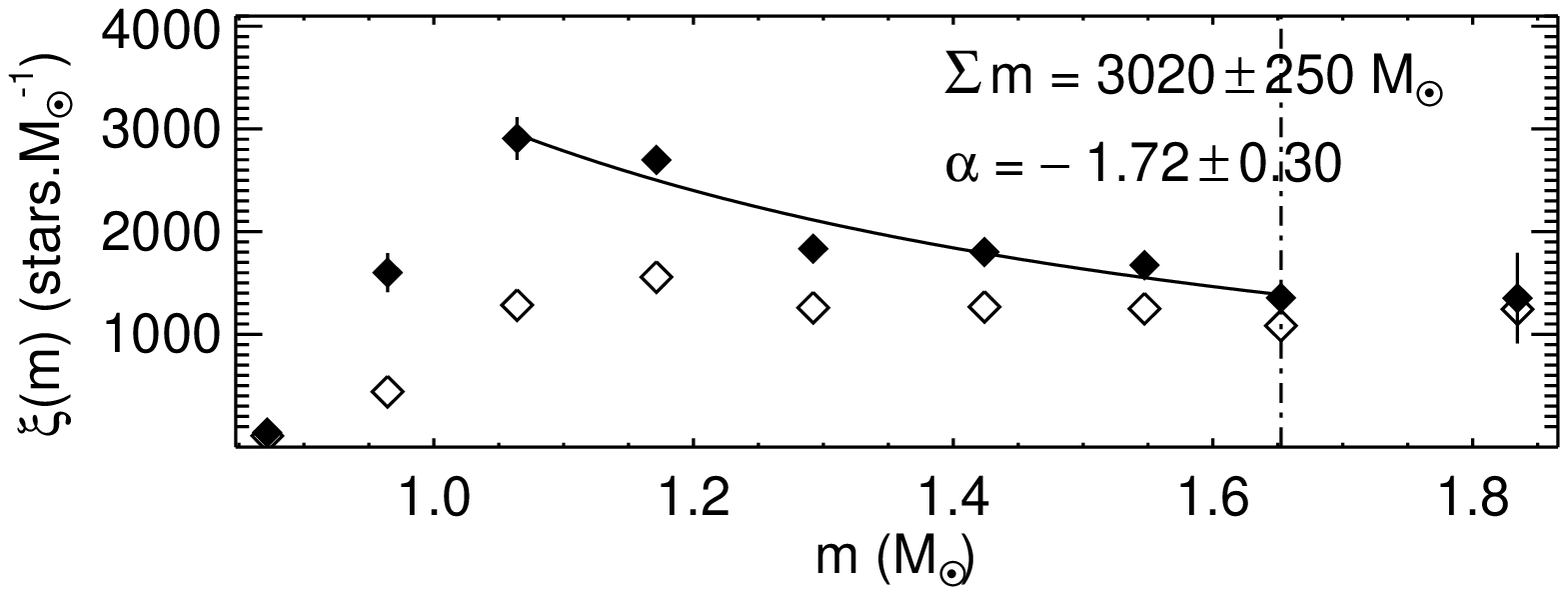} \\
\caption{Observed and completeness corrected LFs (filled and open histograms, respectively) and MFs 
(open and filled symbols, respectively) of LMC clusters. From top left to bottom right: 
LFs (top panels) and MFs (bottom panels) of KMHK228, OHSC3, SL576 and SL61. The vertical dashed lines 
correspond to the turn-offs and the solid lines represent the MF fits.}
\label{fig:MFs_LMC}
\end{figure}

This appendix compiles the figures resulting from the stellar luminosity and mass function derivations
of the studied clusters, as described in Sect.~\ref{sec:analysis_mf}.  
Figs. \ref{fig:MFs_SMC} and \ref{fig:MFs_LMC} show the LFs and the power law fits (Eq.~\ref{eq:mf})
over the resulting cluster MFs for the present sample. Total masses and MF slopes are indicated.

\section{List of clusters observed by the VISCACHA survey.}
\label{sec:clulist}

This appendix lists all the clusters observed by the VISCACHA survey up to the 
2017B observing run. Their names, equatorial coordinates, 
observing dates and location in the Magellanic System, respectively, are shown 
in the columns of Table~\ref{tab:clulist}.

\begin{table}    
\caption{List of observed clusters}
\scriptsize
\begin{tabular}{l c c c c}
\bf Name & \bf RA\,(J2000) & \bf Dec\,(J2000) & \bf Date & \bf Loc. \\ 
 & [h:m:s] & [\,$^\circ$\,:\,\arcmin\,:\,\arcsec\,] & [yyyy-mm-dd] & \\ \hline
SL882       & 06:19:03.2 & -72:23:09.0 & 2015-02-12 & LMC \\
LW458       & 06:19:10.8 & -67:29:37.0 & 2015-02-12 & LMC \\
LW463       & 06:19:45.8 & -71:18:47.0 & 2015-02-13 & LMC \\
LW460       & 06:19:14.7 & -71:43:36.0 & 2015-02-13 & LMC \\
LW459       & 06:19:16.8 & -68:19:39.0 & 2015-02-13 & LMC \\
LW462       & 06:19:39.7 & -72:16:02.0 & 2015-02-13 & LMC \\
KMHK1732    & 06:19:46.4 & -69:47:28.0 & 2015-02-14 & LMC \\
NGC2241     & 06:22:52.4 & -68:55:30.0 & 2015-02-14 & LMC \\
SL883       & 06:19:54.6 & -68:15:09.0 & 2015-02-14 & LMC \\
LW469       & 06:21:33.8 & -72:47:24.0 & 2015-02-23 & LMC \\
OHSC36      & 06:29:40.6 & -70:35:24.0 & 2015-02-23 & LMC \\
SL889       & 06:23:28.4 & -68:59:50.0 & 2015-02-23 & LMC \\
SL897       & 06:33:00.8 & -71:07:40.0 & 2015-02-24 & LMC \\
SL891       & 06:24:48.6 & -71:39:32.0 & 2015-02-24 & LMC \\
SL892       & 06:25:14.3 & -71:06:08.0 & 2015-02-24 & LMC \\
SL28        & 04:44:39.9 & -74:15:36.0 & 2015-12-06 & LMC \\
SL13        & 04:39:41.7 & -74:01:00.0 & 2015-12-06 & LMC \\
LW15        & 04:38:25.4 & -74:27:48.0 & 2015-12-06 & LMC \\
SL788       & 05:55:45.9 & -71:11:30.0 & 2015-12-07 & LMC \\
SL29        & 04:45:12.3 & -75:07:00.0 & 2015-12-07 & LMC \\
LW62        & 04:46:17.4 & -74:09:36.0 & 2015-12-07 & LMC \\
SL36        & 04:46:08.3 & -74:53:18.0 & 2015-12-07 & LMC \\
KMHK1739    & 06:21:02.5 & -71:02:01.0 & 2016-01-10 & LMC \\
SL53        & 04:49:53.4 & -75:37:42.0 & 2016-01-10 & LMC \\
SL61        & 04:50:44.3 & -75:32:00.0 & 2016-01-10 & LMC \\
OHSC1       & 04:52:40.5 & -75:16:36.0 & 2016-01-11 & LMC \\
SL80        & 04:52:21.9 & -74:53:24.0 & 2016-01-11 & LMC \\
SL74        & 04:52:00.4 & -74:50:42.0 & 2016-01-11 & LMC \\
SL886       & 06:21:24.3 & -69:17:56.0 & 2016-01-11 & LMC \\
OHSC2       & 04:53:09.7 & -74:40:54.0 & 2016-01-12 & LMC \\
KMHK228     & 04:53:02.8 & -74:00:14.0 & 2016-01-12 & LMC \\
LW470       & 06:22:23.3 & -72:14:14.0 & 2016-01-12 & LMC \\
SL84        & 04:52:44.4 & -75:04:30.0 & 2016-01-12 & LMC \\
LW472       & 06:23:10.8 & -68:19:08.0 & 2016-01-13 & LMC \\
LW475       & 06:23:22.9 & -70:33:14.0 & 2016-01-13 & LMC \\
SL118       & 04:55:31.6 & -74:40:36.1 & 2016-01-13 & LMC \\
SL890       & 06:23:02.7 & -71:41:11.0 & 2016-01-13 & LMC \\
HW33        & 00:57:23.0 & -70:48:36.0 & 2016-09-24 & SMC \\
BS95-198    & 01:48:00.0 & -73:07:59.9 & 2016-09-24 & SMC \\
HW56        & 01:07:41.2 & -70:56:03.6 & 2016-09-24 & SMC \\
L100        & 01:18:16.0 & -72:00:06.1 & 2016-09-25 & SMC \\
L73         & 01:04:23.7 & -70:21:12.0 & 2016-09-25 & SMC \\
NGC422      & 01:09:35.7 & -71:46:23.0 & 2016-09-25 & SMC \\
HW85        & 01:42:27.3 & -71:16:48.0 & 2016-09-25 & SMC \\
L32         & 00:47:23.3 & -68:55:32.0 & 2016-09-25 & SMC \\
HW38        & 00:59:25.4 & -73:49:01.2 & 2016-09-27 & SMC \\
B94         & 00:58:16.6 & -74:36:28.0 & 2016-09-27 & SMC \\
HW20        & 00:44:48.0 & -74:21:47.0 & 2016-09-27 & SMC \\
HW44        & 01:01:22.0 & -73:47:12.1 & 2016-09-27 & SMC \\
B168        & 01:26:43.0 & -70:46:48.0 & 2016-09-27 & SMC \\
IC1641      & 01:09:36.7 & -71:46:02.8 & 2016-09-27 & SMC \\
L114        & 01:50:19.0 & -74:21:24.1 & 2016-09-28 & SMC \\                
K57         & 01:08:13.8 & -73:15:27.0 & 2016-09-28 & SMC \\
K7          & 00:27:45.2 & -72:46:52.5 & 2016-09-28 & SMC \\
K55         & 01:07:32.6 & -73:07:17.1 & 2016-09-28 & SMC \\
HW67        & 01:13:01.8 & -70:57:47.1 & 2016-09-28 & SMC \\
BS95-75     & 00:54:31.0 & -74:11:06.0 & 2016-11-02 & SMC \\
B1          & 00:19:21.3 & -74:06:24.1 & 2016-11-02 & SMC \\
K6          & 00:25:26.6 & -74:04:29.7 & 2016-11-03 & SMC \\
HW71NW      & 01:15:30.0 & -72:22:36.0 & 2016-11-03 & SMC \\
BS95-187    & 01:31:01.0 & -72:50:48.1 & 2016-11-03 & SMC \\
SL53        & 04:49:54.0 & -75:37:42.0 & 2016-11-03 & LMC \\
L116        & 01:55:33.0 & -77:39:18.0 & 2016-11-04 & SMC \\
KMHK343     & 04:55:55.0 & -75:08:17.0 & 2016-11-04 & LMC \\
L112        & 01:36:01.0 & -75:27:29.9 & 2016-11-04 & SMC \\
SL703       & 05:44:54.0 & -74:50:57.0 & 2016-11-04 & LMC \\
K9          & 00:30:00.3 & -73:22:40.7 & 2016-11-04 & SMC \\
NGC152      & 00:32:56.3 & -73:06:56.6 & 2016-11-05 & SMC \\
AM3         & 23:48:59.0 & -72:56:42.0 & 2016-11-05 & SMC \\
NGC796      & 01:56:44.0 & -74:13:12.0 & 2016-11-05 & SMC \\
L113        & 01:49:30.0 & -73:43:40.0 & 2016-11-05 & SMC \\
HW77        & 01:20:10.0 & -72:37:12.0 & 2016-11-05 & SMC \\
K37         & 00:57:48.5 & -74:19:31.6 & 2016-11-05 & SMC \\
HW5         & 00:31:01.3 & -72:20:30.0 & 2016-11-05 & SMC \\
L114        & 01:50:19.0 & -74:21:24.1 & 2016-11-05 & SMC \\
IC1708      & 01:24:57.3 & -71:10:59.9 & 2016-11-05 & SMC \\
L106        & 01:30:38.0 & -76:03:18.0 & 2016-11-05 & SMC \\
\hline
\label{tab:clulist}
\end{tabular}
\end{table}

\begin{table}    
\contcaption{List of observed clusters}
\scriptsize
\begin{tabular}{l c c c c}
\bf Name & \bf RA\,(J2000) & \bf Dec\,(J2000) & \bf Date & \bf Loc. \\ 
 & [h:m:s] & [\,$^\circ$\,:\,\arcmin\,:\,\arcsec\,] & [yyyy-mm-dd] & \\ \hline
IC2148      & 05:39:12.3 & -75:33:47.0 & 2016-11-30 & LMC \\
SL126       & 04:57:20.0 & -62:32:06.0 & 2016-11-30 & LMC \\
SL192       & 05:02:27.0 & -74:51:51.0 & 2016-11-30 & LMC \\
SL576       & 05:33:13.0 & -74:22:08.0 & 2016-11-30 & LMC \\
SL828       & 06:02:13.0 & -74:11:24.0 & 2016-12-01 & LMC \\
SL835       & 06:04:48.0 & -75:06:09.0 & 2016-12-01 & LMC \\
H4          & 05:32:25.0 & -64:44:11.0 & 2016-12-01 & LMC \\
SL647       & 05:39:35.0 & -75:12:30.0 & 2016-12-02 & LMC \\
SL737       & 05:48:44.0 & -75:44:00.0 & 2016-12-02 & LMC \\
LW141       & 05:07:34.0 & -74:38:06.0 & 2016-12-02 & LMC \\
IC2161      & 05:57:25.0 & -75:08:23.0 & 2016-12-02 & LMC \\
LW75        & 04:50:18.7 & -73:38:55.0 & 2016-12-02 & LMC \\
OHSC4       & 04:59:13.3 & -75:07:58.0 & 2016-12-03 & LMC \\
SL783       & 05:54:39.0 & -74:36:19.0 & 2016-12-03 & LMC \\
OHSC3       & 04:56:36.0 & -75:14:29.0 & 2016-12-03 & LMC \\
NGC1755     & 04:56:55.3 & -70:25:28.0 & 2016-12-03 & LMC \\
SL295       & 05:10:09.0 & -75:32:36.0 & 2016-12-03 & LMC \\
Kron11      & 00:36:27.0 & -72:28:44.0 & 2017-10-20 & SMC \\
Kron16      & 00:40:33.0 & -72:44:23.0 & 2017-10-20 & SMC \\
Kron8       & 00:28:02.0 & -73:18:14.0 & 2017-10-20 & SMC \\
NGC362A     & 01:03:00.0 & -70:51:45.0 & 2017-10-20 & SMC \\
Kron47      & 00:57:47.0 & -74:19:36.0 & 2017-10-20 & SMC \\
Lindsay108  & 01:31:32.0 & -71:57:12.0 & 2017-10-20 & SMC \\
Kron15      & 00:40:13.0 & -72:41:55.0 & 2017-10-20 & SMC \\
BS95-196    & 01:48:02.0 & -70:00:12.0 & 2017-10-20 & SMC \\
NGC643      & 01:35:01.0 & -75:33:26.0 & 2017-10-20 & SMC \\                
ESO51SC9    & 00:58:58.0 & -68:54:54.0 & 2017-10-22 & SMC \\
HW86        & 01:42:22.0 & -74:10:24.0 & 2017-10-22 & SMC \\
HW66        & 01:12:04.0 & -75:11:54.0 & 2017-10-22 & SMC \\
Kron13      & 00:35:42.0 & -73:35:51.0 & 2017-10-22 & SMC \\
Lindsay32   & 00:47:24.0 & -68:55:12.0 & 2017-10-22 & SMC \\
Lindsay93   & 01:12:47.0 & -73:27:58.0 & 2017-10-22 & SMC \\
NGC121      & 00:26:49.0 & -71:31:58.0 & 2017-10-22 & SMC \\
Lindsay109  & 01:33:14.0 & -74:10:00.0 & 2017-10-22 & SMC \\
KMHK19      & 04:37:06.0 & -72:01:11.0 & 2017-12-18 & LMC \\
KMHK6       & 04:32:48.0 & -71:27:30.0 & 2017-12-18 & LMC \\
KMHK44      & 04:43:26.0 & -64:53:05.0 & 2017-12-18 & LMC \\
ESO85SC03   & 04:46:56.0 & -64:50:25.0 & 2017-12-19 & LMC \\
SL2         & 04:24:09.7 & -72:34:13.0 & 2017-12-20 & LMC \\
BSDL1       & 04:39:35.7 & -70:44:47.0 & 2017-12-20 & LMC \\
DES001SC04  & 05:24:30.7 & -64:19:31.0 & 2017-12-20 & LMC \\
KMHK9       & 04:34:55.7 & -68:14:39.0 & 2017-12-20 & LMC \\
KMHK1593    & 06:01:49.0 & -64:07:58.1 & 2017-12-20 & LMC \\
LW7         & 04:35:36.7 & -69:21:46.0 & 2017-12-20 & LMC \\
NGC1629     & 04:29:36.7 & -71:50:18.0 & 2017-12-21 & LMC \\
KMHK15      & 04:36:20.7 & -70:10:22.0 & 2017-12-21 & LMC \\
KMHK3       & 04:29:34.0 & -68:21:22.0 & 2017-12-21 & LMC \\
HS13        & 04:35:28.0 & -67:42:39.0 & 2017-12-21 & LMC \\
LW20        & 04:39:57.3 & -71:37:07.0 & 2017-12-21 & LMC \\
\hline
\end{tabular}
\end{table}


\bsp	
\label{lastpage}
\end{document}